\documentclass[prd,nofootinbib,showpacs,preprint,superscriptaddress]{revtex4}
\pdfoutput=1
\usepackage[T1]{fontenc}
\usepackage{amsmath,amssymb}
\usepackage{epsfig}
\usepackage{subfigure}
\usepackage{float}
\usepackage{graphicx}
\usepackage[usenames,dvipsnames]{color}
\usepackage{slashed}
\usepackage{multirow}
\usepackage[colorlinks,citecolor=blue]{hyperref}
\usepackage{pdfpages}
\usepackage{color}

\begin{document}
\title{Type III Seesaw for Neutrino Masses in $U(1)_{B-L}$ Model with Multi-component Dark Matter}
\author{Anirban Biswas}
\email{anirban.biswas.sinp@gmail.com}
\affiliation{School of Physical Sciences, Indian Association for the
Cultivation of Science, 2A $\&$ 2B Raja S.C. Mullick Road,
Kolkata 700032, India}
\author{Debasish Borah}
\email{dborah@iitg.ac.in}
\affiliation{Department of Physics, Indian Institute of Technology
Guwahati, Assam 781039, India}
\author{Dibyendu Nanda}
\email{dibyendu.nanda@iitg.ac.in}
\affiliation{Department of Physics, Indian Institute of Technology
Guwahati, Assam 781039, India}
\begin{abstract}
We propose a $B-L$ gauged extension of the Standard Model where light neutrino masses arise from type III seesaw mechanism. Unlike the minimal $B-L$ model with three right handed neutrinos having unit lepton number each, the model with three fermion triplets is however not anomaly free. We show that the leftover triangle anomalies can be cancelled by two neutral Dirac fermions having fractional $B-L$ charges, both of which are naturally stable by virtue of a remnant $\mathbb{Z}_2 \times \mathbb{Z}'_2$ symmetry, naturally leading to a two component dark matter scenario without any ad-hoc symmetries. We constrain the model from all relevant phenomenological constraints including dark matter properties. Light neutrino mass and collider prospects are also discussed briefly. Due to additional neutral gauge bosons, the fermion triplets in type III seesaw can have enhanced production cross section in collider experiment. 

\end{abstract}
\pacs{12.60.Fr,12.60.-i,14.60.Pq,14.60.St}
\maketitle

\section{Introduction}
Origin of light neutrino mass and dark matter (DM) in the Universe have been well known mysteries in particle physics for many decades \cite{Tanabashi:2018oca}. While two mass squared differences and three mixing angles in the neutrino sector are very well measured \cite{Esteban:2018azc}, evidence suggests that DM gives rise to around $26\%$ of the present Universe's energy density. In terms of density 
parameter $\Omega_{\rm DM}$ and $h = \text{Hubble Parameter}/(100 \;\text{km} ~\text{s}^{-1} 
\text{Mpc}^{-1})$, the present DM abundance is conventionally reported as \cite{Aghanim:2018eyx}:
$\Omega_{\text{DM}} h^2 = 0.120\pm 0.001$
at 68\% CL. Apart from this evidence from cosmology experiments like Planck, there have been plenty of astrophysical evidences accumulated for several decades. Among them, the galaxy cluster observations by Fritz Zwicky \cite{Zwicky:1933gu} in 1933, 
observations of galaxy rotation curves in 1970's by Rubin and collaborators\cite{Rubin:1970zza}, the observation of the bullet cluster 
by Chandra observatory \cite{Clowe:2006eq} along with several galaxy survey experiments which map the distribution of such matter based on their gravitational lensing effects are noteworthy. While all these evidences are based on purely gravitational interactions of dark matter, there have been significant efforts in hunting for other types of interactions, possibly of weak interaction type, at several dark matter direct detection experiments. Such interactions are typical of weakly interacting massive particle (WIMP) dark matter, the most widely studied beyond standard model (BSM) framework for accommodating dark matter. Here a DM candidate typically with electroweak (EW) scale mass and interaction rate similar to EW interactions can give rise to the correct DM relic abundance, a remarkable coincidence often referred to as the \textit{WIMP Miracle}~\cite{Kolb:1990vq}. The very same interactions could also give rise to dark matter nucleon scattering at an observable rate. However, experiments like LUX \cite{Akerib:2016vxi}, PandaX-II \cite{Tan:2016zwf, Cui:2017nnn} and XENON1T \cite{Aprile:2017iyp, Aprile:2018dbl} have so far produced only negative results putting stringent upper limits on DM interactions with the standard model (SM) particles like quarks. Next generation direct detection experiments like LZ~\cite{Akerib:2015cja}, XENONnT~\cite{Aprile:2015uzo}, DARWIN~\cite{Aalbers:2016jon} and PandaX-30T~\cite{Liu:2017drf} are going to probe DM interaction rates very close to the rates at which coherent neutrino-nucleus elastic scattering takes place beyond which it will be extremely difficult to distinguish the neutrino or DM origin of nuclear recoil events. Similar null results have been reported by collider experiments like the large hadron collider (LHC) \cite{Kahlhoefer:2017dnp} as well as indirect detection experiments putting stricter upper limits on DM annihilation to standard model (SM) particles \cite{Ahnen:2016qkx}, specially the charged ones which can finally lead to excess of gamma rays for WIMP type DM. Though the null results reported by these experiments do not rule out all the parameter space for a single particle WIMP type DM, it may be hinting at a much richer dark sector. Though a single component DM is a very minimal and predictive scenario to begin with, a richer dark sector may in fact be natural given the complicated visible sector. There have been several proposals for 
multi-component WIMP dark matter during last few years, some of which can be found in \cite{Cao:2007fy, Zurek:2008qg, Chialva:2012rq, Heeck:2012bz, Biswas:2013nn, Bhattacharya:2013hva, Bian:2013wna, Bian:2014cja, Esch:2014jpa, Karam:2015jta, Karam:2016rsz, DiFranzo:2016uzc, Bhattacharya:2016ysw, DuttaBanik:2016jzv, Klasen:2016qux, Ghosh:2017fmr, Ahmed:2017dbb, Bhattacharya:2017fid, Ahmed:2017dbb, Borah:2017hgt, Bhattacharya:2018cqu, Bhattacharya:2018cgx, Aoki:2018gjf, DuttaBanik:2018emv, Barman:2018esi, YaserAyazi:2018lrv, Poulin:2018kap, Chakraborti:2018lso, Chakraborti:2018aae, Bernal:2018aon, Elahi:2019jeo, Borah:2019epq, Borah:2019aeq, Bhattacharya:2019fgs}. Going from single component to multi-component DM sector can significantly alter the direct as well as indirect detection rates for DM. Since direct and indirect detection (considering annihilations only, for stable DM) rates of DM are directly proportional to the DM density and DM density squared respectively, multi-component DM models can find larger allowed region of parameter space by appropriate tuning of their relative abundance. In addition to that, such multi-component DM scenarios often give rise to very interesting signatures at direct and indirect detection experiments \cite{Profumo:2009tb, Fukuoka:2010kx, Cirelli:2010nh, Aoki:2012ub, Aoki:2013gzs, Modak:2013jya, Geng:2013nda, Gu:2013iy, Biswas:2013nn, Aoki:2014lha, Geng:2014zea, Geng:2014dea,  Biswas:2015sva, Borah:2015rla, Borah:2016ees,  DuttaBanik:2016jzv, Borah:2017xgm, Borah:2017hgt, Herrero-Garcia:2017vrl, Herrero-Garcia:2018mky, Profumo:2019pob}.

Similar to the BSM proposals of DM, there have been plenty of ways proposed so far, in order to accommodate light neutrino masses and mixing. While the SM can not generate light neutrino mass at renormalisable level, one can generate a tiny Majorana mass for the neutrinos from the SM Higgs field (H) through the non-renormalisable dimension five Weinberg operator \cite{Weinberg:1979sa} $(L L H H)/\Lambda, L \equiv$ lepton doublet, $\Lambda \equiv$ unknown cut-off scale. Dynamical ways to generate such an operator are classified as seesaw mechanism where one or more heavy fields are responsible for tiny light neutrino mass, resulting from the seesaw between electroweak scale and the scale of heavy fields. Popular seesaw models are categorised as type I seesaw \cite{Minkowski:1977sc, GellMann:1980vs, Mohapatra:1979ia, Schechter:1980gr}, type II seesaw \cite{Mohapatra:1980yp, Lazarides:1980nt, Wetterich:1981bx, Schechter:1981cv, Brahmachari:1997cq}, type III seesaw \cite{Foot:1988aq} and so on. However, none of these models and their predictions have found any experimental verifications at ongoing experiments like the LHC \cite{Cai:2017mow}. In view of this, it is not really outlandish to consider frameworks where DM and neutrino finds common origin opening up the possibility to probe such models at different frontiers with more observables. One popular scenario that is built upon such objectives is the scotogenic framework, originally proposed by Ma \cite{Ma:2006km}, where particles odd under an unbroken $\mathbb{Z}_2$ sector take part in radiative generation of light neutrino masses while the lightest $\mathbb{Z}_2$ odd particle is the DM candidate. The same formalism was extended to two component DM scenario by the authors of \cite{Aoki:2017eqn}. While multiple copies of $\mathbb{Z}_2$ symmetries remain ad-hoc in these models, one could find their UV completion with additional gauge symmetries. For example, in a recent work \cite{Bernal:2018aon}, two component fermion DM was proposed as a new anomaly free gauged $B-L$ model, where $B$ and $L$ correspond to baryon and lepton numbers respectively \footnote{Please see \cite{Ellis:2017tkh, FileviezPerez:2019jju} and references therein for general approach to build anomaly free Abelian gauge models with DM.}. The neutrino mass in this model originate from just two right handed neutrinos, often referred to as the littlest seesaw (LS) model \cite{King:2015dvf}, leading to a vanishing lightest neutrino mass. While TeV scale minimal type I or LS model has limited scope of being verified at experiments like the LHC due to the gauge singlet nature of additional fermions, presence of new gauge symmetries like $U(1)_{B-L}$. Motivated by this, we consider another interesting possibility, within a gauged $B-L$ model, where light neutrino masses arise from a type III seesaw scenario, along with the possibility of a two component fermion DM naturally arising as a possible way to make the model anomaly free. While implementing type I seesaw with three or two right handed neutrinos is a matter of choice (as both are allowed due to the unknown lightest neutrino mass) with the former not requiring any additional fermions (possible DM candidates) for anomaly cancellation, the implementation of type III seesaw in a gauged $B-L$ model inevitably brings in new anomalies due to the non-trivial transformation of fermion triplet fields under the $SU(2)_L$ gauge symmetry of the SM. Out of several possible solutions to the anomaly cancellation conditions, we discuss one possible scenario which naturally leads to a two component fermion DM scenario. The $U(1)_{B-L}$ gauge boson apart from dictating the relic abundance of two component DM, also enhances the production cross section of heavy fermion triplets in proton proton collisions at the LHC due to the on-shell nature of mediating neutral gauge boson. We constrain the model parameters from all relevant constraints from DM sector as well as collider bounds and present the available parameter space.

This paper is organised as follows. In section \ref{sec0}, we give an overview of gauged $B-L$ model with different solutions to anomaly conditions including the one we choose to discuss in details in this work. In section \ref{model}, we discuss our model in details followed by section \ref{sec:constraints} where we mention different existing constraints on model parameters. In section \ref{sec:boltzmann} we discuss in details the coupled Boltzmann equations for two component dark matter in our model followed by section \ref{sec:dd} where we mention briefly the details of dark matter direct detection. In section \ref{sec:results}, we discuss our results related to dark matter relic abundance and constraints on the model parameters from all our requirements and constraints. We briefly mention some interesting LHC signatures of our model in section \ref{sec:LHC} and then finally conclude in section \ref{sec:conclude}.

\section{Gauged $B-L$ Symmetry}
\label{sec0}
Gauged $B-L$ symmetric extension of the SM \cite{Davidson:1978pm, Mohapatra:1980qe, Marshak:1979fm, Masiero:1982fi, Mohapatra:1982xz, Buchmuller:1991ce} is one of the most popular and very well motivated BSM frameworks that has been studied quite well for a long time. Since the SM already has an accidental and global $B-L$ symmetry at renormalisable level, it is very straightforward as well as motivating to uplift it to a gauge symmetry and study the consequences. Compared to an arbitrary Abelian gauge extension of the SM, gauged
$B-L$ symmetry emerges as a very natural and minimal possibility as the corresponding charges of all the SM fields under this new symmetry are well known. However, a $U(1)_{B-L}$ gauge symmetry with only the SM fermions is not anomaly free. This is because the triangle anomalies for both $U(1)^3_{B-L}$ and the mixed $U(1)_{B-L}-(\text{gravity})^2$ diagrams are non vanishing. These triangle anomalies for the SM fermion content are given as
\begin{align}
\mathcal{A}_1 \left[ U(1)^3_{B-L} \right] = \mathcal{A}^{\text{SM}}_1 \left[ U(1)^3_{B-L} \right]=-3\,,  \nonumber \\
\mathcal{A}_2 \left[(\text{gravity})^2 \times U(1)_{B-L} \right] = \mathcal{A}^{\text{SM}}_2 \left[ (\text{gravity})^2 \times U(1)_{B-L} \right]=-3\,.
\end{align}
Remarkably, if three right handed neutrinos are added to the model, they contribute $\mathcal{A}^{\text{New}}_1 \left[ U(1)^3_{B-L} \right] = 3, \mathcal{A}^{\text{New}}_2 \left[ (\text{gravity})^2 \times U(1)_{B-L} \right] = 3$ leading to vanishing total of triangle anomalies. This is the most natural and economical $U(1)_{B-L}$ model where the fermion sector has three right handed neutrinos apart from the usual SM fermions and it has been known for a long time. However, there exist non-minimal ways of constructing anomaly free versions of $U(1)_{B-L}$ model. For example, it has been known for a few years that three right handed neutrinos with exotic $B-L$ charges $5, -4, -4$ can also give rise to vanishing triangle anomalies \cite{Montero:2007cd}. It is clear to see how the anomaly cancels, as follows
\begin{align}
\mathcal{A}_1 \left[ U(1)^3_{B-L} \right] = \mathcal{A}^{\text{SM}}_1 \left[ U(1)^3_{B-L} \right]+\mathcal{A}^{\text{New}}_1 \left[ U(1)^3_{B-L} \right]=-3 + \left [ -5^3 - (-4)^3 - (-4)^3 \right]=0\,, \nonumber
\end{align}
\begin{align}
\mathcal{A}_2 \left[(\text{gravity})^2 \times U(1)_{B-L} \right] & = \mathcal{A}^{\text{SM}}_2 \left[ (\text{gravity})^2 \times U(1)_{B-L} \right]+ \mathcal{A}^{\text{New}}_2 \left[ (\text{gravity})^2 \times U(1)_{B-L} \right]\,, \nonumber \\
&=-3 + \left[ -5 - (-4) - (-4) \right]=0\,.
\end{align}
This model was also discussed recently in the context of neutrino mass \cite{Ma:2014qra, Ma:2015mjd} and DM \cite{Sanchez-Vega:2014rka, Sanchez-Vega:2015qva, Singirala:2017see, Nomura:2017vzp, Okada:2018tgy} by several groups. Another solution to the anomaly cancellation conditions with irrational $B-L$ charges of new fermions was proposed by the authors of \cite{Wang:2015saa} where both DM and neutrino mass can have a common origin through radiative linear seesaw.

Very recently, another anomaly free $U(1)_{B-L}$ framework was proposed where the additional right handed fermions possess more exotic $B-L$ charges namely, $-4/3, -1/3, -2/3, -2/3$ \cite{Patra:2016ofq}. The triangle anomalies get cancelled as follows.
\begin{align}
\mathcal{A}_1 \left[ U(1)^3_{B-L} \right] &= \mathcal{A}^{\text{SM}}_1 \left[ U(1)^3_{B-L} \right]+\mathcal{A}^{\text{New}}_1 \left[ U(1)^3_{B-L} \right] \nonumber \\
&=-3 + \left [ -(-4/3)^3 - (-1/3)^3 - (-2/3)^3-(-2/3)^3 \right]=0 \nonumber
\end{align}
\begin{align}
\mathcal{A}_2 \left[(\text{gravity})^2 \times U(1)_{B-L} \right] & = \mathcal{A}^{\text{SM}}_2 \left[ (\text{gravity})^2 \times U(1)_{B-L} \right]+ \mathcal{A}^{\text{New}}_2 \left[ (\text{gravity})^2 \times U(1)_{B-L} \right] \nonumber \\
&=-3 + \left[ -(-4/3) - (-1/3) - (-2/3) - (-2/3) \right]=0
\end{align}
These four chiral fermions constitute two Dirac fermion mass eigenstates, the lighter of which becomes the DM candidate having either thermal \cite{Patra:2016ofq} or non-thermal origins \cite{Biswas:2016iyh}. The light neutrino mass in this model had its origin from a variant of type II seesaw mechanism and hence remained disconnected to the anomaly cancellation conditions. In a follow up work by the authors of \cite{Nanda:2017bmi}, these fermions with fractional charges were also responsible for generating light neutrino masses at one loop level. One can have even more exotic right handed fermions with $B-L$ charges $-17/3, 6, -10/3$ so that the triangle anomalies cancel as
\begin{align}
\mathcal{A}_1 \left[ U(1)^3_{B-L} \right] &= \mathcal{A}^{\text{SM}}_1 \left[ U(1)^3_{B-L} \right]+\mathcal{A}^{\text{New}}_1 \left[ U(1)^3_{B-L} \right] \,,\nonumber \\
&=-3 + \left [ -(-17/3)^3 - (6)^3 - (-10/3)^3 \right]=0 \,,\nonumber
\end{align}
\begin{align}
\mathcal{A}_2 \left[(\text{gravity})^2 \times U(1)_{B-L} \right] & = \mathcal{A}^{\text{SM}}_2 \left[ (\text{gravity})^2 \times U(1)_{B-L} \right]+ \mathcal{A}^{\text{New}}_2 \left[ (\text{gravity})^2 \times U(1)_{B-L} \right] \nonumber \,,\\
&=-3 + \left[ -(-17/3) - (6) - (-10/3)  \right]=0\,.
\end{align}
In the recent work on $U(1)_{B-L}$ gauge symmetry with two component DM \cite{Bernal:2018aon}, the authors considered two right handed neutrinos with $B-L$ number -1 each so that the model still remains anomalous. The remaining anomalies were cancelled by four chiral fermions with fractional $B-L$ charges leading to two Dirac fermion mass eigenstates both of which are stable and hence DM candidates.

To implement type III seesaw in $U(1)_{B-L}$ model, we consider two copies $(n_{\Sigma}=2)$ of fermion triplets $\Sigma_{Ri} (i=1,2)$ into the model having quantum numbers $(1, 3, 0, -1)$ under $SU(3)_c, SU(2)_L, U(1)_Y$ and $U(1)_{B-L}$ gauge groups respectively. Note that, two copies are enough to satisfy the light neutrino data, as in LS models. The non-vanishing anomalies are
\begin{align}
[SU(2)_L]^2 U(1)_{B-L} = 2 n_{\Sigma}=4, \;\; [U(1)_{B-L}]^3=-3+3n_{\Sigma}=3, \;\; [U(1)_{B-L}]=-3+3n_{\Sigma}=3.
\end{align}
Note that the first anomaly is arising only due to the $SU(2)_L$ transformation of newly introduced fermions and was absent in the minimal $B-L$ extension of SM. Obviously, the first anomaly can be cancelled only by introducing an additional field which has non-trivial transformation under both $SU(2)_L$ and $U(1)_{B-L}$. We introduce such additional fields with a goal to keep our setup minimal and connected to the origin of light neutrino mass. Introducing a quintuplet $\Psi (1, 4, 0, n_1)$, the first anomaly becomes
$$ [SU(2)_L]^2 U(1)_{B-L} = 2 n_{\Sigma}-5n_1$$
which can vanish if $n_1 = 2n_{\Sigma}/5=4/5$. The other anomalies can be cancelled by introducing additional fields which do not contribute to the first anomaly and hence $SU(2)_L$ singlets. The remaining anomalies are 
\begin{align}
[U(1)_{B-L}]^3=-3+3n_{\Sigma}-4 n^3_1=\frac{119}{125}, \;\; [U(1)_{B-L}]=-3+3n_{\Sigma}-4 n_1=-\frac{1}{5}.
\end{align}
These remaining anomalies can be cancelled by introducing three $SU(2)_L$ singlet chiral fermions 
\begin{align}
N_{1L} (1,1,0, -\frac{7}{5}), N_{2L} (1,1,0, \frac{2}{5}), N_{3L} (1,1,0,\frac{6}{5}).
\end{align}
This can be seen as follows
$$ [U(1)_{B-L}]^3=-3+3n_{\Sigma}-4 n^3_1 - \left(\frac{7}{5}\right)^3+\left(\frac{2}{5}\right)^3+\left(\frac{6}{5}\right)^3=\frac{119}{125}-\frac{119}{125}=0,$$
$$[U(1)_{B-L}]=-3+3n_{\Sigma}-4 n_1-\frac{7}{5}+\frac{2}{5}+\frac{6}{5}=-\frac{1}{5}+\frac{1}{5}=0.$$
Since the quintuplet and the other singlet fermions have no role to play in generating light neutrino mass, we do not pursue this possibility further.

We can have three fermion triplets: two of them having $B-L$ charge $-1$ and the third having exotic charge $n_1$. We will check if the third fermion can have any possible role in generating light neutrino masses. In such a case, there arises a possibility to get vanishing $[SU(2)_L]^2 U(1)_{B-L}$ anomaly. In this case, 
$$ [SU(2)_L]^2 U(1)_{B-L} = 2 n_{\Sigma}-2n_1$$
which can vanish if $n_1 = n_{\Sigma}=2$. The other anomalies can be cancelled by introducing additional fields which do not contribute to the first anomaly and hence $SU(2)_L$ singlets. The remaining anomalies are 
\begin{align}
[U(1)_{B-L}]^3=-3+3n_{\Sigma}-3 n^3_1=-21, \;\; [U(1)_{B-L}]=-3+3n_{\Sigma}-3 n_1=-3.
\end{align}
We now consider different possible solutions to these anomalies one by one.

\textbf{Solution 2}: \\
The remaining anomalies mentioned above can be cancelled by introducing four $SU(2)_L$ singlet chiral fermions 
\begin{align}
N_{1R} (1,1,0, 2), N_{2R} (1,1,0, -1), N_{3R} (1,1,0,-1), N_{4R} (1,1,0,-3).
\end{align}
This can be seen as follows
$$ [U(1)_{B-L}]^3=-3+3n_{\Sigma}-3 n^3_1 - \left(2 \right)^3-2\left(-1\right)^3-\left(-3\right)^3=-21+21=0,$$
$$[U(1)_{B-L}]=-3+3n_{\Sigma}-3 n_1-2+1+1+3=-3+3=0.$$
However, this solution is not very motivating owing to the existence of singlet fermions having $B-L$ charge $-1$, which will give type I seesaw contribution to light neutrino masses, already discussed by several earlier works.

\textbf{Solution 3}: \\
The remaining anomalies can also be cancelled by the following fermions with fractional $B-L$ charges:
\begin{align}
N_{1R} (1,1,0, \frac{2}{3}), N_{2R} (1,1,0, \frac{1}{3}), N_{3R} (1,1,0, -\frac{4}{3}), N_{4R} (1,1,0,-\frac{8}{3}).
\end{align}
In order to have non-zero masses for all new fermions and sticking to minimal scalar contents having integer $B-L$ charges, we find that there can be one stable dark matter candidate, in terms of one of the singlet fermions. Since single component dark matter in such models have already been discussed in several works, we do not pursue this possibility further.

\textbf{Solution 4}: \\
The most interesting possibility is the solution 
\begin{align}
N_{1R} (1,1,0, \frac{7}{5}), N_{2R} (1,1,0, -\frac{2}{5}), N_{3R} (1,1,0, -\frac{6}{5}), N_{4R} (1,1,0,-\frac{14}{5}).
\end{align}
which can also be recast as 
\begin{align}
N_{1L} (1,1,0, -\frac{7}{5}), N_{1R} (1,1,0, -\frac{2}{5}), N_{2L} (1,1,0, \frac{6}{5}), N_{2R} (1,1,0,-\frac{14}{5}).
\end{align}
We can construct two Dirac fermions from these four chiral ones, just by introducing two singlet scalars $\phi_1, \phi_2$ having $B-L$ charges $1, 4$ respectively. The corresponding mass terms will be 
$$ Y_1 \overline{N_{1L}} N_{1R} \phi^{\dagger}_1 + Y_2 \overline{N_{2L}} N_{2R} \phi_2 + \text{h.c.}$$
Another scalar singlet having $B-L$ charge $2$ is also required in order to give mass to the fermion triplets $\Sigma_{1,2}$. The third fermion triplet acquires mass and also mixes with the other two fermion triplets by virtue of its couplings to the scalars $\phi_1, \phi_2$. Although the third fermion triplet is not stable due to its mixing with the first two, the two Dirac fermions constructed above can be separately stable and hence give rise to multi-component dark matter. Also, the third fermion triplet, through its mixing with the first two, can contribute to the light neutrino mass matrix as we discuss in details in upcoming sections. Due to these interesting possibilities, we pursue this scenario in our work.

\section{The Model}
\label{model}
In this section we have discussed our model elaborately. As mentioned earlier,
in this work our prime motivation is to have a multicomponent dark matter
scenario where both dark matter candidates are stable by virtue of a single
symmetry group. Additionally, we want to address neutrino mass generation as
well. Keeping these two things in our mind, we have extended the SM in all three
sectors namely the gauge sector, the fermionic sector and the scalar sector.
In the gauge sector, we have demanded an additional local $U(1)_{B-L}$
gauge invariance where $B$ and $L$ are denoting baryon and lepton numbers
respectively of a particular field. This introduces anomalies (both axial vector
and gauge-gravitational anomalies) in the theory which can only be evaded by
the inclusion of additional fermionic degrees of freedom. This has elaborately
been discussed in the previous section. We have seen that for the case
of three additional $SU(2)_L$ triplet fermions $\Sigma_{i\:\!R}$, $i=1$ to 3,
two of which with $B-L$ charge $-1$ are necessary to generate neutrino masses
via Type-III seesaw mechanism and the rest having $B-L$ charge 2
is solely required to cancel $\left[SU(2)_L\right]^2\,U(1)_{B-L}$
anomaly. We further need four SM gauge singlet chiral fermions with fractional
$B-L$ charges to cancel both $U(1)^3_{B-L}$ and (gravity)$^2\,U(1)_{B-L}$
anomalies. Moreover, at least three scalar fields $\phi_i$ ($i=1$ to 3) are
also necessary to give masses to all new fermions in the broken
phase of the $U(1)_{B-L}$ symmetry. We have properly adjusted
$B-L$ charges of these $\phi_i$s such that only the Dirac mass terms
among these singlet fermions are possible and more importantly
the Dirac mass matrix is diagonal. This results in two physical
Dirac fermions out of these four chiral fermions which are
simultaneously stable and thus both can be viable dark matter
candidates. In Tables \ref{tab:1} and \ref{tab:2}, we have
listed all fermions as well as scalar fields (including the SM ones)
of the present model and their charges under the $SU(3)_c \times SU(2)_L
\times U(1)_Y \times U(1)_{B-L}$ symmetry.   
\begin{table}
\begin{center}
\begin{tabular}{|c|c|}
\hline
Particles & $SU(3)_c \times SU(2)_L \times U(1)_Y \times U(1)_{B-L} $   \\
\hline
$q_L=\begin{pmatrix}u_{L}\\
d_{L}\end{pmatrix}$ & $(3, 2, \frac{1}{6}, \frac{1}{3})$  \\
$u_R$ & $(3, 1, \frac{2}{3}, \frac{1}{3})$  \\
$d_R$ & $(3, 1, -\frac{1}{3}, \frac{1}{3})$  \\

$\ell_L=\begin{pmatrix}\nu_{L}\\
e_{L}\end{pmatrix}$ & $(1, 2, -\frac{1}{2}, -1)$  \\
$e_R$ & $(1, 1, -1, -1)$ \\
\hline
$\Sigma_{1\:\!R}$ & $(1, 3, 0, -1)$ \\
$\Sigma_{2\:\!R}$ & $(1, 3, 0, -1)$ \\
$\Sigma_{3\:\!R}$ & $(1, 3, 0, 2)$ \\
\hline
$N_{1L}$ & $(1, 1, 0, -\frac{7}{5})$ \\
$N_{1R}$ & $(1, 1, 0, -\frac{2}{5})$ \\
$N_{2L}$ & $(1, 1, 0, \frac{6}{5})$ \\
$N_{2R}$ & $(1, 1, 0, -\frac{14}{5})$ \\
\hline
\end{tabular}
\end{center}
\caption{Fermionic fields of the present Model including
the SM fermions.}
\label{tab:1}
\end{table}
\begin{table}
\begin{center}
\begin{tabular}{|c|c|}
\hline
Particles & $SU(3)_c \times SU(2)_L \times U(1)_Y \times U(1)_{B-L} $   \\
\hline
$H=\begin{pmatrix}H^+\\
H^0\end{pmatrix}$ & $(1,2,\frac{1}{2},0)$  \\
\hline
$\phi_1$ & $(1, 1, 0, 1)$ \\
$\phi_2$ & $(1, 1, 0, 4)$ \\
$\phi_3$ & $(1, 1, 0, 2)$ \\
\hline
\end{tabular}
\end{center}
\caption{Scalar fields and their corresponding charges under all
the symmetry groups.}
\label{tab:2}
\end{table}

The Lagrangian of our present model invariant under the full
symmetry group is given by
\begin{eqnarray}
\mathcal{L}&=&\mathcal{L}_{SM} -\frac{1}{4} {B^{\prime}}_{\alpha \beta}
\,{B^{\prime}}^{\alpha \beta} + \mathcal{L}_{scalar} 
+ \mathcal{L}_{fermion}\;.
\label{LagT}
\end{eqnarray}
Here, $\mathcal{L}_{SM}$ denotes the SM Lagrangian involving quarks,
gluons, charged leptons, left handed neutrinos and electroweak gauge
bosons. The second term is the kinetic term of $B-L$ gauge boson ($Z_{BL}$)
expressed in terms of field strength tensor ${B^\prime}^{\alpha\beta}=
\partial^{\alpha}Z_{BL}^{\beta}-\partial^{\beta}Z_{BL}^{\alpha}$.  
From Table \ref{tab:2}, we have already understood that our
model has a very rich scalar sector and the gauge invariant interactions
among the scalar fields  are described by $\mathcal{L}_{scalar}$ which
contains following terms,
\begin{eqnarray}
\mathcal{L}_{scalar} &=& \left({D_{H}}_{\mu} H \right)^\dagger
\left({D_{H}}^{\mu} H \right) + 
\sum_{i=1}^3 \left({D_{\phi_i}}_{\mu} \phi_i \right)^\dagger
\left({D_{\phi_i}}^{\mu}\,\phi_i \right)
-\Bigg[-\mu_{H}^2 (H^\dagger H) + \lambda_{H} (H^\dagger H)^2
\nonumber \\ && 
+ \sum_{i=1}^{3}\bigg( -\mu_{\phi_{i}}^2 (\phi_{i}^\dagger \phi_{i}) 
+ \lambda_{\phi_i} (\phi_{i}^\dagger \phi_{i})^2\bigg)
+ \sum_{i,j=1(i\neq j)}^{3} \lambda_{\phi_{i}\phi_{j}}
(\phi_{i}^\dagger \phi_{i}) (\phi_{j}^\dagger \phi_{j})
\nonumber \\ && 
+ \sum_{i=1}^{3}\lambda_{H\phi_{i}}(H^\dagger H)(\phi_{i}^\dagger \phi_{i})
+ \bigg(\beta\,\phi_{1} \phi_{1}\phi_{3}\phi_{2}^{\dagger}
+ \delta\,\phi_{1}\phi_{1}\phi_{3}^\dagger 
+\,\,\zeta\,\phi_{3} \phi_{3} \phi_{2}^{\dagger} + h.c. \bigg)\Bigg]\,,
\label{Sca:pot}
\end{eqnarray}
where covariant derivatives for the Higgs doublet $H$
and singlet scalars $\phi_i$s are defined as
\begin{eqnarray}
{D_{H}}_{\mu}\,H &=& \left(\partial_{\mu} + i\,\dfrac{g}{2}\,\sigma_a\,W^a_{\mu}
+ i\,\dfrac{g^\prime}{2}\,B_{\mu}\right)H \,, \nonumber \\
{D_{\phi}}_{\mu}\,\phi_i &=& \left(\partial_{\mu} + i\,g_{BL}\,n_{\phi_i}
{Z_{BL}}_{\mu}\right)\phi_i\,.
\end{eqnarray}
The quantity ${D_H}_{\mu}$ is the usual covariant derivative of
the SM Higgs doublet with $g$ and $g^\prime$ are
gauge couplings of $SU(2)_L$ and $U(1)_Y$ respectively
and the corresponding gauge bosons are denoted by $W^a_{\mu}$
($a=1,\,3$) and $B_{\mu}$. The covariant derivative of $H$ does not include
$B-L$ gauge boson $Z_{BL}$ as the corresponding gauge
charge of $H$ is zero. On the other hand, being a
SM gauge singlet, covariant derivative of $\phi_i$ only
contains $Z_{B-L}$ with $n_{\phi_i}$ is the respective $B-L$
charge of $\phi_i$ and $g_{BL}$ is the new gauge coupling.
After breaking of both $B-L$ symmetry and electroweak symmetry
by the VEVs of $H$ and $\phi_i$s, the doublet and all three singlets
are given by
\begin{eqnarray}
H=\begin{pmatrix}H^+\\
\dfrac{h^{\prime} + v + i z}{\sqrt{2}}\end{pmatrix}\,,\,\,\,\,\,\,
\phi_i = \dfrac{s^{\prime}_i +u_i+ A^{\prime}_i}{\sqrt{2}}\,\,\,\,(i=1,\,2,\,3)\,\,,
\label{H&phi_broken_phsae}
\end{eqnarray}
where $v$ and $u_i$s ($i=1,\,2,\,3$) are VEVs of $H$
and $\phi_i$s respectively. For calculational simplicity
we have assumed all three VEVs of singlet scalars
are equal i.e. $u_1=u_2=u_3=u$. Therefore, substituting
Eq.\,(\ref{H&phi_broken_phsae}) in Eq.\,(\ref{Sca:pot})
we have a $4\times 4$ mixing matrix for the real scalar
fields in the basis $\frac{1}{\sqrt{2}}\left(h^\prime\,\, s^\prime_1\,\,s^\prime_2
\,\,s^\prime_3\right)^T$ which has the following form,
{\footnotesize{
\begin{eqnarray}
\mathcal{M}_{rs} =
\left(
\begin{array}{cccc} 
 2 \lambda_H v^2 & \lambda_{H\phi_1} u\,v & \lambda_{H\phi_2} u\,v & \lambda_{H\phi{_3}} u\,v \\
\lambda_{H\phi_1} u\,v & 2 \lambda_{\phi_1} u^2 & u^2 (\beta +\lambda_{\phi_1 \phi_2})&
u \left(\sqrt{2}\,\delta +u (\beta + \lambda_{\phi_1 \phi_3})\right) \\
\lambda_{H\phi_2} u\,v & u^2 (\beta +\lambda_{\phi_1\phi_2}) & -\frac{1}{2} u
\left(\sqrt{2}\,\zeta +u (\beta -4 \lambda_{\phi_2})\right) & \frac{1}{2} u
\left(2 \sqrt{2}\,\zeta + u (\beta +2 \lambda_{\phi_2\phi_3})\right) \\
\lambda_{H\phi_3} u\,v & u \left(\sqrt{2}\,\delta +u (\beta +\lambda_{\phi_1\phi_3})\right)
& \frac{1}{2} u \left(2 \sqrt{2}\,\zeta +u (\beta +2\lambda_{\phi_2\phi_3})\right)
& -\frac{1}{2} u \left(\sqrt{2}\,\delta +u (\beta -4 \lambda_{\phi_3})\right) \\
\end{array}
\right) \,.\nonumber \\
\label{mrs}
\end{eqnarray}}}
The physical scalars ($h,\,s_1,\,s_2,\,s_3$) are obtained by
diagonalising the above real symmetric 
mass matrix and they are related by an
orthogonal transformation to the unphysical scalars
(i.e.\,\,the basis state before diagonalisation) as
{\small{\begin{eqnarray}
\left(
\begin{array}{cccc}
 h \\
 s_1 \\
 s_2 \\
s_3 \\
\end{array}
\right)=
\mathcal{O}^T
\left(
\begin{array}{cccc}
 h^\prime \\
 s_1^\prime \\
 s_2^\prime \\
s_3^\prime \\
\end{array}
\right)\,,
\end{eqnarray}}}
where $\mathcal{O}$ is a $4 \times 4$ orthogonal
matrix which makes $\mathcal{M}_{rs}$ diagonal
i.e.\,\,$\mathcal{O}^T \mathcal{M}_{rs} \mathcal{O} \Rightarrow$
a diagonal matrix containing all the masses of
four physical scalars as diagonal elements. In Appendix
\ref{rs_matrix_diagonalisation}, we have
expressed all the elements of $\mathcal{O}$ matrix
in terms of four mixing angles (assuming mixing among
the three singlet scalars are identical). Additionally,
for simplicity we also set $\beta=0$, \footnote{$\beta$
is the coupling of quartic interaction among all
four singlet scalars which has less significant
impact in dark matter phenomenology compared
to other cubic scalar interaction terms like $\delta\,
\phi_{1}\phi_{1}\phi_{3}^\dagger$ and $\zeta\,\phi_{3} \phi_{3}
\phi_{2}^{\dagger}$.} and $\zeta=\delta$ in Eq.\,(\ref{Sca:pot}).

On the other hand, in our model we have four pseudo
scalars as well, which are $z$, $A^{\prime}_1$, $A^\prime_2$
and $A^\prime_3$ (see Eq.\,(\ref{H&phi_broken_phsae})).
Out of these four pseudo scalars, $z$ does not
mixes with others and becomes the Goldstone boson
corresponding to the SM $Z$ boson after EWSB. However,
the pseudo scalars of complex singlet $\phi_i$s
mix among each other when $B-L$ symmetry is
broken by the VEVs of $\phi_i$s. Therefore, unlike
the case of real scalars, here we have a $3\times3$
mixing matrix in the basis $\frac{1}{\sqrt{2}}$($A^\prime_1$
\,\,$A^\prime_2$\,\,$A^\prime_3$)$^T$ which is written below, 
{\footnotesize{
\begin{eqnarray}
\mathcal{M}_{ps} =
\left(
\begin{array}{ccc} 
-2u\,\left(u\,\beta + \sqrt{2}\,\delta\right) &  u^2\,\beta &
u\left(-u\,\beta +\sqrt{2}\,\delta\right)\\
 u^2\,\beta & -\dfrac{u}{2}\left(u\,\beta + \sqrt{2}\,\zeta \right) &
 \dfrac{u}{2}\left(u\,\beta + 2\sqrt{2}\,\zeta \right)\\
u\left(-u\,\beta +\sqrt{2}\,\delta\right) & \dfrac{u}{2}\left(u\,\beta
+ 2\sqrt{2}\,\zeta \right) & -\dfrac{u}{2}\left(u\,\beta + \sqrt{2}(\delta + 4\,\zeta)
\right)\\
\end{array}
\right)\,.
\label{mps}
\end{eqnarray}}}
After diagonalising, this matrix we get two physical pseudo
scalars $A_2$, $A_3$ and a massless Goldstone boson $A_1$ corresponding
to the extra neutral gauge boson $Z_{BL}$. This can easily be checked
as the pseudo scalar mass matrix has a null determinant. The
eigenvalues of the pseudo scalar mass matrix for $\zeta=\delta$, $\beta=0$
are\footnote{Eigenvalues of $\mathcal{M}_{ps}$ for $\zeta\neq\delta\neq\beta$
are given in Appendix \ref{ps_matrix_diagonalisation}.}
\begin{eqnarray}
m^2_{A_2} = -\dfrac{3\,\zeta\,u}{\sqrt{2}}\,,\nonumber \\
m^2_{A_3} = -\dfrac{7\,\zeta\,u}{\sqrt{2}}\,,\
\label{ps_mass}
\end{eqnarray} 
where $\zeta$ must be less than zero to ensure $m_{A_2}$ and $m_{A_3}$
are real. Further, the physical CP-odd scalars ($A_i$s) and the Goldstone boson $A_1$
are related to the unphysical basis states ($A^\prime_1$, $A^\prime_2$, $A^\prime_3$)
as follows
{\small{\begin{eqnarray}
\left(
\begin{array}{ccc}
 A_1 \\
 A_2 \\
 A_3 \\
\end{array}
\right)=
\left(
\begin{array}{ccc}
 \dfrac{1}{\sqrt{21}} & \dfrac{4}{\sqrt{21}} & \dfrac{2}{\sqrt{21}} \\
 \dfrac{2}{\sqrt{3}} & -\dfrac{1}{\sqrt{6}} & \dfrac{1}{\sqrt{6}} \\
 -\dfrac{2}{\sqrt{7}} & -\dfrac{1}{\sqrt{14}} & \dfrac{3}{\sqrt{14}} \\
\end{array}
\right)
\left(
\begin{array}{ccc}
 A_1^\prime \\
 A_2^\prime \\
 A_3^\prime \\
\end{array}
\right)\,.
\end{eqnarray}}}
Moreover, the neutral gauge boson $Z_{BL}$ becomes
massive after the breaking of $U(1)_{B-L}$ and the corresponding
mass term of $Z_{B-L}$ is given by
\begin{eqnarray}
M_{Z_{BL}} &=& g_{BL}\sqrt{\left(\sum_{i=1}^3 n^2_{\phi_i}u^2_i\right)}\,, \nonumber \\
&=& \sqrt{21}\,g_{BL}\,u\,.
\end{eqnarray}

Now, let us concentrate on the fermionic sector of the present model.
Here, in addition to the usual SM fermions, we have four gauge singlet
fermions and three $SU(2)_{L}$ triplet fermions. All these new
fermions have appropriate $B-L$ charges.  In Eq.\,(\ref{LagT}), $\mathcal{L}_f$
is the Lagrangian for these newly added fermionic fields and it is composed
of two parts as
\begin{eqnarray}
\mathcal{L}_f =  \mathcal{L}_{Singlet} + \mathcal{L}_{Triplet}\,, 
\end{eqnarray}
where, the Lagrangian for the singlet fields are given below
\begin{eqnarray}
\mathcal{L}_{Singlet} &=& i \sum_{\kappa=1}^{2}[\overline{{N_{\kappa}}_L}
\slashed{D}(Q^L_{\kappa}){N_{\kappa}}_L 
+\overline{ {{N_{\kappa}}_R}}\slashed{D}(Q^R_{\kappa}) {{N_{\kappa}}_R}]
-\bigg(\mathcal{Y}_1\,\overline{{{N_1}_L}} {N_1}_R\,\phi^{\dagger}_1
+ \mathcal{Y}_2\,\overline{{N_2}_L} {N_2}_R\,\phi_2 + \text{h.c.} \bigg) \,.\,\nonumber
\label{Lfsinglet} \\
\end{eqnarray}
In the above $\mathcal{Y}_i$ ($i=1$, 2) are the dimensionless Yukawa couplings
and the covariant derivative is defined as
\begin{eqnarray}
\slashed{D}(Q^{L(R)}_{\kappa})\,{N_{\kappa}}_{L(R)} =
\gamma^{\mu}\left(\partial_{\mu}
+ i g_{BL}\,Q^{L(R)}_{\kappa}\,{Z_{BL}}_{\mu}\right) {N_{\kappa}}_{L(R)}\,, 
\end{eqnarray}
where $Q^{L(R)}_{\kappa}$ is the corresponding $B-L$ charge of ${N_{\kappa}}_{L(R)}$
which is listed in Table \ref{tab:1}. As mentioned earlier, due to
special choice of $B-L$ charges of scalar fields $\phi_1$,
$\phi_2$ and $\phi_3$, the Yukawa interactions in Eq.\,(\ref{Lfsinglet})
are exactly diagonal in the basis $\xi_1 = {N_1}_L + {N_1}_R$
and $\xi_2 = {N_2}_L + {N_2}_R$. In this basis above
Lagrangian can be rewritten as,
\begin{eqnarray}
\mathcal{L}_{\rm Singlet} &=& i\,\overline{\xi_1}\,\slashed{\partial}\xi_1
+ i\,\overline{\xi_2}\,\slashed{\partial}\xi_2
-g_{BL}\left(-\dfrac{7}{5}\right)\,\overline{\xi_1}\,\slashed{Z}_{BL}\,P_L\,\xi_1\,
-g_{BL}\left(\dfrac{6}{5}\right)\,\overline{\xi_2}\,\slashed{Z}_{BL}\,P_L\,\xi_2\, \nonumber \\
&&-g_{BL}\left(-\dfrac{2}{5}\right)\,\overline{\xi_1}\,\slashed{Z}_{BL}\,P_R\,\xi_1\,
-g_{BL}\left(-\dfrac{14}{5}\right)\,\overline{\xi_2}\,\slashed{Z}_{BL}\,P_R\,\xi_2\,
-\mathcal{Y}_1\,\overline{\xi_1}\,P_R\,{\xi_1}\,\phi^{\dagger}_1 \nonumber \\
&&
- \mathcal{Y}_2\,\overline{\xi_2}\,P_R\,{\xi_2}\,\phi_2
-\mathcal{Y}_1\,\overline{\xi_1}\,P_L\,{\xi_1}\,\phi_1
- \mathcal{Y}_2\,\overline{\xi_2}\,P_L\,{\xi_2}\,\phi^\dagger_2 
\,,\nonumber \\
&& \hspace{-2.7cm} \text{which can be further simplified as} \nonumber \\
&=& i\,\overline{\xi_1}\,\slashed{\partial}\xi_1
+ i\,\overline{\xi_2}\,\slashed{\partial}\xi_2
+\dfrac{g_{BL}}{10}\,\overline{\xi_1}\,\slashed{Z}_{BL}\left(9-5\gamma_5\right)\,\xi_1\,
+\dfrac{2\,g_{BL}}{5}\,\overline{\xi_2}\,\slashed{Z}_{BL}\left(2+5\gamma_5\right)\,\xi_2\,
\nonumber \\ &&
-\mathcal{Y}_1\,\overline{\xi_1}\,P_R\,{\xi_1}\,\phi^{\dagger}_1 
- \mathcal{Y}_2\,\overline{\xi_2}\,P_R\,{\xi_2}\,\phi_2
-\mathcal{Y}_1\,\overline{\xi_1}\,P_L\,{\xi_1}\,\phi_1
- \mathcal{Y}_2\,\overline{\xi_2}\,P_L\,{\xi_2}\,\phi^\dagger_2\,, 
\label{eq:DMZBL}
\end{eqnarray}
where $P_{L,R} = \dfrac{1 \pm \gamma_5}{2}$, left and right chiral projection
operators. Besides, in the above we have assumed the Yukawa couplings
$\mathcal{Y}_i$s are real. Therefore, from the above Lagrangian one
can easily notice that both $\xi_1$ and $\xi_2$ are decoupled from
each other and thus can be stable simultaneously. Hence,
they naturally form a two component dark matter system stabilises
by the $B-L$ symmetry only \footnote{Actually, we consider the $B-L$
charges of $\phi_1$, $\phi_2$ and $\phi_3$ in such a way that $U(1)_{B-L}$ breaks
into a $\mathbb{Z}_2\times\mathbb{Z}^{\prime}_2$ symmetry where $\xi_1$
and $\xi_2$ have following charges $(-,\,+)$ and $(+,\,-)$ under
the $\mathbb{Z}_2 \times \mathbb{Z}^{\prime}_2$ symmetry respectively.}
. On the other hand, $SU(2)_L\times U(1)_{B-L}$ \footnote{Triplet
fermion fields have no colour charge and hypercharge.}
invariant Lagrangian for the triplet fields are given by, 
\begin{eqnarray}\nonumber
\mathcal{L}_{\rm Triplet} &=& \frac{i}{2}\sum_{k=1}^{3}
\left({\rm Tr}\big[\overline{\Sigma_{kR}} \slashed{D} \Sigma_{kR}\big] +
{\rm Tr}\big[\overline{{\Sigma_{kR}}^c} \slashed{D}^\prime {\Sigma_{kR}}^c\big]\right)
- \frac{1}{2}\left({\rm Tr}[\overline{{\Sigma_{1R}}^c}\sqrt{2}\
Y_{\Sigma_{1}\phi_{3}}\Sigma_{1R}]\phi_{3} \right. \nonumber \\ 
&& \hspace{-2cm} \left. + \ {\rm Tr}[\overline{{\Sigma_{2R}}^c}\sqrt{2}\
Y_{\Sigma_{2}\phi_{3}}\Sigma_{2R}]\phi_{3} 
+ {\rm Tr}[\overline{{\Sigma_{3R}}^c}\sqrt{2}\ 
Y_{\Sigma_{3}\phi_{2}}\Sigma_{3R}]\phi_{2}^{\dagger} + h.c. \right)
-\frac{1}{2}\bigg\{\left({\rm Tr}[\overline{{\Sigma_{1R}}^c}\sqrt{2}
\ Y_{\Sigma_{13}\phi_{1}}\Sigma_{3R}]
\right. \nonumber \\ && \hspace{-2cm} \left.
+ {\rm Tr}[\overline{{\Sigma_{3R}}^c}\sqrt{2}
\ Y_{\Sigma_{13}\phi_{1}}\Sigma_{1R}]\right)\phi_{1}^{\dagger} 
+ \left({\rm Tr}[\overline{{\Sigma_{2R}}^c}\sqrt{2}
\ Y_{\Sigma_{23}\phi_{1}}\Sigma_{3R}]
+ {\rm Tr}[\overline{{\Sigma_{3R}}^c}\sqrt{2}
\ Y_{\Sigma_{23}\phi_{1}}\Sigma_{2R}] \right)
\phi_{1}^{\dagger} \nonumber \\ 
&& \hspace{-2cm} 
+ \left({\rm Tr}[\overline{{\Sigma_{1R}}^c}\sqrt{2}
\ Y_{\Sigma_{12}\phi_{3}}\Sigma_{2R}]
+ {\rm Tr}[\overline{{\Sigma_{2R}}^c}\sqrt{2}
\ Y_{\Sigma_{12}\phi_{3}}\Sigma_{1R}] \right)
\phi_{3} +h.c. \bigg\}\,, 
\end{eqnarray}
where, we consider fermion triplets $\Sigma_{kR}$ and its CP conjugate
${\Sigma_{kR}}^c$ in $2 \times 2$ representation as
\begin{eqnarray}
\Sigma_{k\:\!R}
~=~\left(\begin{array}{cc}
   {\Sigma_{k\:\!R}^0}/{\sqrt{2}} & \Sigma^{+}_{k\:\!R} \\
   \Sigma^{-}_{k\:\!R}  & -\,{\Sigma_{k\:\!R}^0}/{\sqrt{2}} \\
  \end{array} \right)\,,~~
  {\Sigma_{k\:\!R}}^c= \mathbb{C}\,\overline{\Sigma_{k\:\!R}}^T
~=~\left(\begin{array}{cc}
  {\Sigma_{k\:\!R}^0}^c/{\sqrt{2}} & {\Sigma^{-}_{k\:\!R}}^c \\
   {\Sigma^{+}_{k\:\!R}}^c  & -{\,{\Sigma_{k\:\!R}^0}}^c/{\sqrt{2}} \\
  \end{array} \right),
\label{sigmar}  
\end{eqnarray}
where $\mathbb{C}$ is the charge conjugation matrix. The covariant
derivatives used in the kinetic terms of $\Sigma_{k\:\!R}$ and
$\Sigma^c_{k\:\!R}$ can be defined as
\begin{eqnarray}
D^{(\prime)}_{\mu} {\Sigma_{k\:\!R}}^{(c)} =
\left(\partial_{\mu} + i\,\dfrac{g}{2}\,\sigma_a\,W^a_{\mu} +(-) i\,g_{BL}\,
n^k_{\Sigma}\,{Z_{BL}}_{\mu}\right) {\Sigma_{k\:\!R}}^{(c)} \,,
\end{eqnarray}
where $n^k_{\Sigma}$ is the $B-L$ charge of $\Sigma_{k\:\!R}$ and
there is a sign flip as the $B-L$ charges of $\Sigma_{k\:\!R}$ and
its CP conjugate are equal but opposite in sign. Now, we define
$\psi^0_k = \Sigma^0_{k\:\!R} + {\Sigma^0_{k\:\!R}}^c$, a Majorana
fermion and a Dirac fermion $\psi^-_{k} = \Sigma^-_{k\:\!R} + {\Sigma^+_{k\:\!R}}^c$.
Following \cite{Biswas:2018ybc}, we have written the triplet Lagrangian
in terms of $\psi_k^0$ and $\psi_k^-$ as
\begin{eqnarray}
\mathcal{L}_{Triplet}
&& =\sum_{k=1}^{3}\left \{ \frac{i}{2}\overline{\psi_k^0}\,\slashed{\partial}\psi_k^0
+ i\,\overline{\psi_k^-}\,\slashed{\partial}\psi_k^{-}
- g\Big( \overline{\psi_k^-}\slashed{W}^- \psi_k^0 + h.c. \Big)
+ g\sin \theta_W  \overline{\psi_k^-}\slashed{A} \psi_k^-
\right. \nonumber \\ && \left.
+ g\cos \theta_W  \overline{\psi_k^-}\slashed{Z} \psi_k^- 
- g_{BL}\,n^k_{\Sigma} \left(\dfrac{1}{2}\,\overline{\psi_k^0}\slashed{Z}_{BL}\gamma_{5}\psi_k^0
+ \overline{\psi_k^-}\slashed{Z}_{BL}\gamma_{5}\psi_k^- \right) \right \} 
\nonumber \\ && 
-\sum_{i=1}^{2}Y_{\Sigma_i\phi_3} \bigg\{\left(\dfrac{1}{2} \overline{\psi^0_i}\psi^0_i
+ \overline{\psi^-_i}\psi^-_i\right)s^\prime_3 
+ i\left(\dfrac{1}{2} \overline{\psi^0_i}\gamma_5\psi^0_i
+  \overline{\psi^-_i}\gamma_5\psi^-_i\right)A^\prime_3 \bigg \}
\nonumber \\ && 
-Y_{\Sigma_3\phi_2} \bigg \{ \left(\dfrac{1}{2} \overline{\psi^0_3}\psi^0_3
+ \overline{\psi^-_3}\psi^-_3\right)s^\prime_2 
- i\left(\dfrac{1}{2} \overline{\psi^0_3}\gamma_5\psi^0_3 
+ \overline{\psi^-_3}\gamma_5\psi^-_3\right)A^\prime_2 \bigg \}
\nonumber \\ && 
-Y_{\Sigma_{13\phi_1}} \left \{\left(\dfrac{1}{2} \overline{\psi_1^0}\psi_3^0
+ \overline{\psi_1^-}\psi^-_3\right) s^\prime_1 - i\left(\dfrac{1}{2} \overline{\psi_1^0}\gamma_5\psi_3^0
+ \overline{\psi_1^-}\gamma_5\psi^-_3 \right)A^\prime_1 + h.c.\right \}
\nonumber \\ && 
-Y_{\Sigma_{23\phi_1}} \left \{\left(\dfrac{1}{2} \overline{\psi_2^0}\psi_3^0
+ \overline{\psi_2^-}\psi^-_3\right) s^\prime_1 
- i\left(\dfrac{1}{2} \overline{\psi_2^0}\gamma_5\psi_3^0
+ \overline{\psi_2^-}\gamma_5\psi^-_3 \right)A^\prime_1 + h.c.\right \} 
\nonumber \\ &&
-Y_{\Sigma_{12\phi_3}} \left \{\left(\dfrac{1}{2} \overline{\psi_1^0}\psi_2^0
+ \overline{\psi_1^-}\psi^-_2\right) s^\prime_3 
+ i\left(\dfrac{1}{2} \overline{\psi_1^0}\gamma_5\psi_2^0
+ \overline{\psi_1^-}\gamma_5\psi^-_2 \right)A^\prime_3 + h.c.\right \} 
\,,
\label{Ltriplet}
\end{eqnarray}
$\theta_W = \tan^{-1} \dfrac{g^\prime}{g}$ is the weak mixing angle
(Weinberg angle). The last two terms of the above Lagrangian introduce
off-diagonal elements in the mass matrices of both $\psi^0_k$ and
$\psi^{-}_k$ ($k$ runs from 1 to 3) when $s_1^\prime$ gets a
nonzero VEV. As a result, one needs to diagonalise both the 
mass matrices using bi-unitary transformations in order to get
the physical fermionic states. However, in this work we have not
considered this. We have worked in a limit when $Y_{\Sigma_1\phi_3}$,
$Y_{\Sigma_2\phi_3}$ and $Y_{\Sigma_3\phi_2} >> Y_{\Sigma_{13}\phi_1}$
and $Y_{\Sigma_{23}\phi_1}$, so that mass matrices of both
charged and neutral fermions are effectively diagonal and thus,
there is no need for basis transformation. Here, two triplet fermions
having $B-L$ charge $-1$ will play crucial role in generating
observed neutrino masses and mixings via Type-III seesaw mechanism.
The Yukawa Lagrangian involving the leptons is given by 
\begin{equation}
\mathcal{L}_{\rm Yukawa} \supset \left(\sum_{\alpha=1,2,3,\,\beta=1,2}\sqrt{2} y^{\alpha\beta}_{\Sigma}\,\overline{{l_{\alpha}}_{L}}
{\Sigma_{\beta}}_{R} \tilde{\Phi} + h.c.\right)\,
\end{equation}
The light neutrino mass matrix is generated from the type III seesaw mechanism as 
\begin{equation}
M_{\nu} = -M_D M^{-1}_R M^T_D
\end{equation}
where the Dirac neutrino mass matrix $M_D$ and neutral triplet fermion mass matrix $M_R$ are given by 
\begin{equation}
M_D = \left(
\begin{array}{ccc}
 y^{11}_{\Sigma} v & y^{12}_{\Sigma} v & 0 \\
 y^{21}_{\Sigma} v & y^{22}_{\Sigma} v & 0 \\
y^{31}_{\Sigma} v & y^{32}_{\Sigma} v & 0 \\
\end{array}
\right), \; 
M_R = \left(
\begin{array}{ccc}
Y_{\Sigma_{1}\phi_{3}} u_3 & Y_{\Sigma_{12}\phi_{3}} u_3 & Y_{\Sigma_{13}\phi_{1}} u_1 \\
Y_{\Sigma_{12}\phi_{3}} u_3 & Y_{\Sigma_{2}\phi_{3}} u_3 & Y_{\Sigma_{23}\phi_{1}} u_1 \\
Y_{\Sigma_{13}\phi_{1}} u_1 & Y_{\Sigma_{23}\phi_{1}} u_1 & Y_{\Sigma_{3}\phi_{2}} u_2 \\
\end{array}
\right).
\end{equation}
Diagonalisation of the light neutrino mass matrix using the above forms of $M_D, M_R$ gives one vanishing mass eigenvalue. This is same as the prediction of the recent work \cite{Bernal:2018aon} as well as the littlest seesaw model \cite{King:2015dvf} mentioned earlier. If we simplify our neutral fermion triplet mass matrix $M_R$ by incorporating the smallness on off-diagonal Yukawa couplings mentioned earlier $Y_{\Sigma_1\phi_3}$, $Y_{\Sigma_2\phi_3}$, $Y_{\Sigma_3\phi_2} >> Y_{\Sigma_{13}\phi_1}$, $Y_{\Sigma_{23}\phi_1}$, $Y_{\Sigma_{12}\phi_3}$ and equality of singlet VEVs, we can approximate $M_R$ to be a diagonal matrix leading to a relatively simple light neutrino mass matrix given as 
\begin{equation}
M_{\nu} = -\frac{v^2}{u}\left(
\begin{array}{ccc}
 \frac{(y^{11}_{\Sigma})^2}{Y_{\Sigma_{1}\phi_{3}}} +  \frac{(y^{12}_{\Sigma})^2}{Y_{\Sigma_{2}\phi_{3}}}  &  \frac{y^{11}_{\Sigma} y^{21}_{\Sigma}}{Y_{\Sigma_{1}\phi_{3}}} +  \frac{y^{12}_{\Sigma} y^{22}_{\Sigma}}{Y_{\Sigma_{2}\phi_{3}}} & \frac{y^{11}_{\Sigma} y^{31}_{\Sigma}}{Y_{\Sigma_{1}\phi_{3}}} +  \frac{y^{12}_{\Sigma} y^{32}_{\Sigma}}{Y_{\Sigma_{2}\phi_{3}}} \\
\frac{y^{11}_{\Sigma} y^{21}_{\Sigma}}{Y_{\Sigma_{1}\phi_{3}}} +  \frac{y^{12}_{\Sigma} y^{22}_{\Sigma}}{Y_{\Sigma_{2}\phi_{3}}} &   \frac{(y^{21}_{\Sigma})^2}{Y_{\Sigma_{1}\phi_{3}}} +  \frac{(y^{22}_{\Sigma})^2}{Y_{\Sigma_{2}\phi_{3}}}  &  \frac{y^{21}_{\Sigma} y^{31}_{\Sigma}}{Y_{\Sigma_{1}\phi_{3}}} +  \frac{y^{22}_{\Sigma} y^{32}_{\Sigma}}{Y_{\Sigma_{2}\phi_{3}}} \\
\frac{y^{11}_{\Sigma} y^{31}_{\Sigma}}{Y_{\Sigma_{1}\phi_{3}}} +  \frac{y^{12}_{\Sigma} y^{32}_{\Sigma}}{Y_{\Sigma_{2}\phi_{3}}} &  \frac{y^{21}_{\Sigma} y^{31}_{\Sigma}}{Y_{\Sigma_{1}\phi_{3}}} +  \frac{y^{22}_{\Sigma} y^{32}_{\Sigma}}{Y_{\Sigma_{2}\phi_{3}}} &   \frac{(y^{31}_{\Sigma})^2}{Y_{\Sigma_{1}\phi_{3}}} +  \frac{(y^{32}_{\Sigma})^2}{Y_{\Sigma_{2}\phi_{3}}}  \\
\end{array}
\right).
\end{equation}
This simplified light neutrino mass matrix also leads to a vanishing lightest neutrino mass while the three mixing angles can be satisfied by suitable tuning of the Yukawa couplings. Since for TeV scale triplet fermions, the Dirac Yukawa couplings $y^{\alpha \beta}_{\Sigma}$ have to be fine tuned at the level of $\leq 10^{-4}$, they do not impact the dark matter analysis discussed in this work. Hence, we do not take such couplings into our subsequent discussions.

\section{Constraints on the model parameters}
\label{sec:constraints}
Before going into the detailed calculation of DM relic abundance and relevant parameter scan, we note the existing theoretical as well as experimental constraints on the model parameters.

In order to keep the scalar potential (given in Eq.\,(\ref{Sca:pot}) within the square brackets)
bounded from below, the quartic couplings must satisfy
the following inequalities: 
\begin{eqnarray}\nonumber
\lambda_{H},\lambda_{\phi_1},\lambda_{\phi_2},\lambda_{\phi_3}\geq 0\,,\\ \nonumber
\lambda_{H\phi_1}+\sqrt{\lambda_{H} \lambda_{\phi_1}}\geq 0\,,
\lambda_{H\phi_2}+\sqrt{\lambda_{H} \lambda_{\phi_2}}\geq 0\,,\\ 
\lambda_{H\phi_3}+\sqrt{\lambda_{H} \lambda_{\phi_3}}\geq 0\,,
\lambda_{\phi_1\phi_2}+\sqrt{\lambda_{\phi_1} \lambda_{\phi_2}}\geq 0\,,\\ \nonumber
\lambda_{\phi_1\phi_3}+\sqrt{\lambda_{\phi_1} \lambda_{\phi_3}}\geq 0\,,
\lambda_{\phi_2\phi_3}+\sqrt{\lambda_{\phi_2} \lambda_{\phi_3}}\geq 0\,. \nonumber
\end{eqnarray}
To prevent perturbative breakdown of the model, all quartic, Yukawa and gauge
couplings should obey the following limits at any energy scale:
\begin{eqnarray}
|\lambda_H| < 4 \pi,~|\lambda_{\phi_{1,2,3}}| < 4 \pi,~|\lambda_{H\phi_{1,2,3}}| < 4 \pi,~\nonumber\\
|\lambda_{\phi_1 \phi_2}| < 4 \pi,~|\lambda_{\phi_1 \phi_3}| < 4 \pi,~|\lambda_{\phi_2 \phi_3}| < 4 \pi,~\nonumber\\
|\mathcal{Y}_i| < \sqrt{4 \pi},~ |Y_{\Sigma_{1,2}\phi_{3}}| < \sqrt{4 \pi},~ |Y_{\Sigma_{3}\phi_{2}}| < \sqrt{4 \pi},~\nonumber \\
|Y_{\Sigma_{13}\phi_{1}}| < \sqrt{4 \pi},~|Y_{\Sigma_{23}\phi_{1}}|< \sqrt{4 \pi}, ~|Y_{\Sigma_{12}\phi_{3}}|< \sqrt{4 \pi} \nonumber \\
|g, g'| < \sqrt{4\pi},~|g_{BL}| < \sqrt{4\pi} ,
\label{eq:PerC}
\end{eqnarray}

Experimental limits from LEP II constrains such new gauge sector by putting a lower bound on the ratio of new gauge boson mass to the new gauge coupling $M_{Z'}/g' \geq 7$ TeV \cite{Carena:2004xs, Cacciapaglia:2006pk}. The corresponding bounds from the LHC experiment have become stronger than this by now. Search for high mass dilepton resonances have put strong bounds on mass of such gauge boson coupling to first two generations of leptons with couplings similar to electroweak ones. The latest bounds from the ATLAS experiment \cite{Aaboud:2017buh, Aad:2019fac} and the CMS experiment \cite{Sirunyan:2018exx} at the LHC rule out such gauge boson masses below 4-5 TeV from analysis of 13 TeV data. Such bounds get weaker, if the corresponding gauge couplings are weaker \cite{Aaboud:2017buh} than the electroweak gauge couplings. Also, if the $Z'$ gauge boson couples only to the third generation of leptons, all such collider bounds become much weaker, as explored in the context of DM and collider searches in a recent work \cite{Barman:2019aku}.

Similarly, the additional scalars in the model also face stringent constraints which typically arise due to their mixing with the SM Higgs boson which enable them to couple with the SM particles. The bound on such scalar mixing angles would come from both theoretical and experimental constraints
\cite{Robens:2015gla,Chalons:2016jeu}. In case of scalar singlet extension of SM, the strongest bound on scalar-SM Higgs mixing angle ($\theta_{1j}, j=2,3,4$) comes form $W$ boson mass correction \cite{Lopez-Val:2014jva} at NLO for $250 {\rm ~ GeV} \lesssim M_{s_i} \lesssim 850$ GeV as ($0.2 \lesssim \sin\theta_{1j} \lesssim 0.3$) where $M_{s_i}$ is the mass of other physical Higgs. Whereas, for $M_{s_i}>850$ GeV, the bounds from the requirement of perturbativity and unitarity of the theory turn dominant  which gives $\sin\theta_{1j}\lesssim 0.2$. For lower values {\it i.e.} $M_{s_i}<250$ GeV, the LHC and LEP direct search \cite{Khachatryan:2015cwa,Strassler:2006ri} and measured Higgs signal strength \cite{Strassler:2006ri} restrict the mixing angle $\sin\theta_{1j}$ dominantly ($\lesssim 0.25$). The bounds from the measured value of EW precision parameter are mild for $M_{s_i}< 1$ TeV. While these constraints restrict the singlet scalar mixing with SM Higgs denoted by ($\theta_{1j}, j=2,3,4$), the other three angles ($\theta_{23}, \theta_{24}, \theta_{34}$) remain unconstrained. We choose our benchmark values of singlet scalar masses and their mixing with SM Higgs boson in such a way that these constraints are automatically satisfied.


\section{The Boltzmann Equations for Two component DM} 
\label{sec:boltzmann}
\begin{figure}[h!]
\centering
\includegraphics[height=2cm,width=4cm]{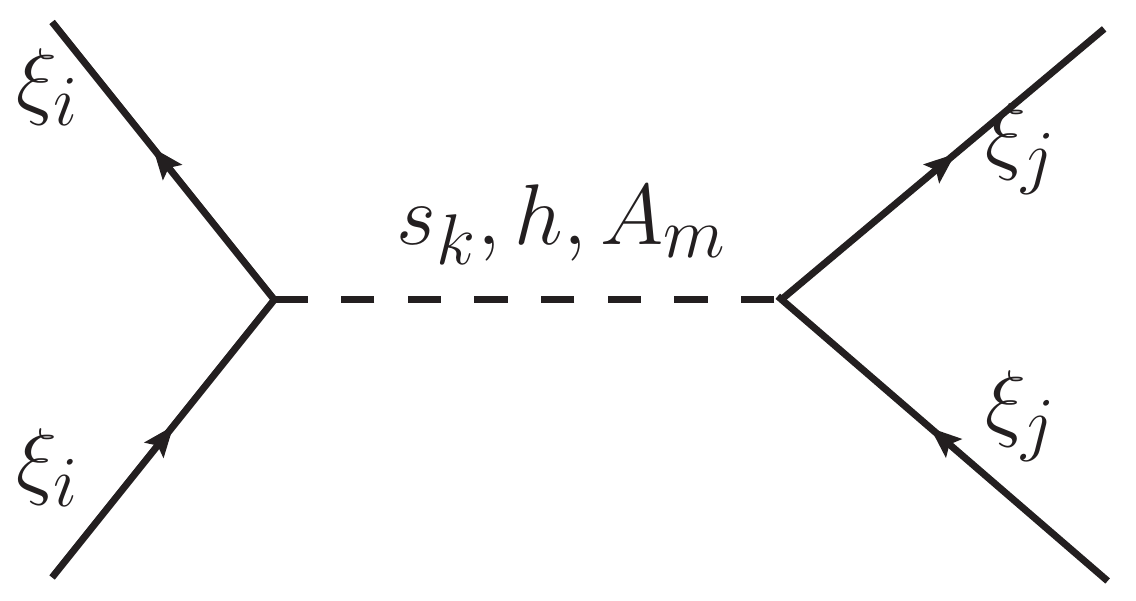}
\includegraphics[height=2cm,width=4cm]{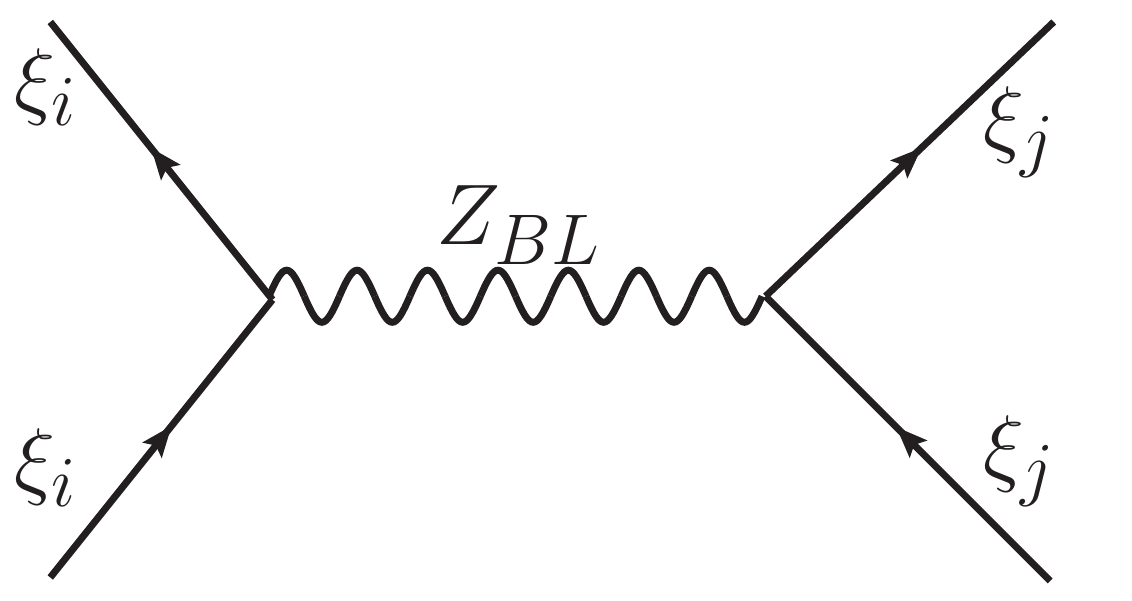}
\includegraphics[height=2cm,width=4cm]{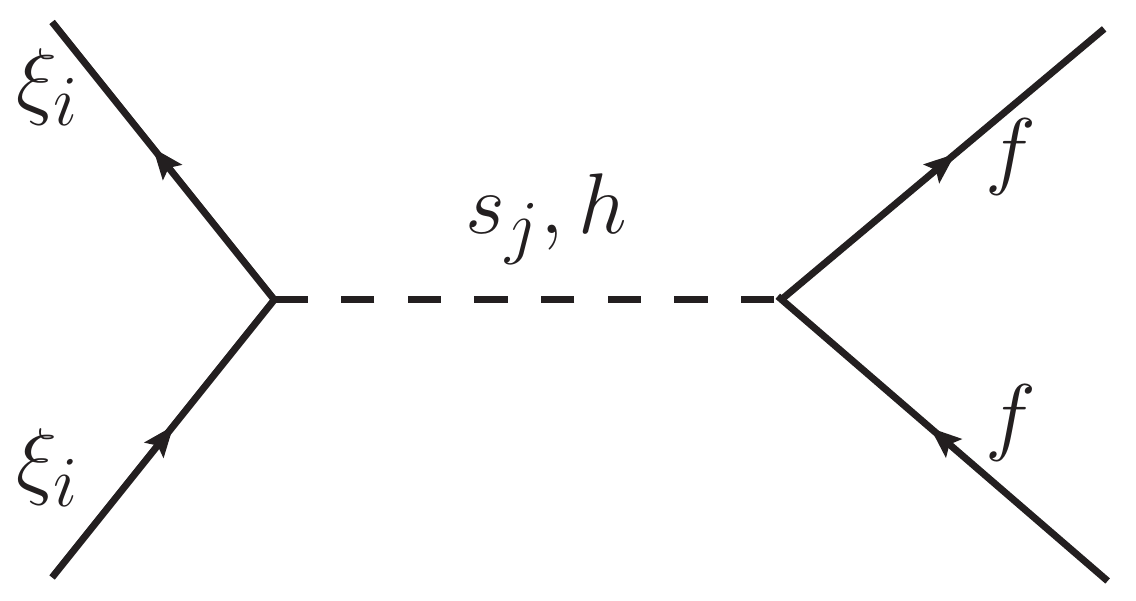}
\includegraphics[height=2cm,width=4cm]{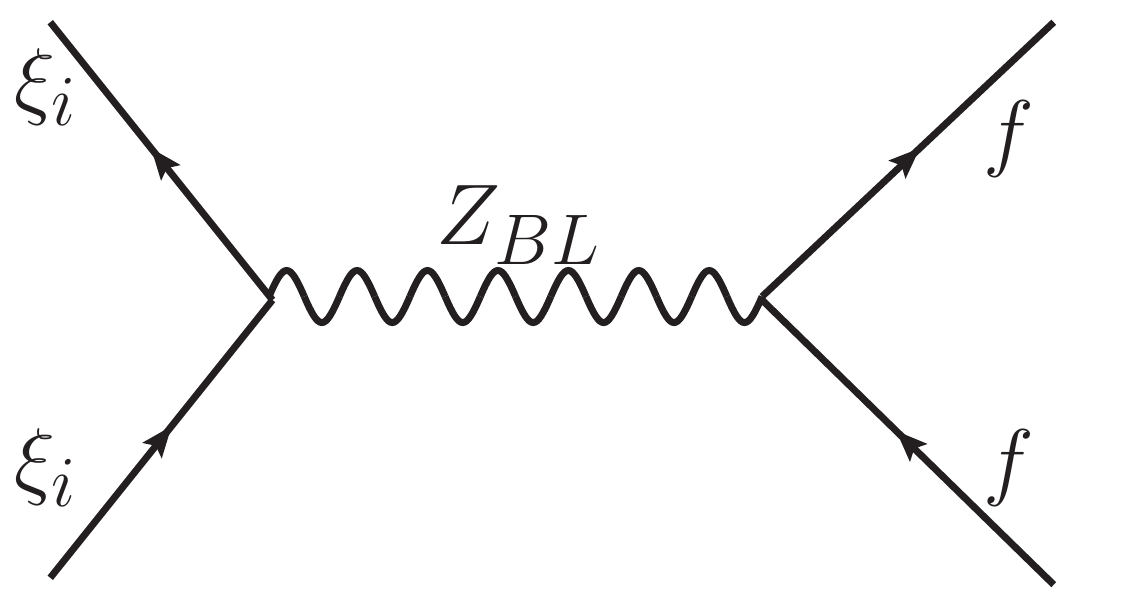}\\
\vskip 0.1in
\includegraphics[height=2cm,width=4cm]{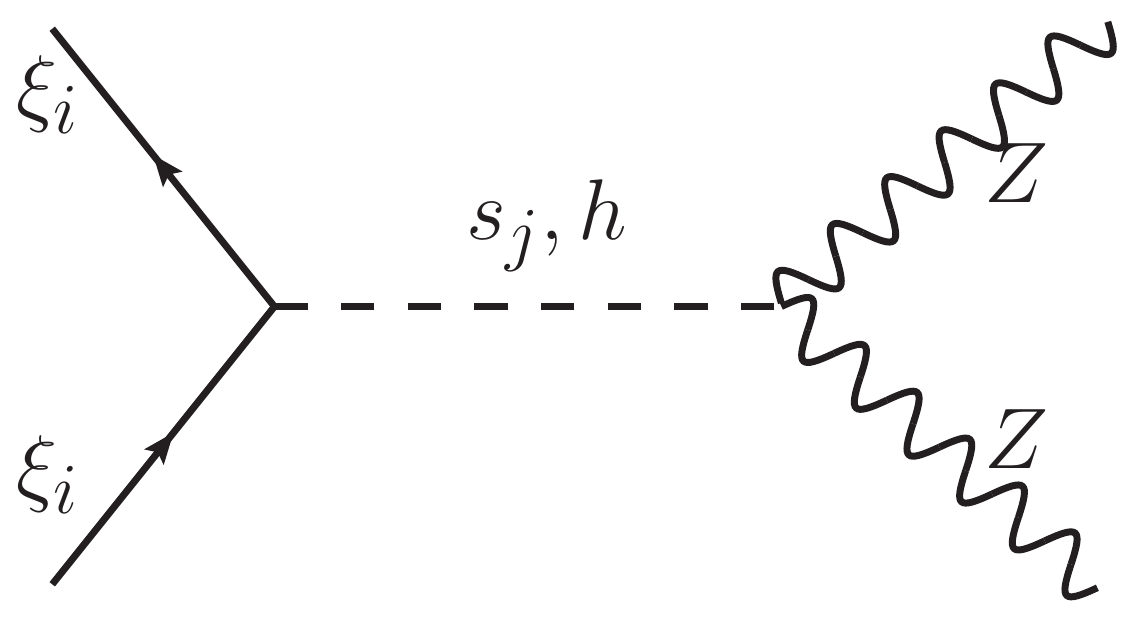}
\includegraphics[height=2cm,width=4cm]{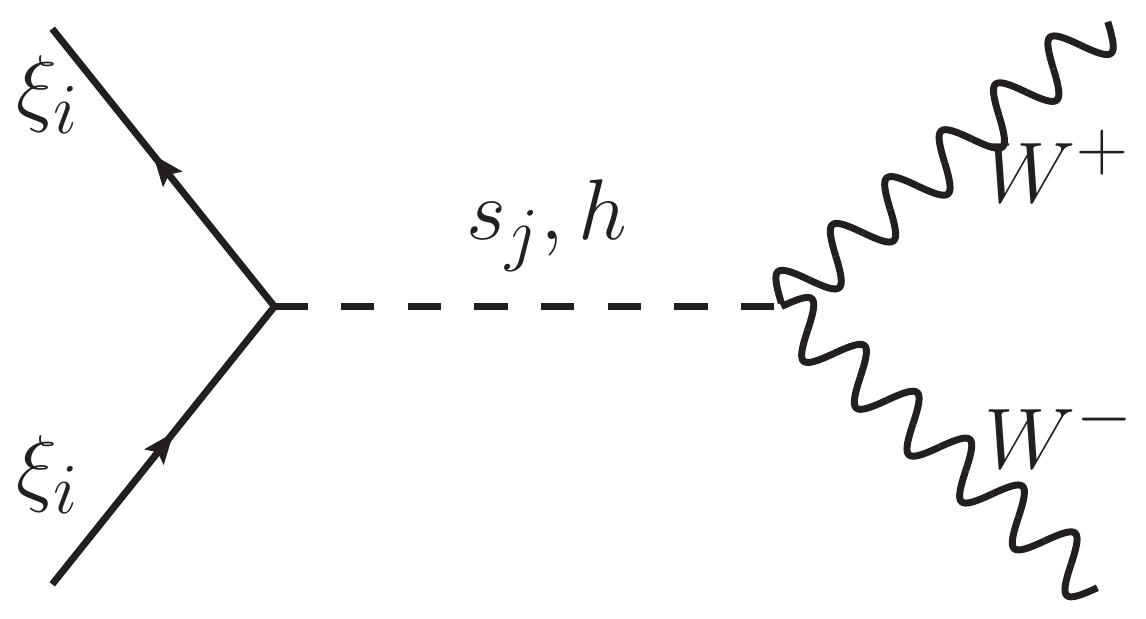}
\includegraphics[height=2cm,width=4cm]{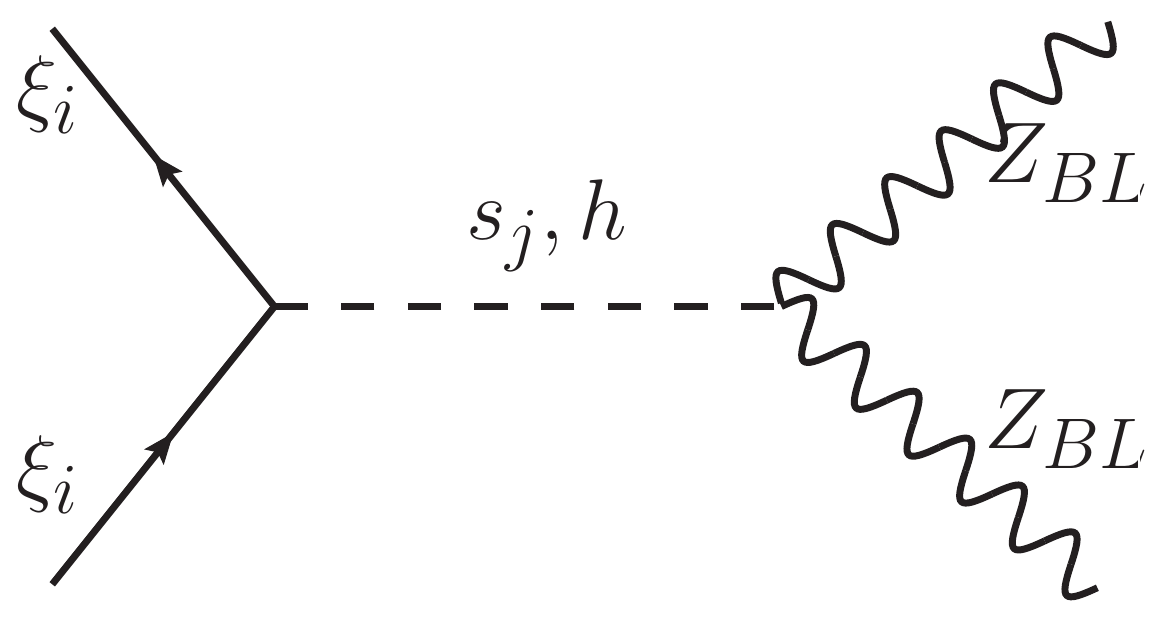}
\includegraphics[height=2cm,width=4cm]{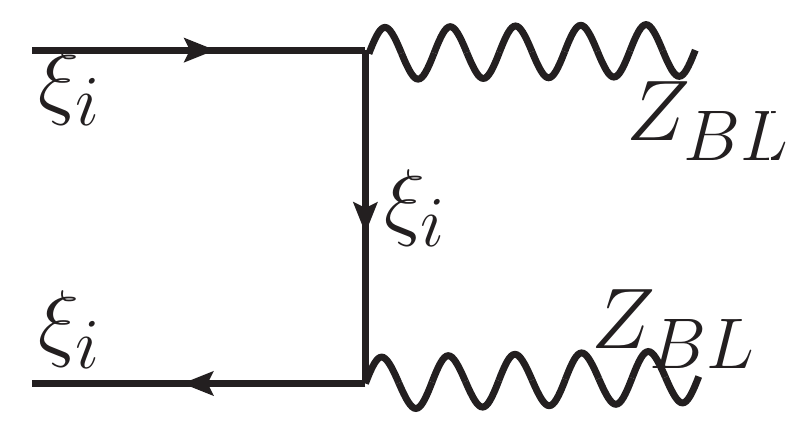}\\
\vskip 0.1in
\includegraphics[height=2cm,width=4cm]{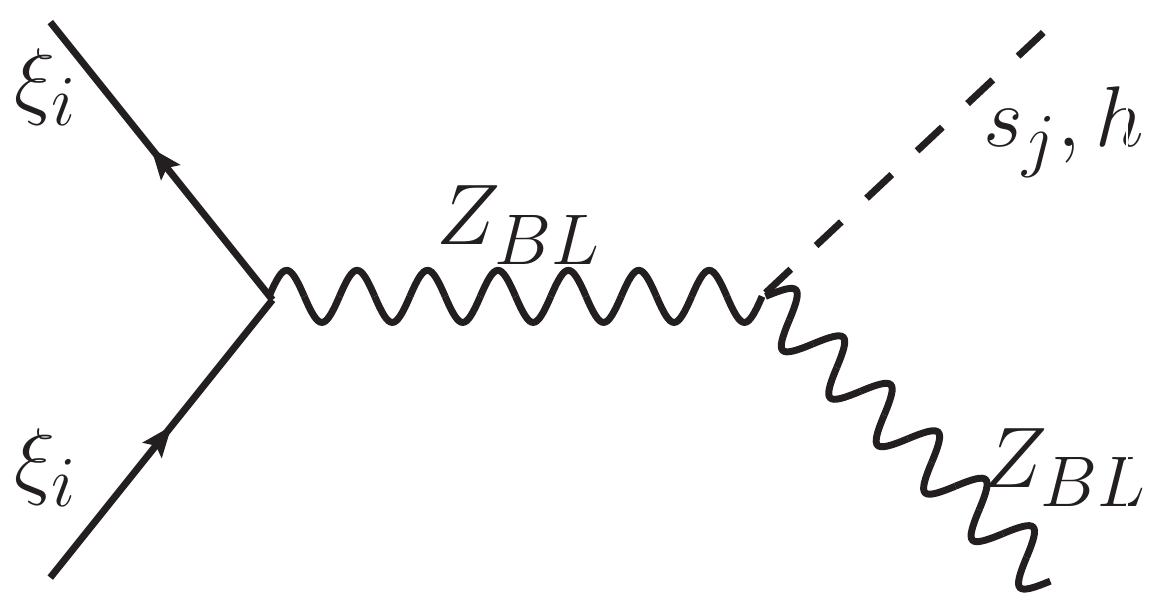}
\includegraphics[height=2cm,width=4cm]{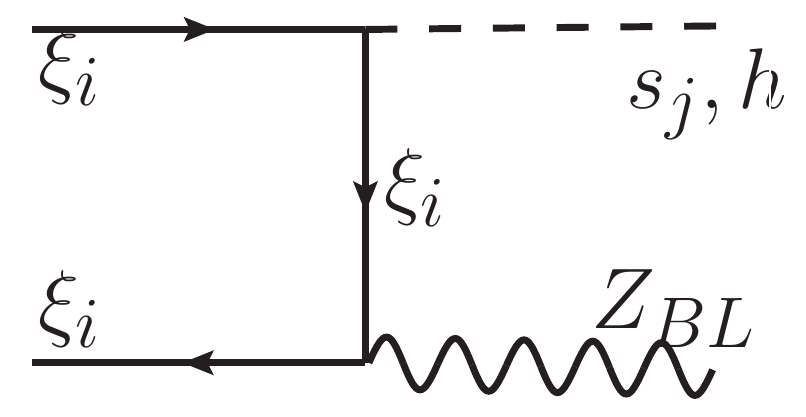}
\includegraphics[height=2cm,width=4cm]{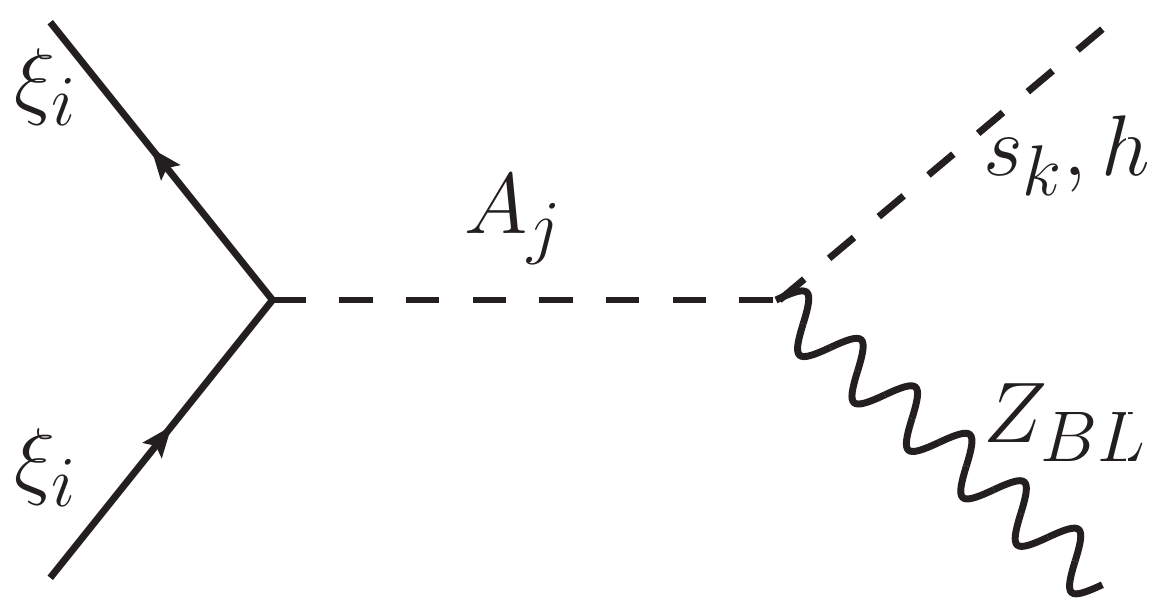}
\includegraphics[height=2cm,width=4cm]{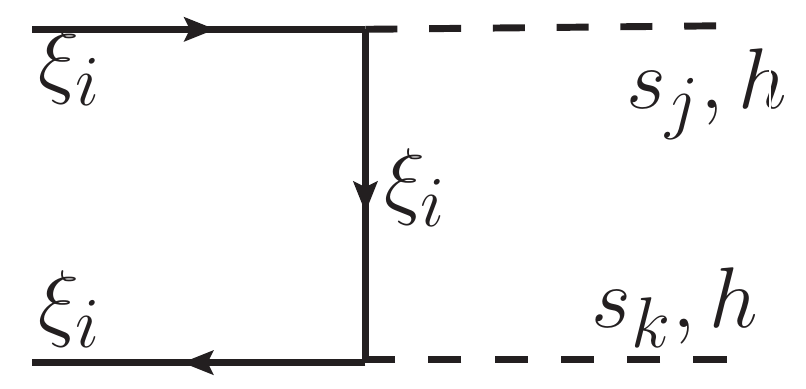}\\
\vskip 0.1in
\includegraphics[height=2cm,width=4cm]{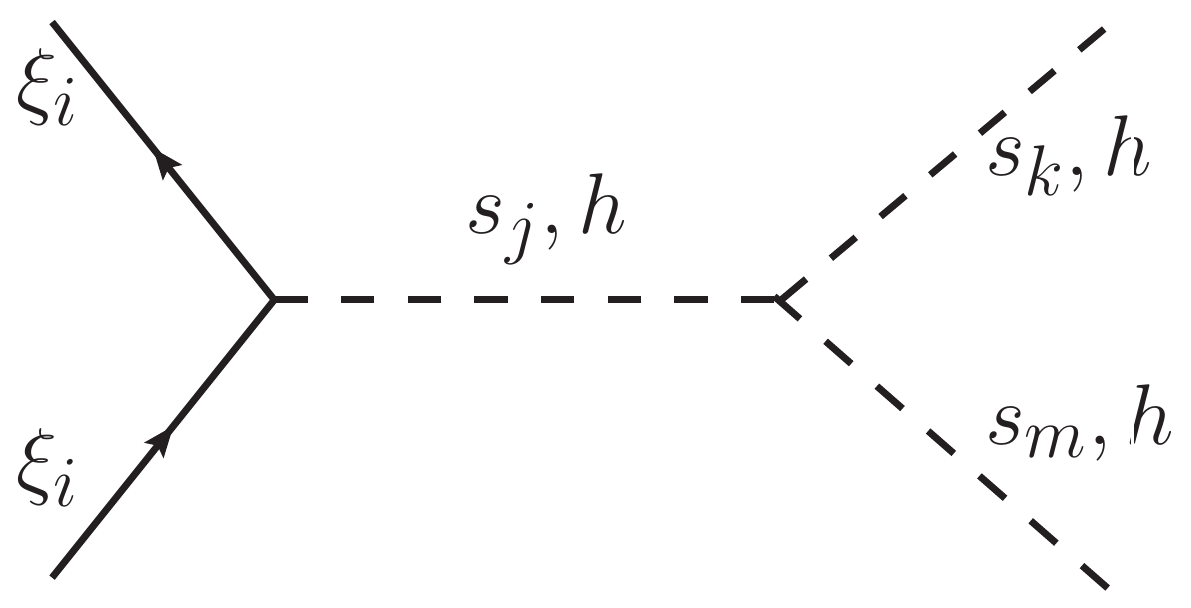}
\includegraphics[height=2cm,width=4cm]{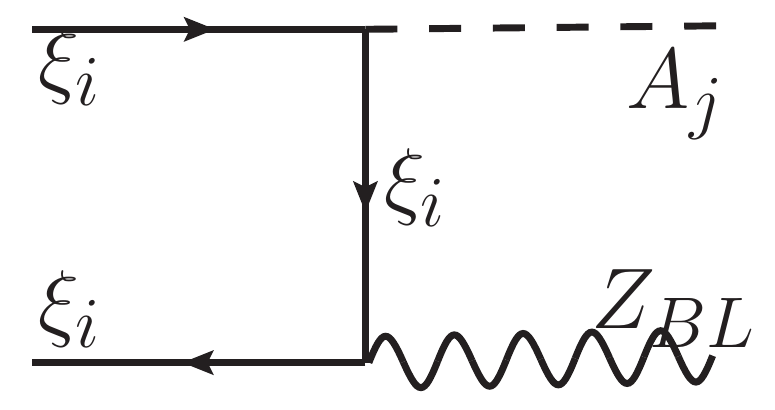}
\includegraphics[height=2cm,width=4cm]{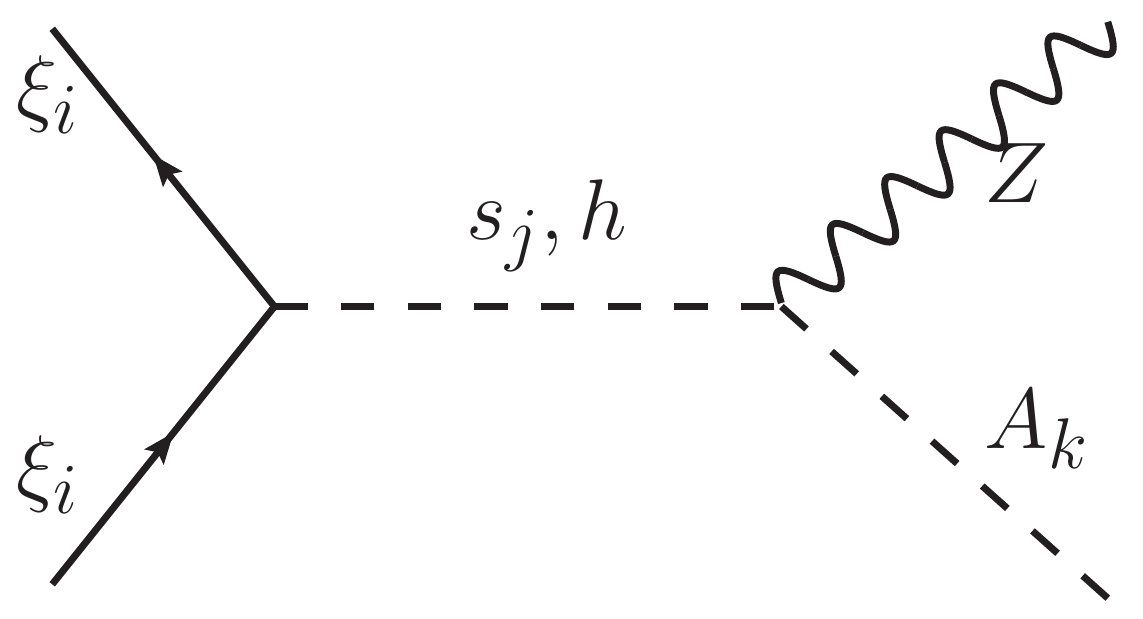}
\includegraphics[height=2cm,width=4cm]{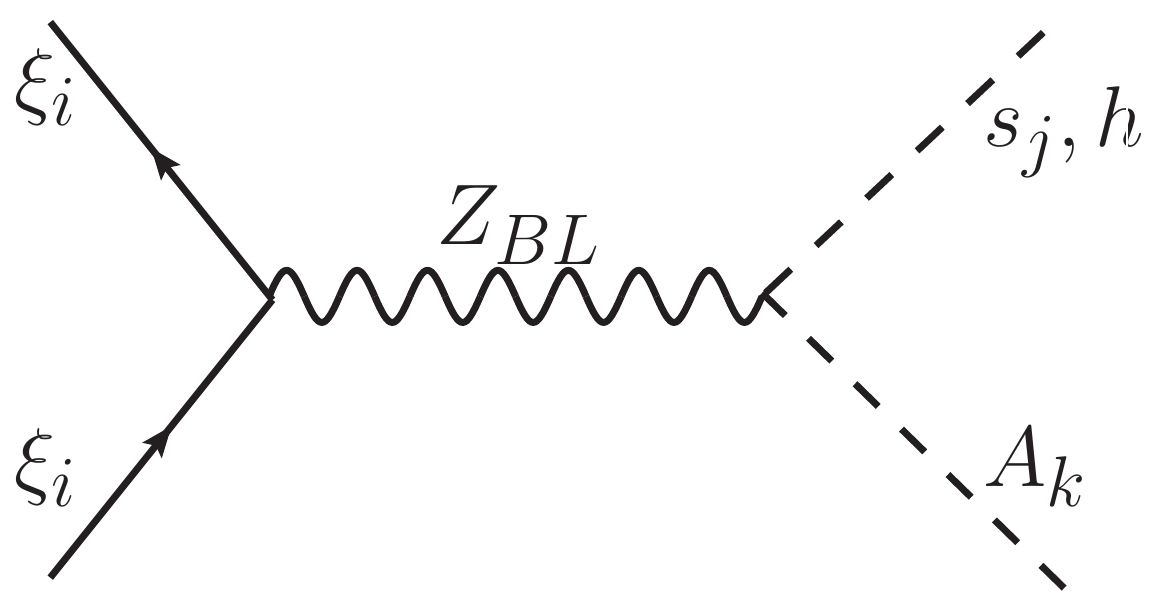}\\
\vskip 0.1in
\includegraphics[height=2cm,width=4cm]{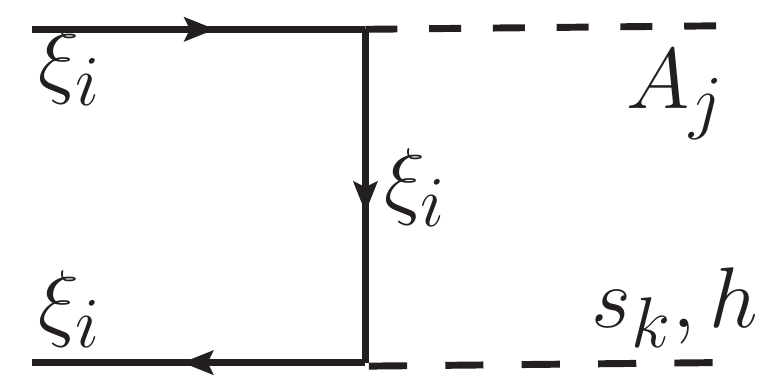}
\includegraphics[height=2cm,width=4cm]{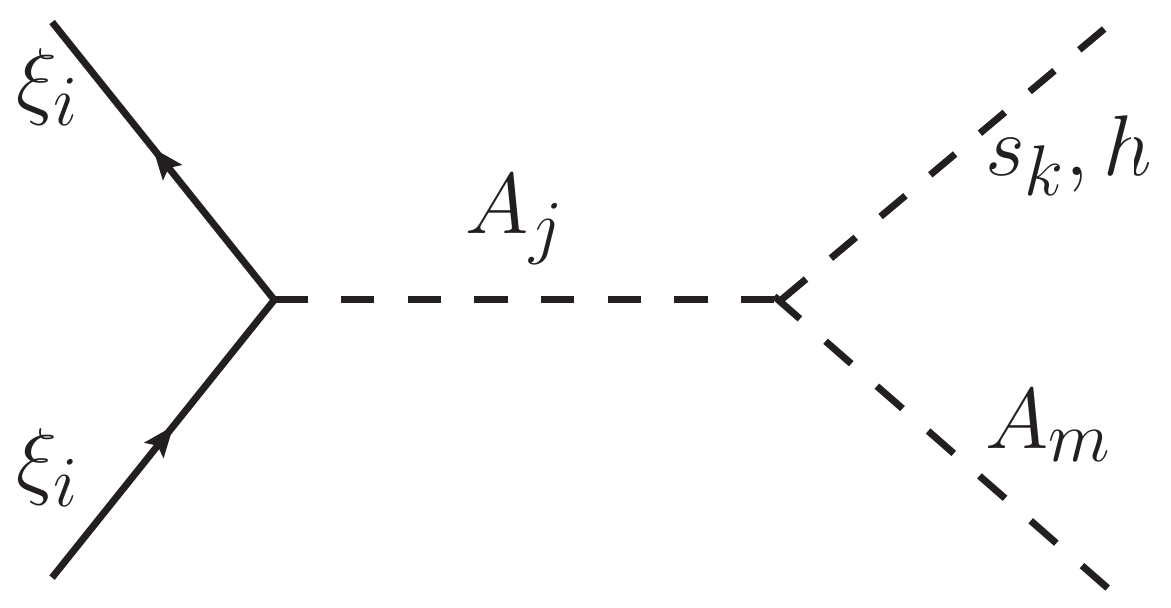}
\includegraphics[height=2cm,width=4cm]{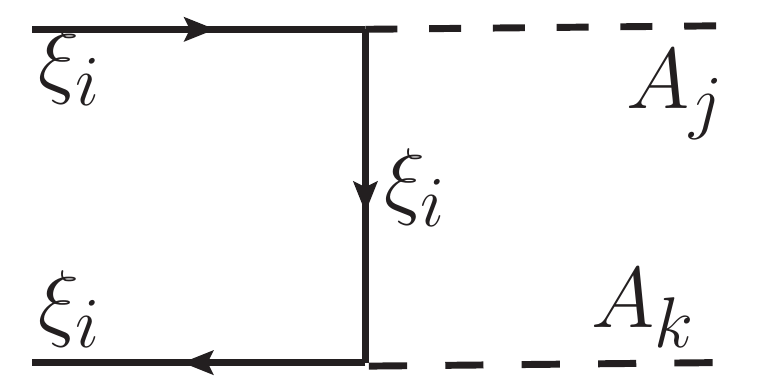}
\includegraphics[height=2cm,width=4cm]{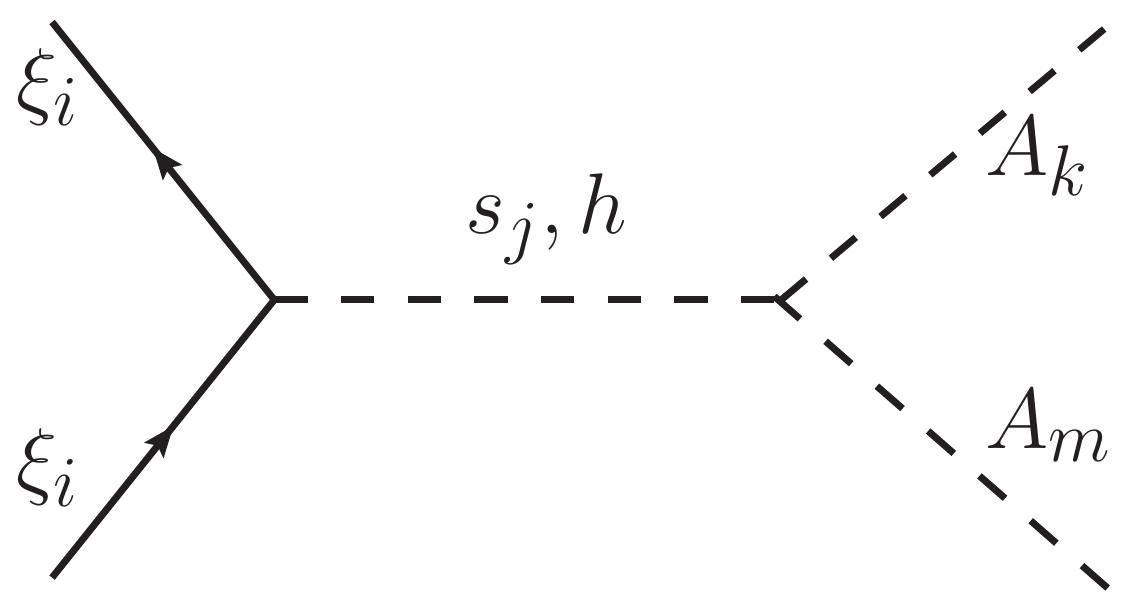}

\caption{Feynman diagrams for all possible annihilation channels of two DM candidates.}
\label{Fig:feyn_ann}
\end{figure}

In this work, as we already know that we are dealing with two
stable dark matter candidates $\xi_1$ and $\xi_2$. To find
the present number densities of dark matter candidates we
need to solve two Boltzmann equations one for each candidate.
The collision term of each Boltzmann equation contains all
possible number changing interactions of that particular dark matter
candidate allowed by the symmetries. In the present model, there are
two types of number changing interactions for a dark matter candidate.
First one is the pair annihilation where $\xi_i$ and $\bar{\xi_i}$
annihilate in a pair into all possible final states ($X$)
except a pair of other dark matter candidate $\xi_j \bar{\xi_j}$
($j\neq i$). These processes reduce the number of $\xi_i$
and $\bar{\xi_i}$ by one unit (assuming there is no asymmetry
in the number densities of dark matter and its anti-particle).
The other type of number changing process is $\xi_i\bar{\xi_i}\rightarrow
\xi_j\bar{\xi_j}$ ($i\neq j$). This is actually the conversion
process where one type of dark matter converts into another. It
increases the number of lighter dark matter candidate by
two unit while reducing the number of heaver one by the same
amount. This conversion process acts as a coupling between
the individual Boltzmann equations of $\xi_1$ and $\xi_2$.
Let $n_2 = n_{\xi_2} + n_{\bar{\xi}_2}$ and
$n_1=n_{\xi_1} + n_{\bar{\xi}_1}$ are the total
number densities of two dark matter
candidates respectively. Assuming there
is no asymmetry in number densities of $\xi_i$
and $\bar{\xi}_i$, the two coupled Boltzmann
equations in terms of $n_2$ and $n_1$ are given below
\cite{Belanger:2011ww, Biswas:2013nn, Biswas:2014hoa},   
\begin{eqnarray}
\frac{dn_{2}}{dt} + 3n_{2}{\mathbf H} &=& 
-\dfrac{1}{2}\langle{\sigma {\rm{v}}}_{\xi_2 \bar{\xi_2} \rightarrow {X \bar{X}}}\rangle 
\left(n_{2}^2 -(n_{2}^{\rm eq})^2\right)
- \dfrac{1}{2}{\langle{\sigma {\rm{v}}}_{\xi_2 \bar{\xi_2}
\rightarrow \xi_1 \bar{\xi_1}}\rangle} \bigg(n_{2}^2 - 
\frac{(n_{2}^{\rm eq})^2}{(n_{1}^{\rm eq})^2}n_{1}^2\bigg) \,,
%
\label{boltz-eq1} \\
\frac{dn_{1}}{dt} + 3n_{1}{\mathbf H} &=& -\dfrac{1}{2}\langle{\sigma {\rm{v}}}
_{\xi_1 \bar{\xi_1} \rightarrow {X \bar{X}}}\rangle \left(n_{1}^2 -
(n_{1}^{\rm eq})^2\right) 
+ \dfrac{1}{2}{\langle{\sigma {\rm{v}}}_{\xi_2 \bar{\xi_2} \rightarrow {\xi_1} \bar{\xi_1}}\rangle} 
\bigg(n_{2}^2 - \frac{(n_{2}^{\rm eq})^2}{(n_{1}^{\rm eq})^2}
n_{1}^2\bigg)\,,
\label{boltz-eq2} 
\end{eqnarray}
where, $n^{\rm eq}_i$ is the equilibrium number density
of dark matter species $i$. 
The second term in the left hand side involving the Hubble
parameter $\mathbf{H}$ represents dilution of number density
due to the expansion of the Universe. 
The extra half factors in the collision term
are due to non-self-conjugate (Dirac fermion) nature
of our both dark matter candidates $\xi_1$ and $\xi_2$
\cite{Gondolo:1990dk}. 
Here we are more interested
to study the evolution of number densities of two component WIMP
system due to all possible interactions which change particle
numbers of either $\xi_1$ or $\xi_2$ or both. Hence, in stead of
actual number density $n_{i}$, it is convenient to describe
the Boltzmann equation for a species in terms of comoving
number density $Y_{i} = n_i/s$ with $s$ representing the
entropy density of the Universe. The quantity $Y_i$ is an useful
one as it absorbs the effect of the expansion on $n_{i}$.
In terms of comoving number densities, the above two coupled
Boltzmann equations can be written as,
\begin{eqnarray}
\frac{dY_{2}}{dx_{\xi_2}} &=&
-\dfrac{1}{2}\sqrt{\dfrac{\pi}{45\,G}}\,
\frac{m_{\xi_2}}{x_{\xi_2}^2}{\sqrt{g_\star}}
\bigg({\langle{\sigma {\rm{v}}}_
{\xi_2 \bar{\xi_2} \rightarrow {X \bar{X}}}\rangle}
\left(Y_{2}^2-(Y_{2}^{\rm eq})^2\right) + 
{\langle{\sigma {\rm{v}}}}_{\xi_2 \bar{\xi_2}
\rightarrow \xi_1 \bar{\xi_1}} \rangle \bigg(Y_{2}^2 - 
\frac{(Y_{2}^{\rm eq})^2}{(Y_{1}^{\rm eq})^2}Y_{1}^2\bigg)
\bigg)
\,\, , \nonumber \\  
\label{boltz-eq-y2}\\
\frac{dY_{1}}{dx_{\xi_1}} &=&-
\dfrac{1}{2}\sqrt{\dfrac{\pi}{45\,G}}
\frac{m_{\xi_1}}{x_{\xi_1}^2}\sqrt{g_\star}\bigg({\langle{\sigma {\rm{v}}}_
{\xi_1 \bar{\xi_1} \rightarrow {X \bar{X}}}\rangle}
\left(Y_{1}^2-(Y_{1}^{\rm eq})^2\right) 
 - {\langle{\sigma {\rm{v}}}_{\xi_2 \bar{\xi_2}
\rightarrow \xi_1 \bar{\xi_1}}\rangle} \bigg(Y_{2}^2 - 
\frac{(Y_{2}^{\rm eq})^2}{(Y_{1}^{\rm eq})^2}Y_{1}^2\bigg)
\bigg)
\,\, , \nonumber\\ 
\label{boltz-eq-y1}
\end{eqnarray} 
where, $G$ is the Gravitational constant and $x_{\xi_i} = \frac{m_{\xi_i}}{T}$,
is a dimensionless variable with $T$ being the temperature of the Universe.
The quantity $g_{\star}$ is expressed as,
\begin{eqnarray}
\sqrt{g_\star} = \frac{h_{{\rm{eff}}}(T)}{\sqrt{g_{{\rm{eff}}}(T)}}
\left(1 + \frac{1}{3}\frac{d\,{\rm ln}(h_{{\rm{eff}}}(T))}{d\,{\rm ln}(T)}\right) \,\, ,
\label{gstar}
\end{eqnarray}
where, $h_{{\rm{eff}}}(T)$ and $g_{{\rm{eff}}}(T)$ are the effective
degrees of freedom related to the entropy and energy densities of
radiation. In the collision of term of the Boltzmann equation,
the first term represents pair annihilations of $\xi_i$ and $\bar{\xi}_i$
into particles which are in thermal equilibrium (including the
SM particles also) while the second term is due to the dark matter
conversion process $\xi_2 \bar{\xi}_2 \rightarrow \xi_1 \bar{\xi}_1$
where none of them are in thermal equilibrium during freeze-out. 
Detailed derivation of this term when both initial as well as
final state particles are not in thermal contact with the
visible world is given in Appendix \ref{coupled-term-BE}.
In the collision term, the thermal averaged annihilation cross
section of a particular process $AA^\prime \rightarrow BB^\prime$ has been
denoted by $\langle {\sigma {\rm v}}_{AA^\prime \rightarrow BB^\prime} \rangle$. The possible annihilation channels of both the DM candidates in our model are shown in figure \ref{Fig:feyn_ann}. This figure not only contains the Feynman diagrams for individual DM annihilations into other particles, but also the conversion of one particular DM pair into a pair of the other DM. We have solved these two coupled Boltzmann equations using 
\texttt{micrOMEGAs} \cite{Belanger:2014vza} where the model information has
been supplied to \texttt{micrOMEGAs} using \texttt{FeynRules}
\cite{Alloul:2013bka}. All the relevant annihilation cross sections
of dark matter number changing  processes required to solve
the coupled equations are calculated using \texttt{CalcHEP} \cite{Belyaev:2012qa}. 
Finally, after solving the Boltzmann equations we get the comoving
number density $Y_{i} (T_0)$ of each dark matter candidate
at the present epoch (at $T=T_0$). Thereafter, one can easily calculate
the total dark matter relic density which is sum of relic densities of
all dark matter candidates,
\begin{eqnarray}
\Omega_{\rm DM} h^2 = 2.755\times 10^8 \sum_{i=1}^{2}
\left(\frac{m_{\xi_i}}{\rm GeV}\right) Y_{i}(T_0) \,.
\end{eqnarray}
\begin{figure}[h!]
\includegraphics[height=6cm,width=8cm,angle=0]{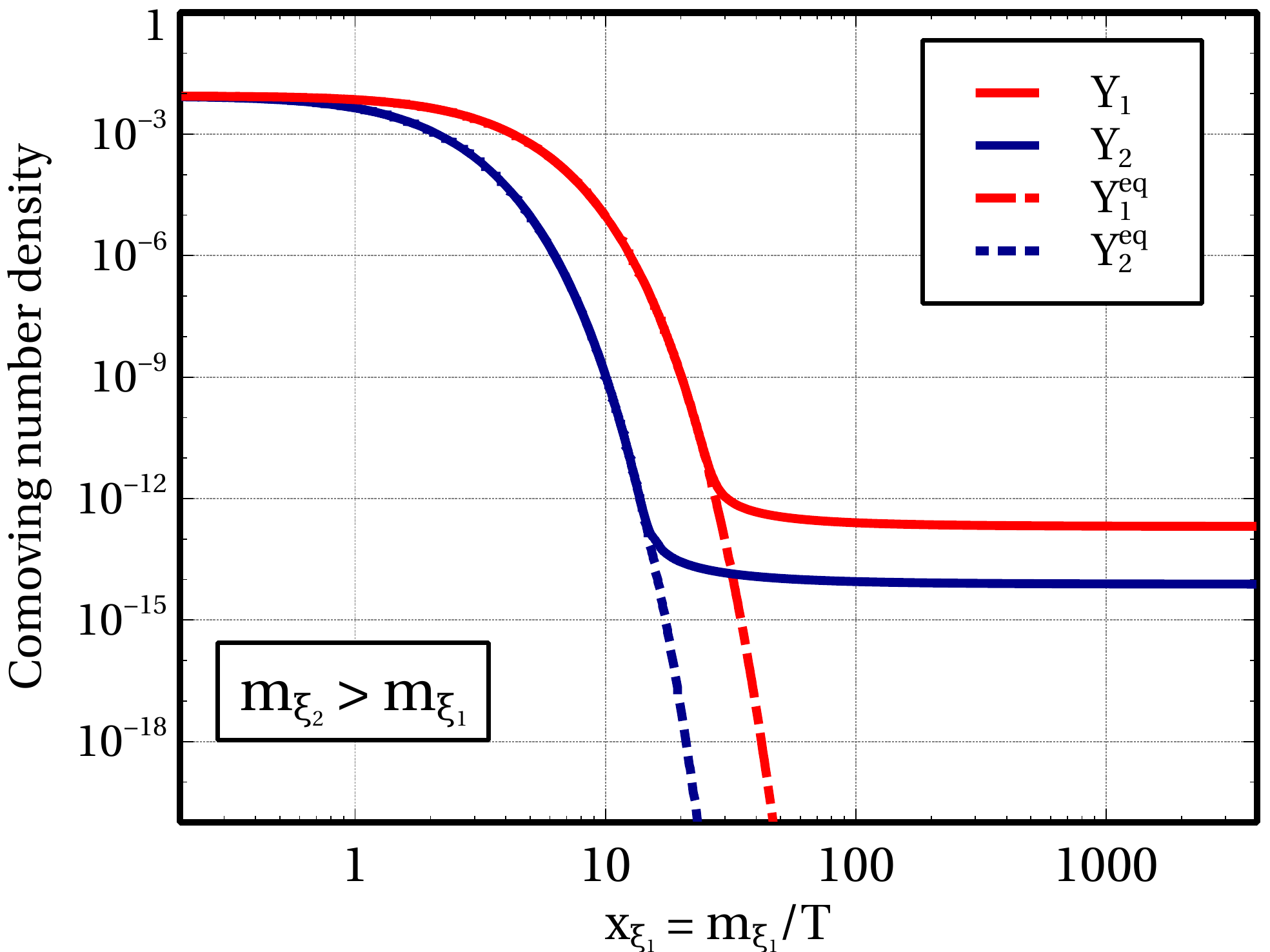}
\includegraphics[height=6cm,width=8cm,angle=0]{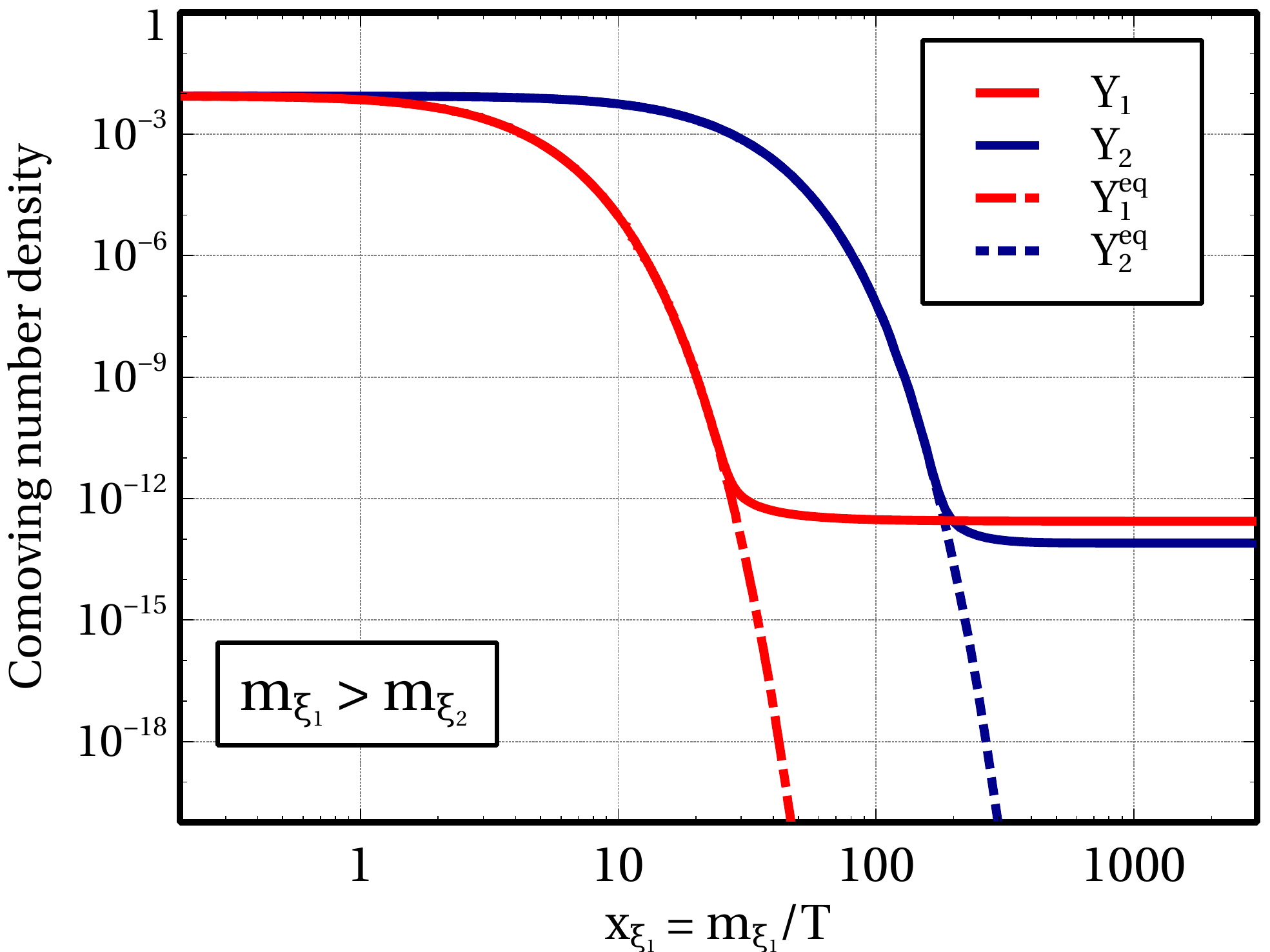}
\caption{Evolution of comoving number densities of both $\xi_1$
and $\xi_2$ for two cases (a) $m_{\xi_2}>m_{\xi_1}$ (left panel) and
(b) $m_{\xi_1}>m_{\xi_2}$
(right panel).}
\label{Yvsx1}
\end{figure}
In order to understand how the comoving number densities are varying
with respect to the temperature $T$, we have shown two plots
in the both panels of figure\,\ref{Yvsx1} illustrating the thermal
evolution of $Y_{1}$ and $Y_{2}$ for $m_{\xi_2}>m_{\xi_1}$ (left panel) and 
$m_{\xi_1}>m_{\xi_2}$ (right panel) respectively. In the
left panel, we consider $m_{\xi_2}=2 m_{\xi_1}$ with $m_{\xi_1}=2110$ GeV.
Here, the red solid line represents the variation of $Y_{1}$ with $x_{\xi_1} =
\frac{m_{\xi_1}}{T}$ while the same for heavier component $\xi_2$
has been indicated by the blue solid line. The corresponding equilibrium
number densities computed using the Maxwell-Boltzmann distribution
are shown by red dash-dotted and blue dotted lines respectively.
Form this plot it is seen that initially for low $x_{\xi_1}$
(i.e.\,\,for high $T$) the comoving number density of each
component follows their respective
equilibrium number density upto a certain temperature $T$
(different for $\xi_1$ and $\xi_2$) and thereafter $Y_{i}$ departs
significantly from the corresponding equilibrium distribution
function $Y^{\rm eq}_{i}$ and remains constant with
respect to the variation of $T$. This is nothing
but the well known freeze-out point of a WIMP and
it depends on when the interaction rate corresponding
to the number changing processes of a particular dark matter
component goes below the expansion rate of the Universe
governed by the Hubble parameter $\mathbf{H}$. In this
particular situation, freeze-out of $\xi_1$ occurs at
$x_{\xi_1} \sim 25$ while that for the heavier component
is $\frac{m_{\xi_2}}{T} \sim 30$ ($x_{\xi_2}=2x_{\xi_1}$).
Furthermore, the interactions of $\xi_1$ and $\xi_2$ are
such that $\xi_2$ which freezes-out earlier has less
relic abundance compared to its lighter counter part
$\xi_1$. The opposite situation is presented in the right panel of
figure\,\ref{Yvsx1}, where $\xi_1$ is the heavier dark matter
component. Here we consider the model parameters in such a way
that although the heavier component $\xi_1$ has earlier
freeze-out, ends up with a greater abundance.

\section{Direct Detection}
\label{sec:dd}
\begin{figure}[h!]
\centering
\includegraphics[height=3cm,width=5cm]{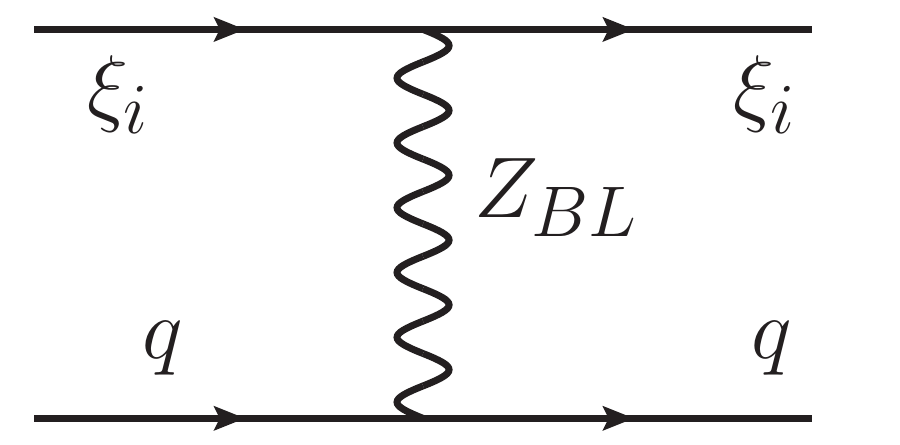}
\includegraphics[height=3cm,width=5cm]{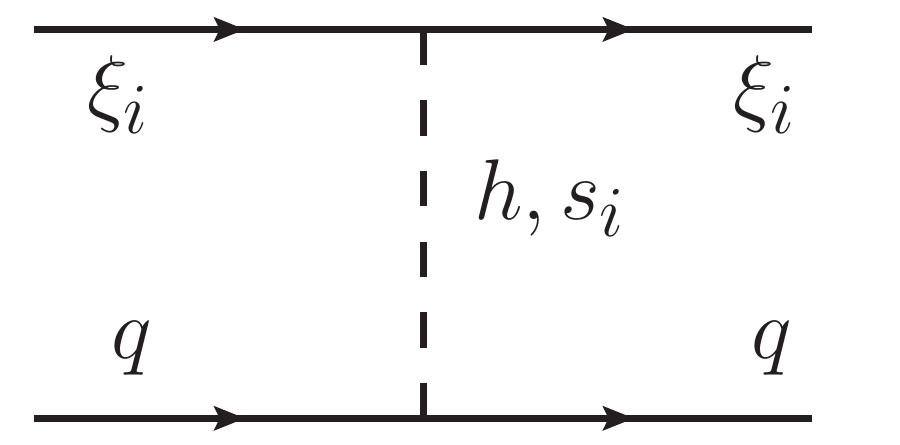}
\caption{Feynman diagrams for spin-independent elastic scattering processes of DM with nucleons (or quarks) in the model.}
\label{Fig:feyn_DD}
\end{figure}

Since each of the DM candidates in our model is a Dirac fermion, there exists $Z_{BL}$ as well as scalar mediated spin independent elastic scattering processes off nucleons. The relevant Feynman diagrams are shown in figure \ref{Fig:feyn_DD}. Since several ongoing experiments like LUX \cite{Akerib:2016vxi}, PandaX-II \cite{Tan:2016zwf, Cui:2017nnn} and Xenon1T \cite{Aprile:2017iyp, Aprile:2018dbl} are looking for such processes, regularly giving stringent upper bounds on DM-nucleon scattering cross section, we can further constrain our model parameters from these data. Parametrising DM and quark interactions with $Z_{BL}$ as 
$$ \mathcal{L} \supset [\overline{\xi_i} \gamma^{\mu} (g_{\xi_i v}+ g_{\xi_1 a} \gamma^5) \xi_i +\overline{q} \gamma^{\mu} (g_{q v}+ g_{q a} \gamma^5) q](Z_{BL})_{\mu}, $$
the dominant spin independent DM-nucleus scattering cross section can be written down as \cite{Berlin:2014tja}
\begin{equation}
\sigma^{\rm SI}_{\xi_i} = \frac{\mu^2_{\xi_i N} g^2_{\xi_i v}}{\pi M^4_{Z_{BL}}} \big [ Z(2\tilde{b}_u + \tilde{b}_d) +(A-Z) (\tilde{b}_u+2\tilde{b}_d) \big ]^2
\end{equation}
where $\mu_{\xi_i N}$ is the reduced mass of DM nucleus system, $\tilde{b}_q$ are the quark-$Z_{BL}$ couplings which is same for all quarks in $B-L$ gauge model. Also, $A, Z$ are mass number and atomic number of the nucleus. From the Lagrangian of DM given in \eqref{eq:DMZBL}, we can write 
$$ g_{\xi_1 v} = \frac{9}{10} g_{BL}, \; g_{\xi_2 v} = \frac{4}{5} g_{BL} $$
which can be used in the expression above to find the DM-nucleus scattering cross section numerically for different values of $g_{BL}, M_{Z_{BL}}$. Similarly, if the scalar $(s_j)$ mediated interactions are parametrised as
$$ \mathcal{L} \supset [\overline{\xi_i}  (\lambda_{\xi_i s}+ \lambda_{\xi_1 p} \gamma^5) \xi_i +\overline{q}  (\lambda_{q s}+ \lambda_{q p} \gamma^5) q] s_j $$
the corresponding spin-independent DM-nucleus scattering cross section can be written as
\begin{equation}
\sigma^{\rm SI}_{\xi_i} = \frac{\mu^2_{\xi_i N} \lambda^2_{\xi_i s}}{\pi M^4_{s_j}} \big [ Z \tilde{f}_p +(A-Z) \tilde{f}_n \big ]^2.
\end{equation}
Here $ \tilde{f}_{p,n}$ are defined as
\begin{equation}
\frac{ \tilde{f}_{p,n}}{m_{p,n}} = \sum_{q=u,d,s} f^{p,n}_{T_q} \frac{ \tilde{f}_q}{m_q} + \frac{2}{27} f_{TG} \sum_{q=c,b,t} \frac{ \tilde{f}_q}{m_q}
\end{equation}
with $\tilde{f}_q = \lambda_{s_i} m_q/{\rm GeV}$ and $f_{TG} = 1-f^{p,n}_{T_u}-f^{p,n}_{T_d}-f^{p,n}_{T_d}$. Here $\lambda_{s_i}$ denotes the quark-singlet scalar couplings which can be derived by using the singlet scalar-SM Higgs mixing shown in Appendix \ref{rs_matrix_diagonalisation}. We take the standard values of other parameters appearing in the above formula as $ f^p_{T_u}=0.020, f^p_{T_d}=0.026$ (and opposite for $f^n_{T_{u,d}} $), $f_{T_s}=0.043$ which further gives $f_{TG} \approx 0.91$. We extract the spin independent elastic scattering cross section for both the DM candidates off nucleons from \texttt{micrOMEGAs}. Keeping the fact in mind that we are analysing a two-component DM scenario we have multiplied the elastic scattering cross-section by the relative number density of each DM candidate to find the individual effective DM-nucleon scattering cross section. 
{As mentioned earlier, both gauge interactions as well as Yukawa interactions will contribute in the direct detection cross sections. However, the gauge mediated interaction is proportional to $g_{BL}^4$ whereas the scalar mediated processes are mixing suppressed as well as Yukawa coupling suppressed (the Yukawa couplings of first generation of quarks are $\mathcal{O}(10^{-6})$). Hence, in our present model DM scattering off nucleons through $Z_{BL}$ will have dominant contribution.}

\section{Results}
\label{sec:results}
\begin{figure}[h!]
\centering
\includegraphics[scale=0.5]{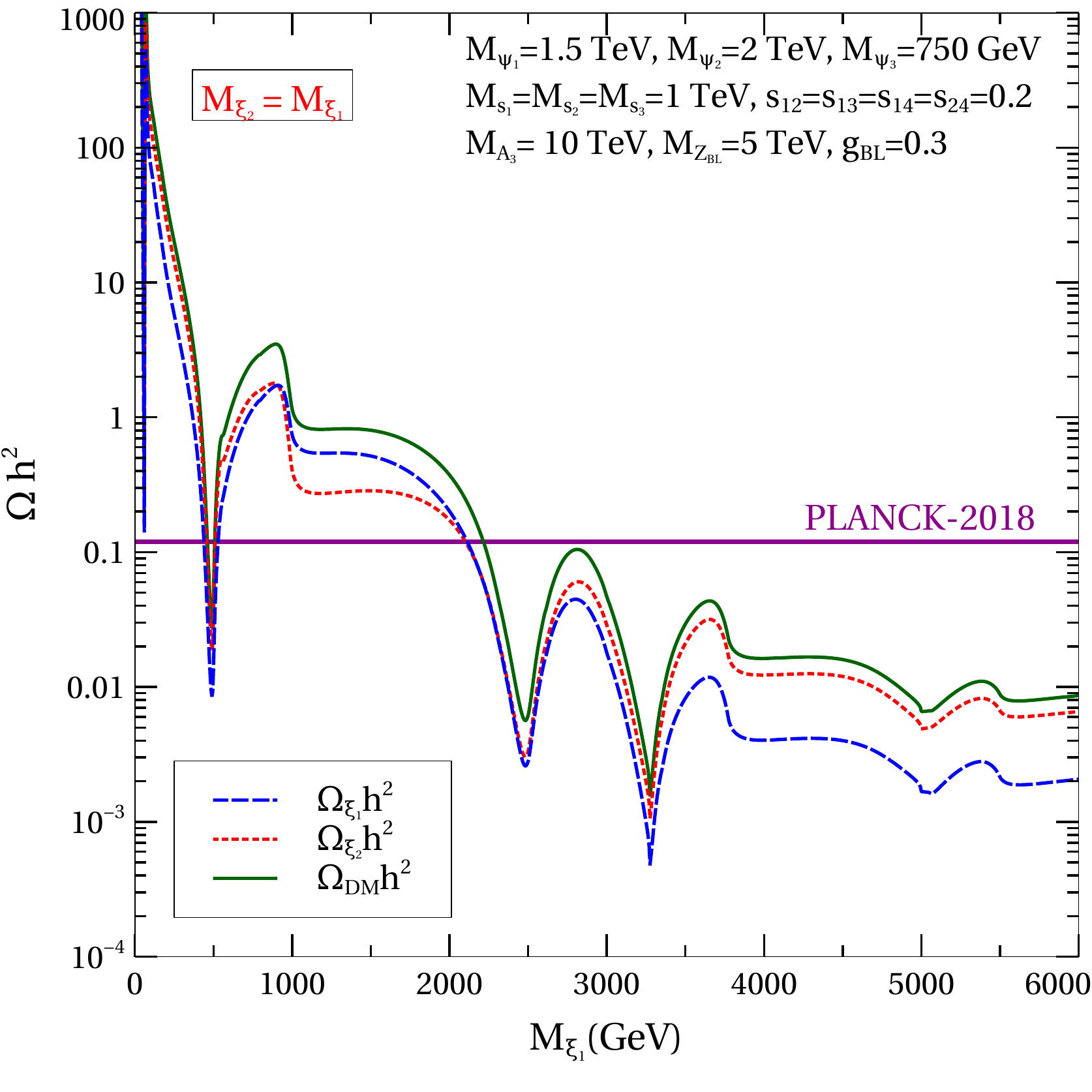}
\caption{Relic abundance of two DM candidates with degenerate masses keeping
all other model parameters fixed to benchmark values.}
\label{Fig:omega-vs-m1-1}
\end{figure}

As we have discussed in the previous section, this model predicts two automatically stable DM candidates $\xi_1$ and $\xi_2$. The total relic density of DM can be expressed as, the sum of the relic densities of $\xi_1$ and $\xi_2$, $\rm {\Omega_{DM} h^2= \Omega_{\xi_1} h^2 + \Omega_{\xi_2} h^2}$  where $\rm {\Omega_{\xi_1} h^2}$ and $\rm {\Omega_{\xi_2} h^2}$ are the relic abundances of $\xi_1$ and $\xi_2$ respectively. Let us now investigate the dependence of the relic densities, $\Omega_{\xi_1}$ and $\Omega_{\xi_2}$, on the parameters of the model. In figure \ref{Fig:omega-vs-m1-1}, we have shown the variation of relic density with the DM mass by assuming $ \rm{M_{\xi_2} = M_{\xi_1}}$. The other parameters of the model were chosen as $\rm {M_{\psi_1}=1.5\ TeV, M_{\psi_2}=2\ TeV, M_{\psi_3}=750\ GeV, M_{s_1}=M_{s_2}=M_{s_3}= \ 1\ TeV,}$ $\rm{M_{A_3}=10\ TeV,\\ s_{12}=s_{13}=s_{14}=s_{24}=0.2, M_{Z_{BL}}=5\ TeV, g_{BL}=0.3}$. The dashed (blue) and dotted (red) lines denote the relic densities of each DM
candidate,$\Omega_{\xi_1}h^2$ and $\Omega_{\xi_2}h^2$, whereas the solid (green) line is their sum ($\Omega_{DM}h^2$). The horizontal magenta line represents the relic density bound from PLANCK data \cite{Aghanim:2018eyx}. Some important features of the model can be indicated here. The resonances due to the s-channel annihilation through $Z_{BL}$ and the different scalars ($\rm{s_1,s_2,s_3, A_2, A_3}$) has reduced the relic density as expected. Figure \ref{Fig:omega-vs-m1-1}  clearly shows four different scalars and Z$_{BL}$ resonances, as we have assumed all the scalars have the same mass, at a DM mass of 500 GeV, 2.5 TeV, 3.27 TeV, and 5 TeV respectively. At DM mass of 1 TeV, DM particles starts annihilating into the two scalars final state giving a sudden reduction in the relic density. Similar behaviour can be found at the DM mass around 3.77 TeV, and 5.5 TeV where DM starts annihilating into one scalar ( $\rm{ M_{s_i} = 1 \ TeV}$ ) plus one pseudo scalar ( $\rm{ M_{A_2} = \sqrt{\frac{3}{7}} M_{A_3} = 6.54 \ TeV}$ and $\rm {M_{A_3}=10 \ TeV}$ ) final states(Feynman diagrams in figure \ref{Fig:feyn_ann}). One can expect similar reduction for the annihilation into one scalar ( $\rm{ M_{s_i} = 1 \ TeV}$ ) and one Z$_{BL}$ ( $\rm{ M_{Z_{BL}} = 5 \ TeV}$ ) final state near 3 TeV mass. However, because of the resonance that effect is not visible. Another important point to note here is that, $\Omega_{\xi_2}$ is subdominant throughout the whole mass range. That can be explained as follows, $\xi_2$ has formed by combining $\rm{N_{2L}}$ and $\rm{N_{2R}}$ where as $\xi_{1}$ has formed from $\rm{N_{1L}}$ and $\rm{N_{1R}}$. The $\rm{B-L}$ quantum number assigned for 
$\rm{N_{2L}}$ and $\rm{N_{2R}}$ is greater than the quantum number for $\rm{N_{1L}}$ and $\rm{N_{1R}}$ and that increases the annihilation cross section of $\xi_2$ by some numerical factor which results to smaller abundance.

\begin{figure}[h!]
\centering
\subfigure[]
{\includegraphics[scale=0.45]{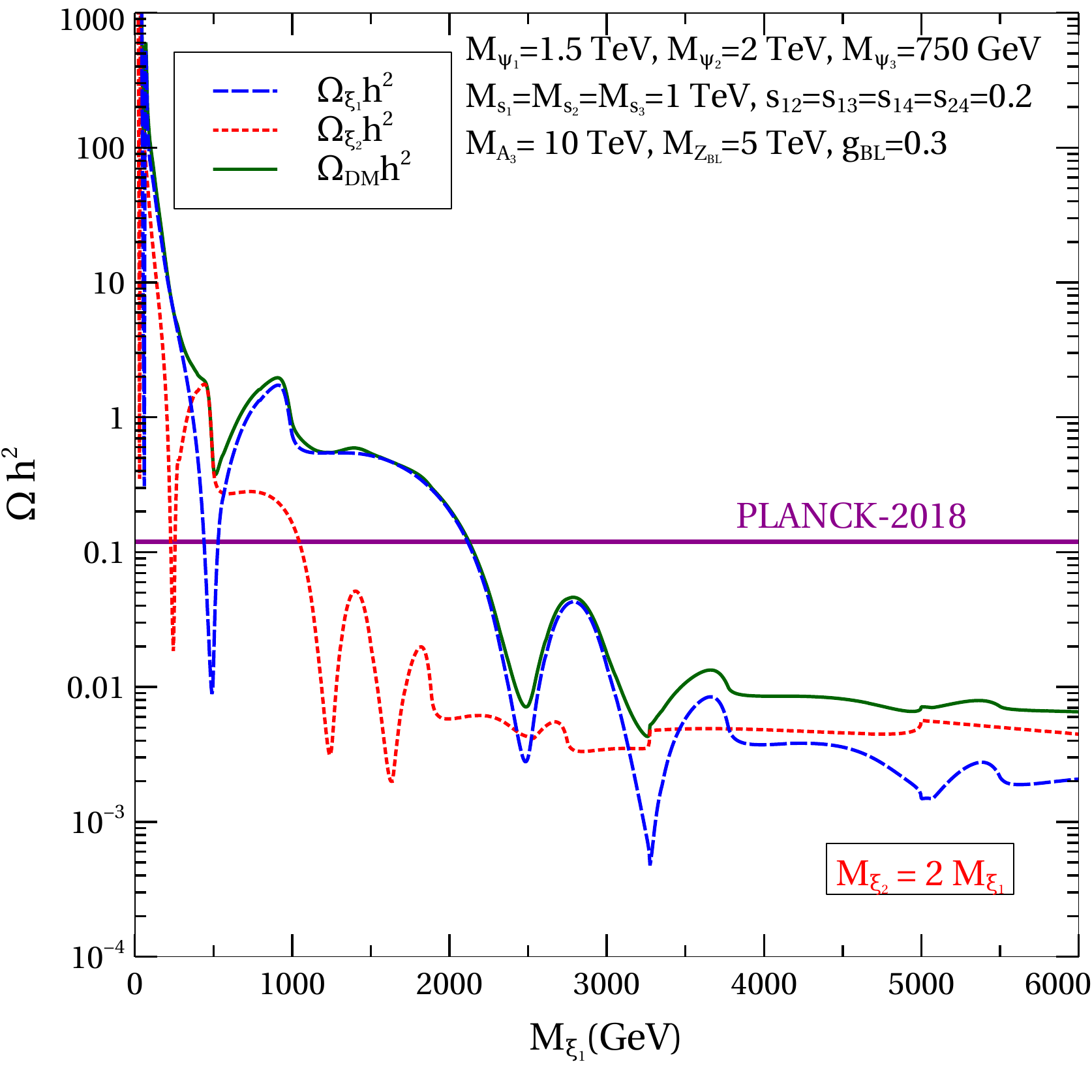}}
\,\,
\subfigure[]
{\includegraphics[scale=0.45]{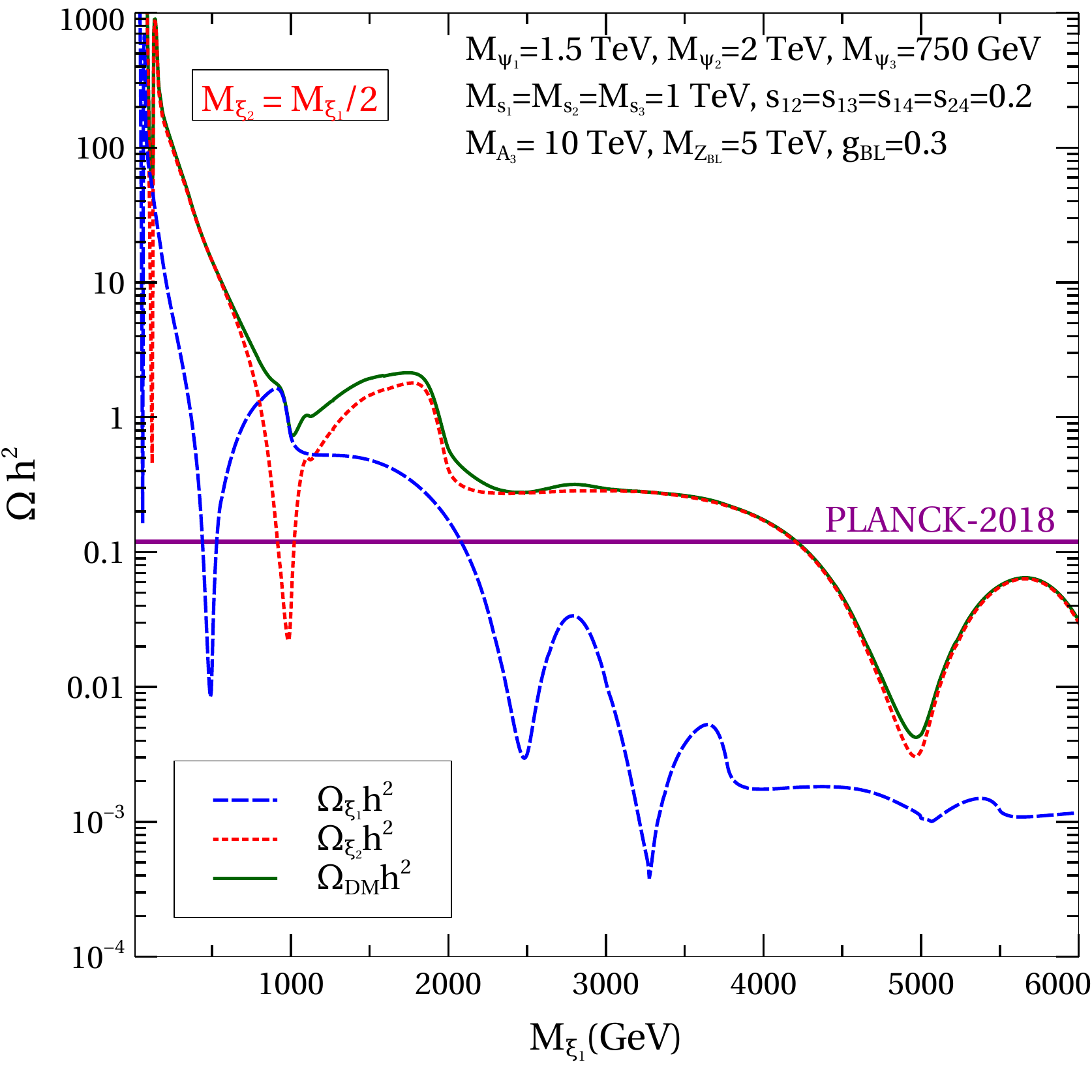}}
\caption{Relic abundance of two DM candidates with non-degenerate masses: $M_{\xi_2}=2M_{\xi_1}$ (left panel) and $M_{\xi_2}=M_{\xi_1}/2$ (right panel), keeping all other model parameters fixed to benchmark values.}
\label{Fig:omega-vs-m1-2}
\end{figure}

{One point to note here is that both the gauge coupling $g_{BL}$ as well as Yukawa couplings $\mathcal{Y}_1\, \, ,\mathcal{Y}_2 $ play important roles in the DM annihilation. The gauge coupling $g_{BL}$ is involved in the annihilation channels mediated by gauge boson $Z_{BL}$ while the scalar mediated annihilation cross sections depend on Yukawa couplings $\mathcal{Y}_1\, \, ,\mathcal{Y}_2 $.  In the lower mass region of DM $g_{BL}$ plays the main role whereas in the high mass region $\mathcal{Y}_i$ also gives a significant contribution as $\mathcal{Y}_i = \sqrt{2}M_{\xi_i}/u $}. In figure \ref{Fig:omega-vs-m1-2}, we have shown the variation of the relic density for two different relations between $\rm{M_{\xi_1}}$ and $\rm{M_{\xi_2}}$. In left panel, we have assumed $\rm{M_{\xi_2}}= \rm{2 \ M_{\xi_1}}$ and in right panel $\rm{M_{\xi_2}} = \rm{M_{\xi_1}/2}\ -$ other parameters remain same as in figure \ref{Fig:omega-vs-m1-1}. In figure \ref{Fig:omega-vs-m1-2}a, the resonances have occurred at different positions for $\Omega_{\xi_2}$, at 250 GeV, 1.25 TeV, 1.63 TeV, and 2.5 TeV, whereas, $\rm{\Omega_{\xi_1}}$  has the same behaviour as in \ref{Fig:omega-vs-m1-1} which is expected as we have assumed the $\rm{M_{\xi_2}} = \rm{2 \ M_{\xi_1}}$. Figure \ref{Fig:omega-vs-m1-2}b can be explained in a similar way. 

\begin{figure}[h!]
\centering
\includegraphics[scale=0.45]{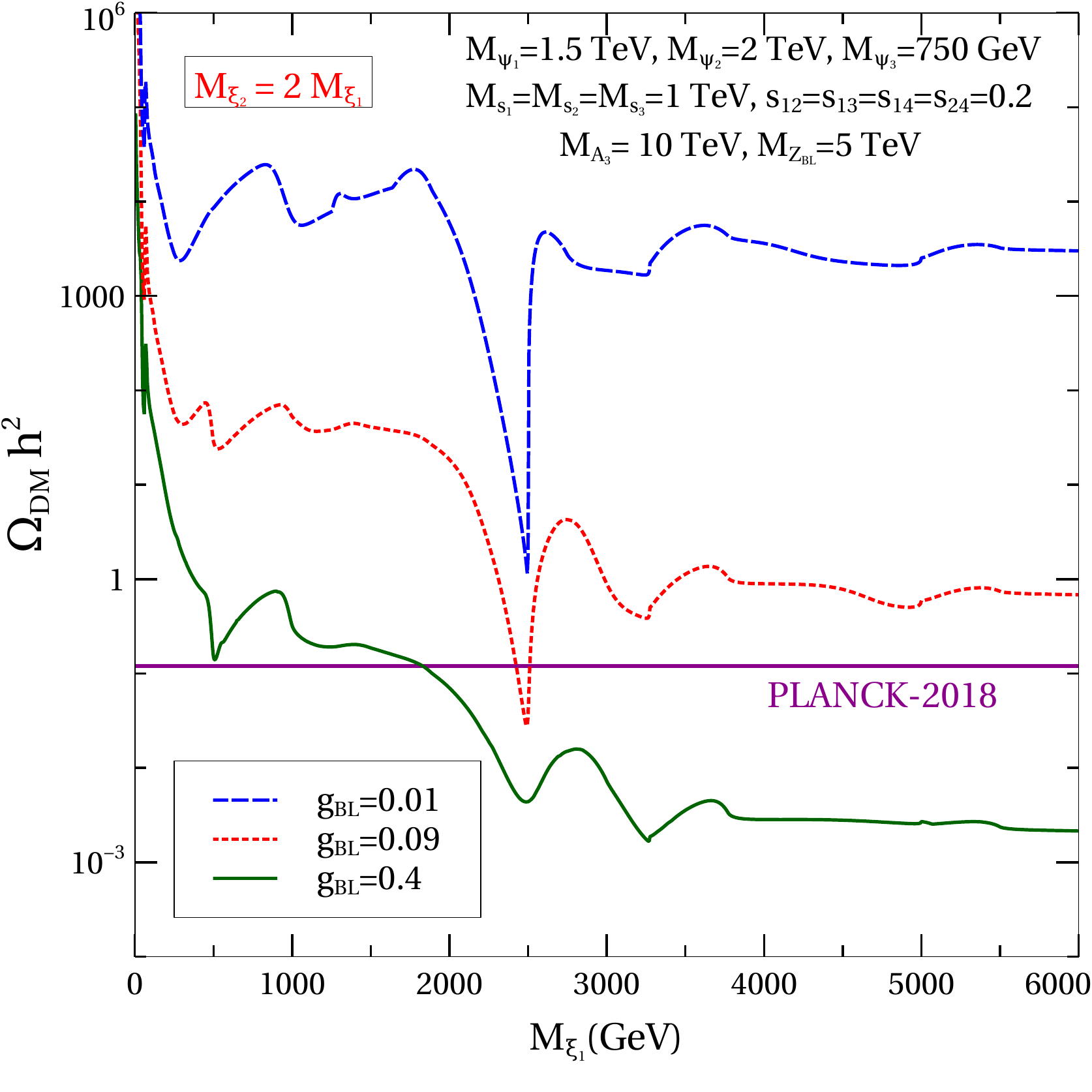}
\,\,
\includegraphics[scale=0.45]{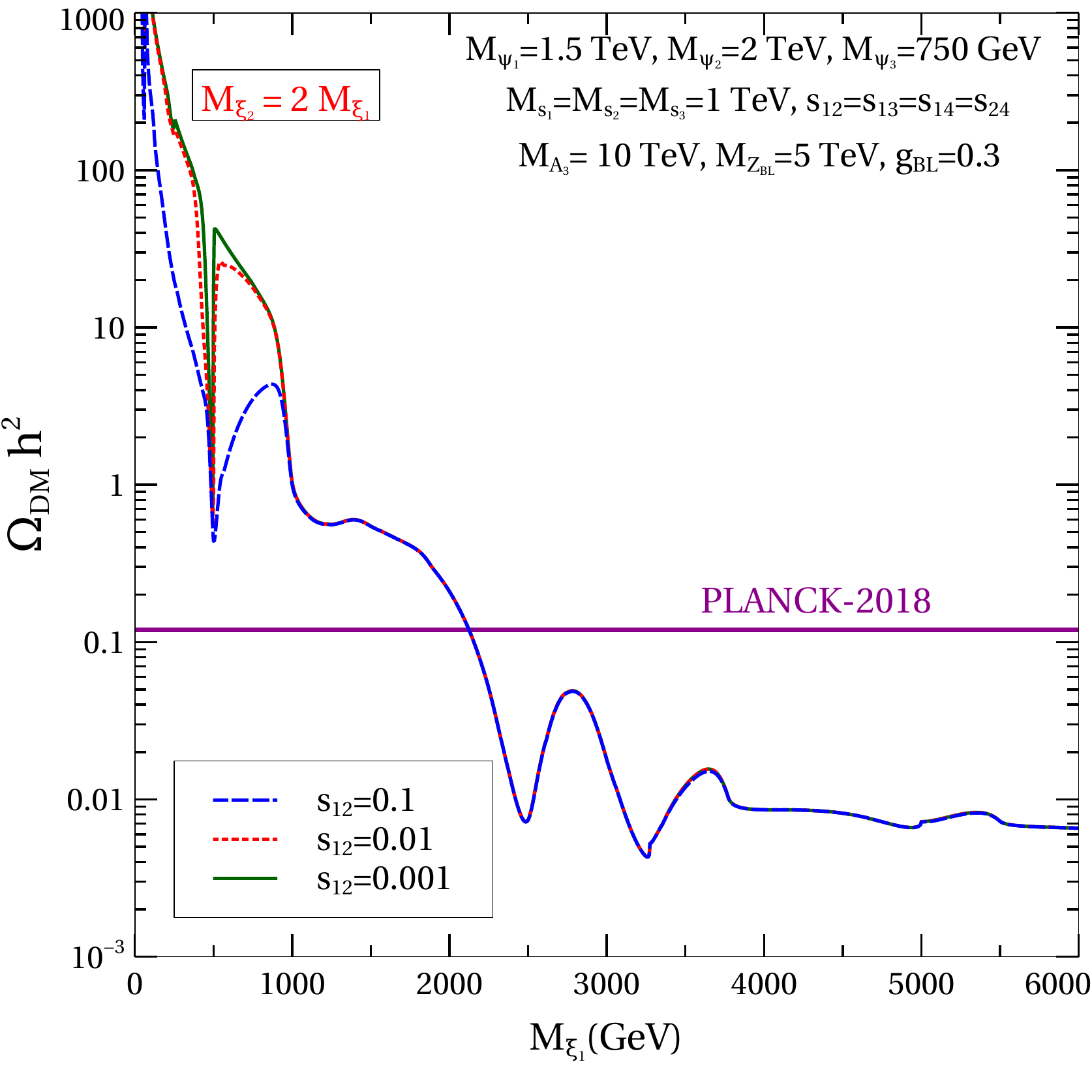}
\caption{Relic abundance of two DM candidates with non-degenerate masses ($M_{\xi_2}=2M_{\xi_1}$) for different benchmark values of: gauge coupling $g_{BL}$ (left panel), singlet scalar-SM Higgs mixing (right panel).}
\label{Fig:gBL-mixing}
\end{figure}

In figure \ref{Fig:gBL-mixing} we have shown the variation of the total relic density as a function of m$_{\xi_1}$ for three benchmark values of g$_{BL}$ (0.01, 0.09, 0.4) and mixing angle (0.001, 0.01, 0.1). The left panel shows that the total DM abundance increases as we choose smaller values of gauge coupling. It is because small g will decrease the annihilation cross section and eventually increase the DM abundance. The right panel shows that the relic abundance hardly depends on the mixing angle. Here for simplicity we have assumed all the mixing angle to be same. In both the cases, we have assumed $\rm{M_{\xi_2}} = \rm{2 \ M_{\xi_1}}$ and the other parameters remain same as in figure \ref{Fig:omega-vs-m1-1}.   

\begin{figure}[h!]
\centering
\includegraphics[scale=0.45]{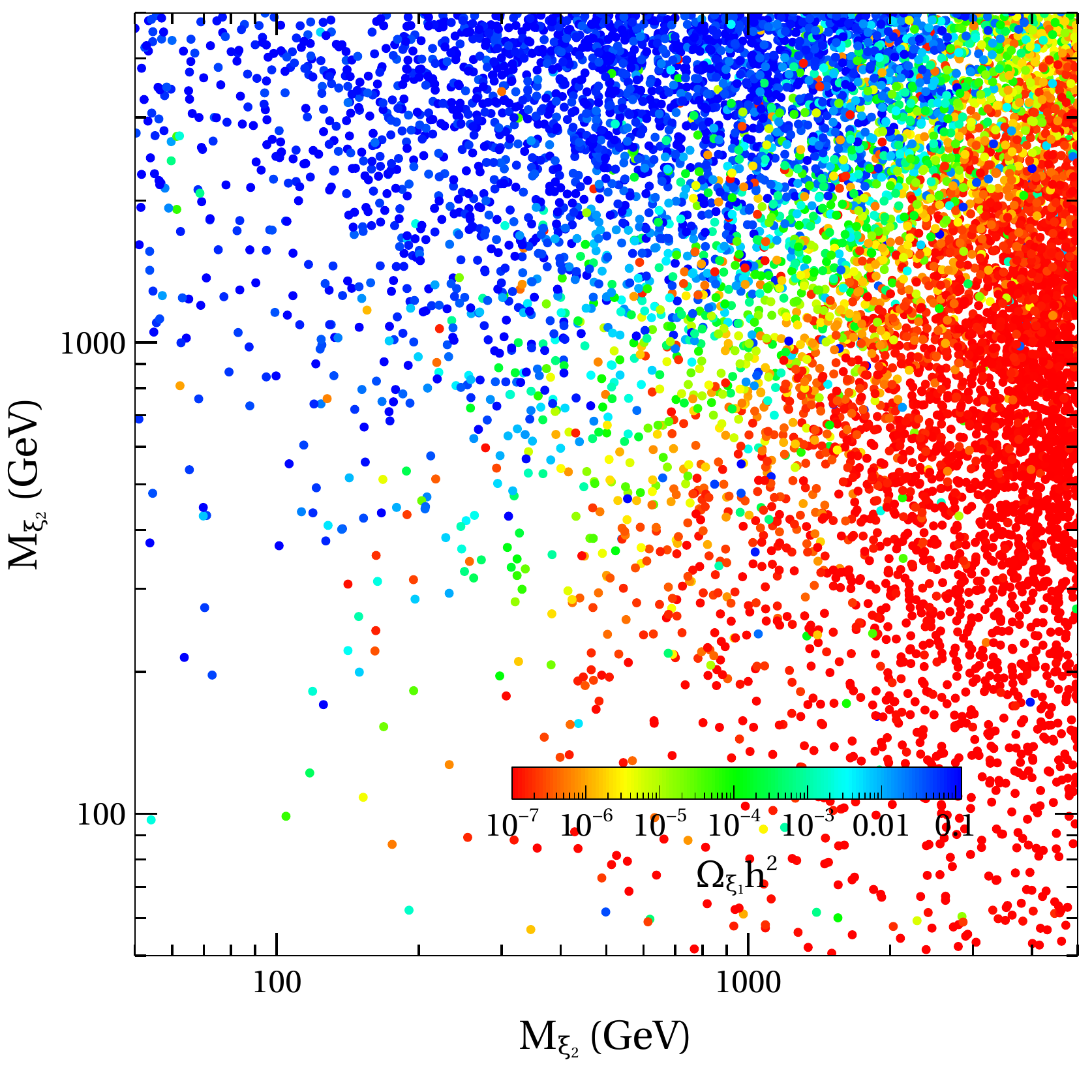}
\,\,
\includegraphics[scale=0.45]{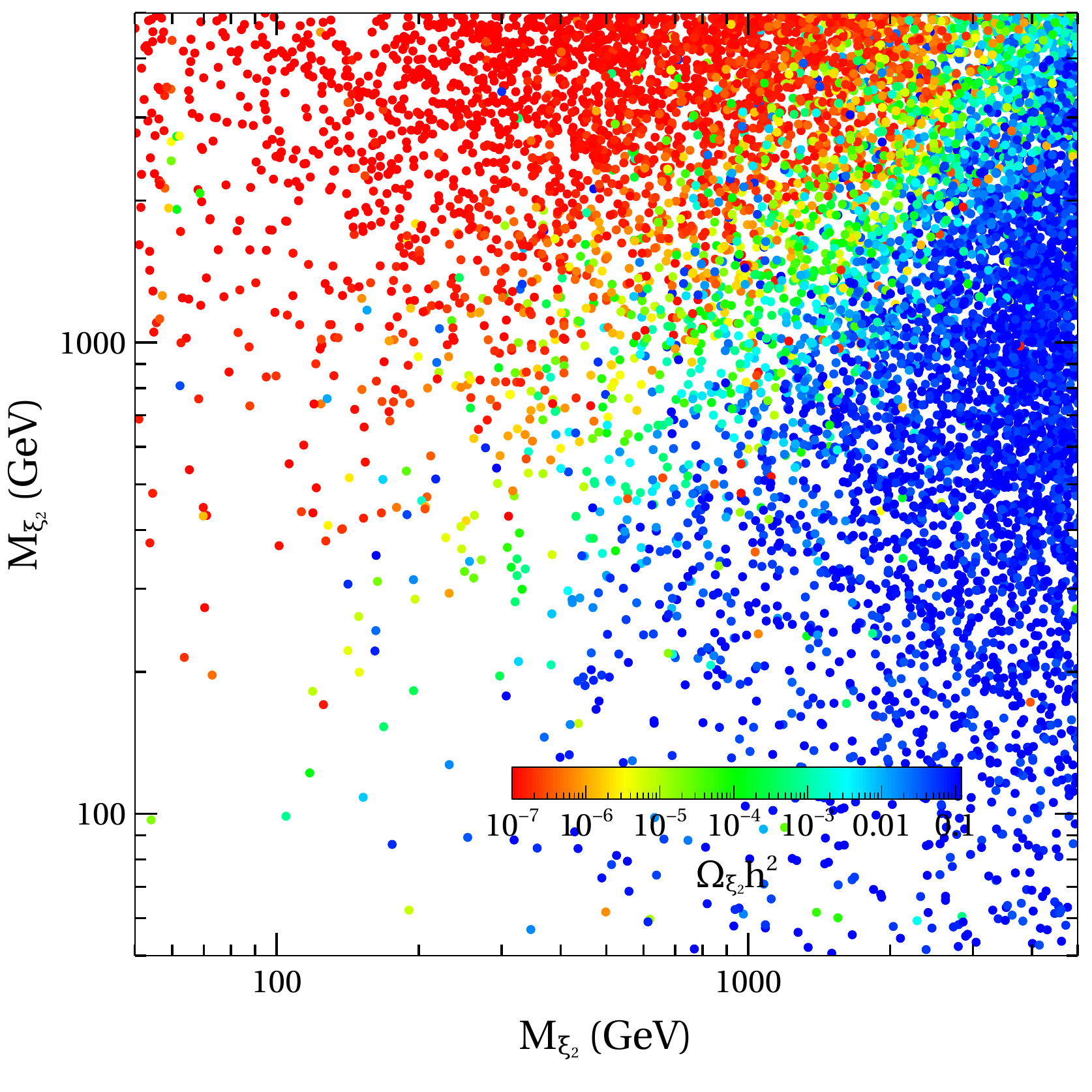}
\caption{Scan plot showing the parameter space in $M_{\xi_1}-M_{\xi_2}$ plane
allowed from total DM relic abundance, perturbativity, bounded from below
criteria of the scalar potential.}
\label{Fig:scan:mDM1:mDM2}
\end{figure}

\begin{figure}[h!]
\centering
\includegraphics[scale=0.45]{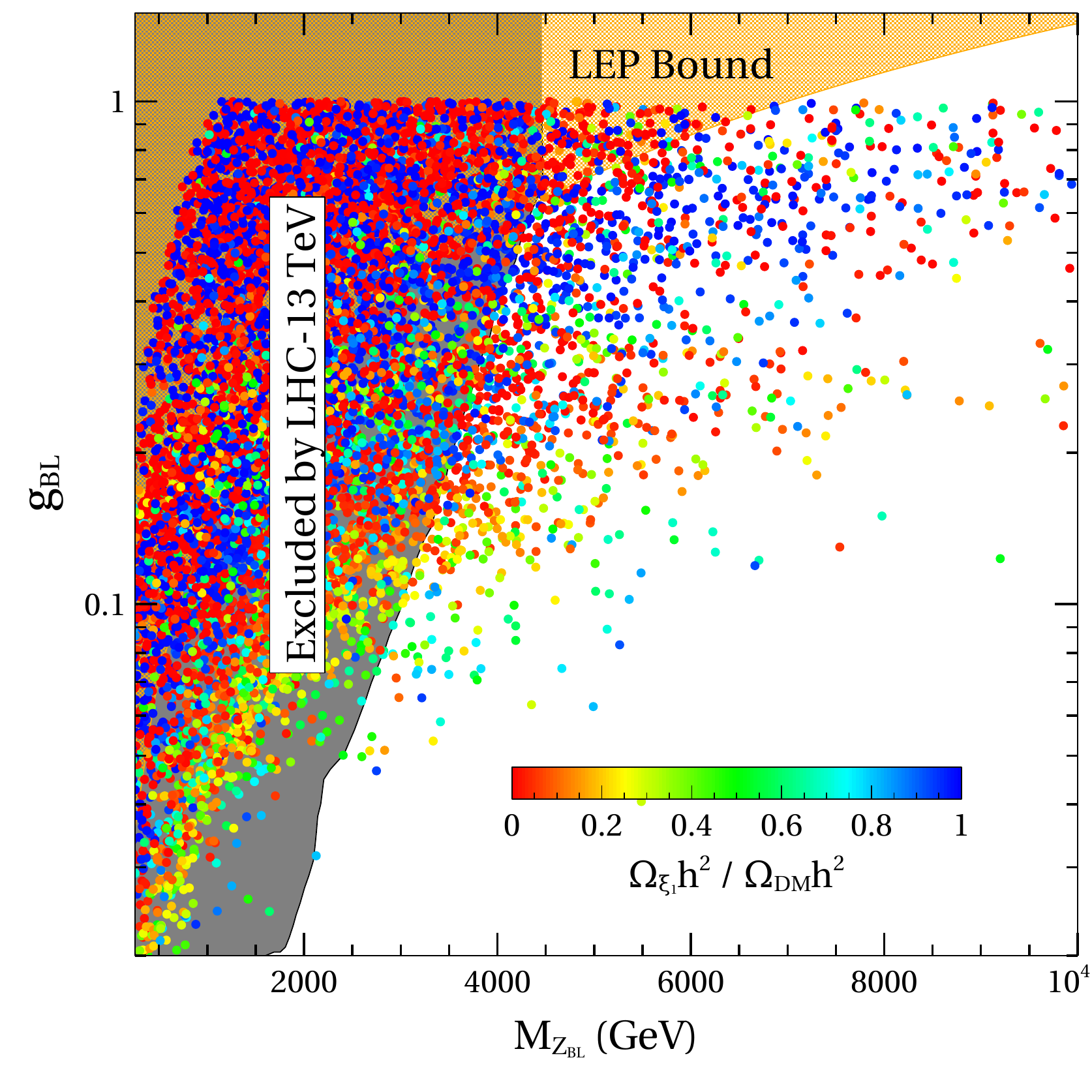}
\,\,
\includegraphics[scale=0.45]{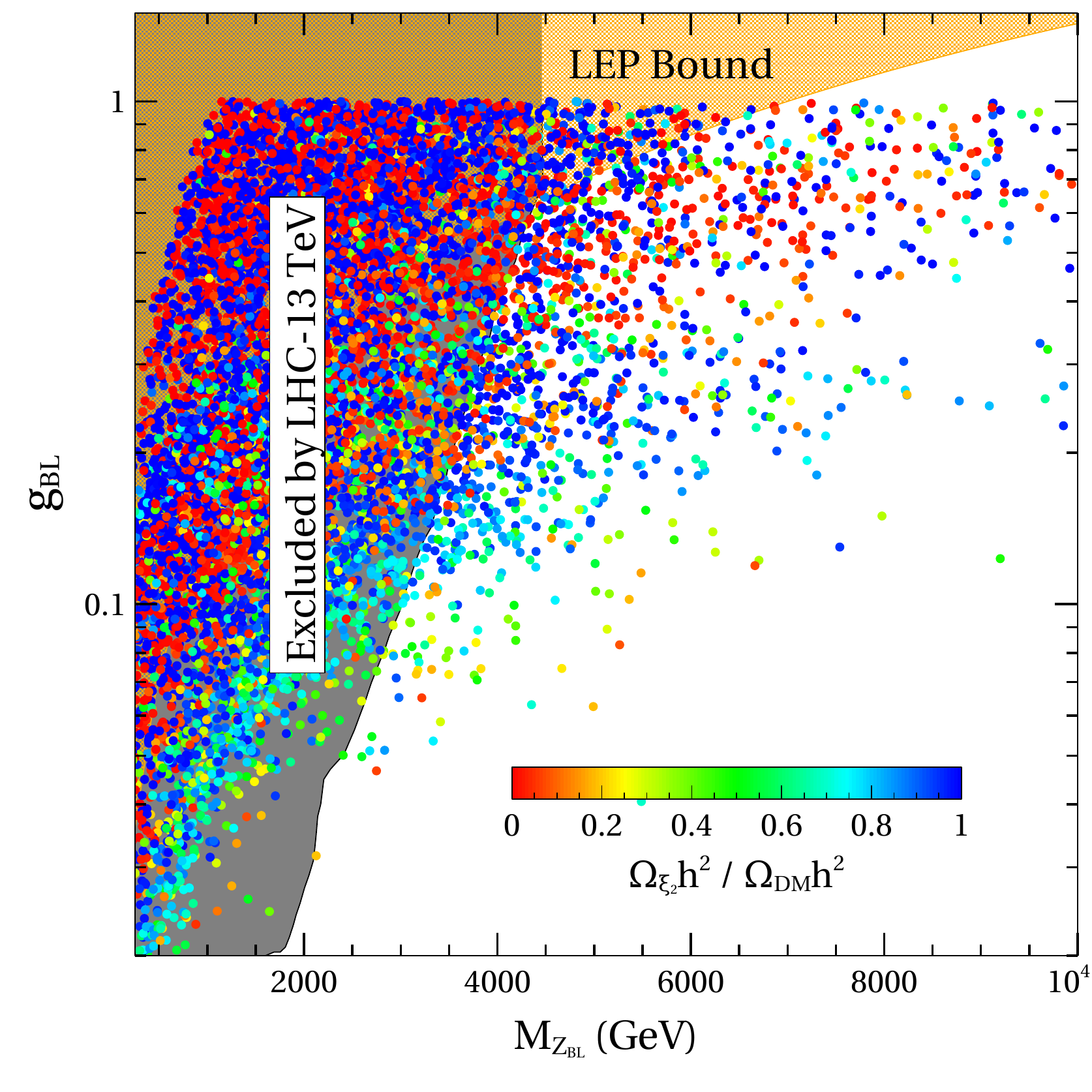}
\caption{Scan plot showing the parameter space in $g_{BL}-M_{Z_{BL}}$ plane
allowed from total DM relic abundance, perturbativity, bounded from below
criteria of the scalar potential.}
\label{Fig:scan:gBL-mZBL}
\end{figure}

As seen from the above plots, the relic abundance of both the DM candidates primarily depend upon the strength of their annihilation cross sections via scalar or gauge portal interactions. After understanding this behaviour from the plots with benchmark choices of model parameters, we now move onto performing a random scan over a set of free parameters mentioned in table \ref{tab:scan}. While the physical masses of DM and all new particles in the model are varied in the mentioned range along with $g_{BL}$, the singlet scalar mixing angles with SM Higgs boson as well as among themselves are kept fixed at $0.2$. We then apply the relevant constraints one by one to arrive at the final allowed parameter space from all relevant bounds. We first show the parameter space in $M_{\xi_1}-M_{\xi_2}$ plane allowed from total DM relic abundance, perturbativity, bounded from below criteria of the scalar potential in figure \ref{Fig:scan:mDM1:mDM2}. In figure \ref{Fig:scan:gBL-mZBL}, we then show the parameter space in $g_{BL}-M_{Z_{BL}}$ plane allowed from total DM relic abundance, perturbativity, bounded from below criteria of the scalar potential. The bounds from the LEP and the LHC\footnote{Note that we have used the LHC bound in $M_{Z_{BL}}-g_{BL}$ plane from the ATLAS collaboration results \cite{Aaboud:2017buh}. The more recent limits from the ATLAS collaboration with increased luminosity but with same centre of mass energy \cite{Aad:2019fac} have small deviations from the limits we apply here.} are separately shown by the respective shaded regions which are excluded. The coloured bar in the two plots shown in figure  \ref{Fig:scan:gBL-mZBL} correspond to the relative abundances of two DM candidates $\xi_1$ (left panel) and $\xi_2$ (right panel) respectively. 

\begin{table}
\begin{center}
\begin{tabular}{|c|c|}
\hline
Parameters & Range   \\
\hline
\hline
$\rm{M_{\xi_1}}$ &  (10 GeV, 8 TeV)\\
$\rm{M_{\xi_2}}$ &  (10 GeV, 8 TeV)\\
\hline
$\rm{M_{Z_{BL}}}$ &  (100 GeV, 10 TeV) \\
$\rm{g_{BL}}$ &  (0.0001, 1)\\
\hline
$\rm{M_{s_1}}$ &  (100 GeV, 10 TeV)\\
$\rm{M_{s_2}}$ &  (100 GeV, 10 TeV)\\
$\rm{M_{s_3}}$ &  (100 GeV, 10 TeV)\\
\hline
$\rm{M_{A_3}}$ & (1 TeV, 20 TeV)\\
$\rm{M_{A_2}}$ & $\sqrt{\frac{3}{7}} \ \rm{M_{A_3}}$\\
\hline
$\rm{M_{\psi_3}}$ & (1 TeV, 2.5 TeV)\\
$\rm{M_{\psi_2}}$ & 750 GeV + $\rm{M_{\psi_3}}$\\
$\rm{M_{\psi_1}}$ & 1.5 TeV + $\rm{M_{\psi_3}}$\\
\hline
\end{tabular}
\end{center}
\caption{The parameters of our model and ranges used in the random scan}
\label{tab:scan}
\end{table}

\begin{figure}[h!]
\centering
\includegraphics[scale=0.45]{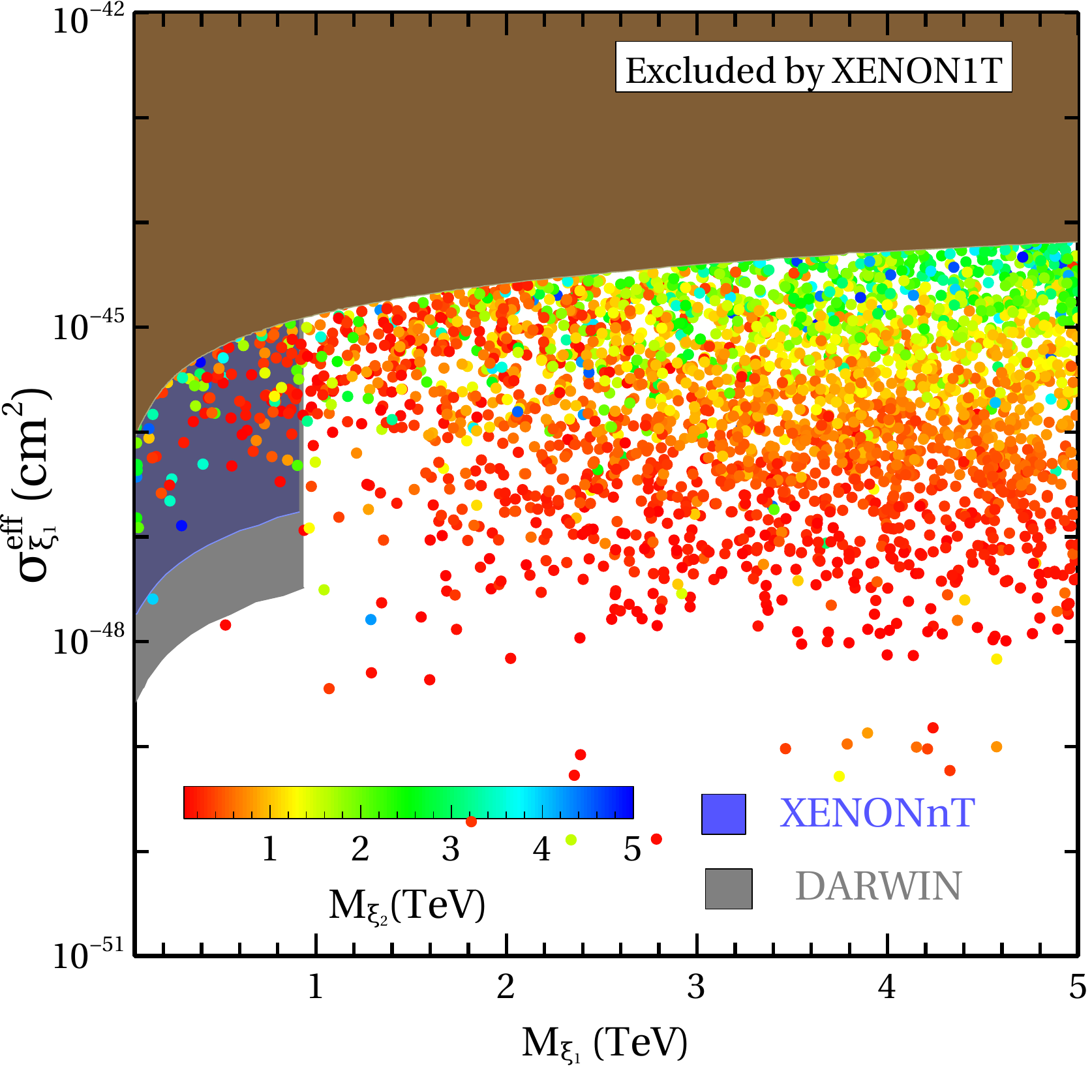}
\,\,
\includegraphics[scale=0.45]{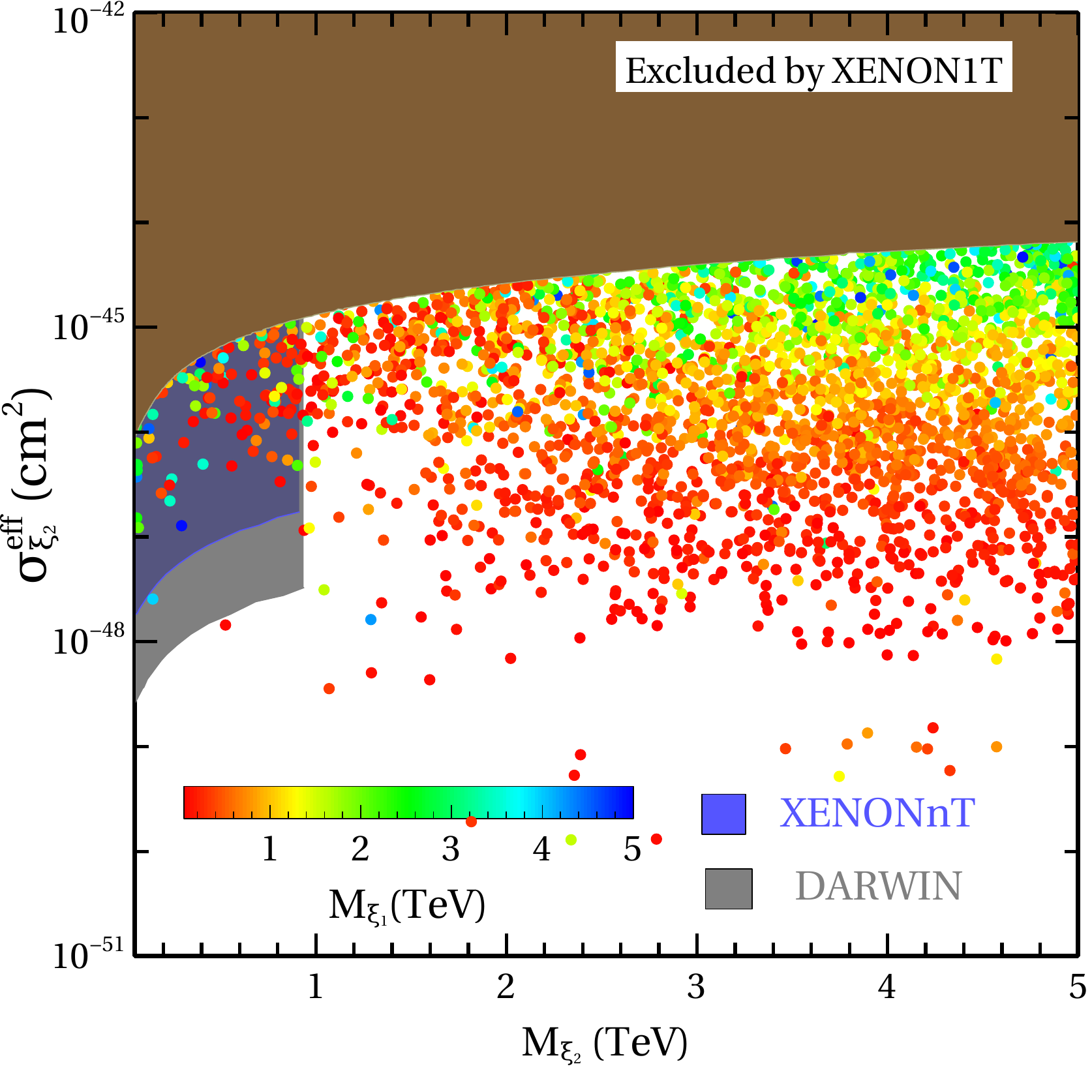}
\caption{Effective spin-independent scattering cross section off
nucleons for individual DM candidates. All the points satisfy total
DM relic, perturbativity, bounded from below criteria of the scalar potential.
{The projected sensitivities of XENONnT (blue region) and
DARWIN (gray region) experiments have been shown in both the plots.}}
\label{Fig:DD}
\end{figure}

\begin{figure}[h!]
\centering
\includegraphics[scale=0.5]{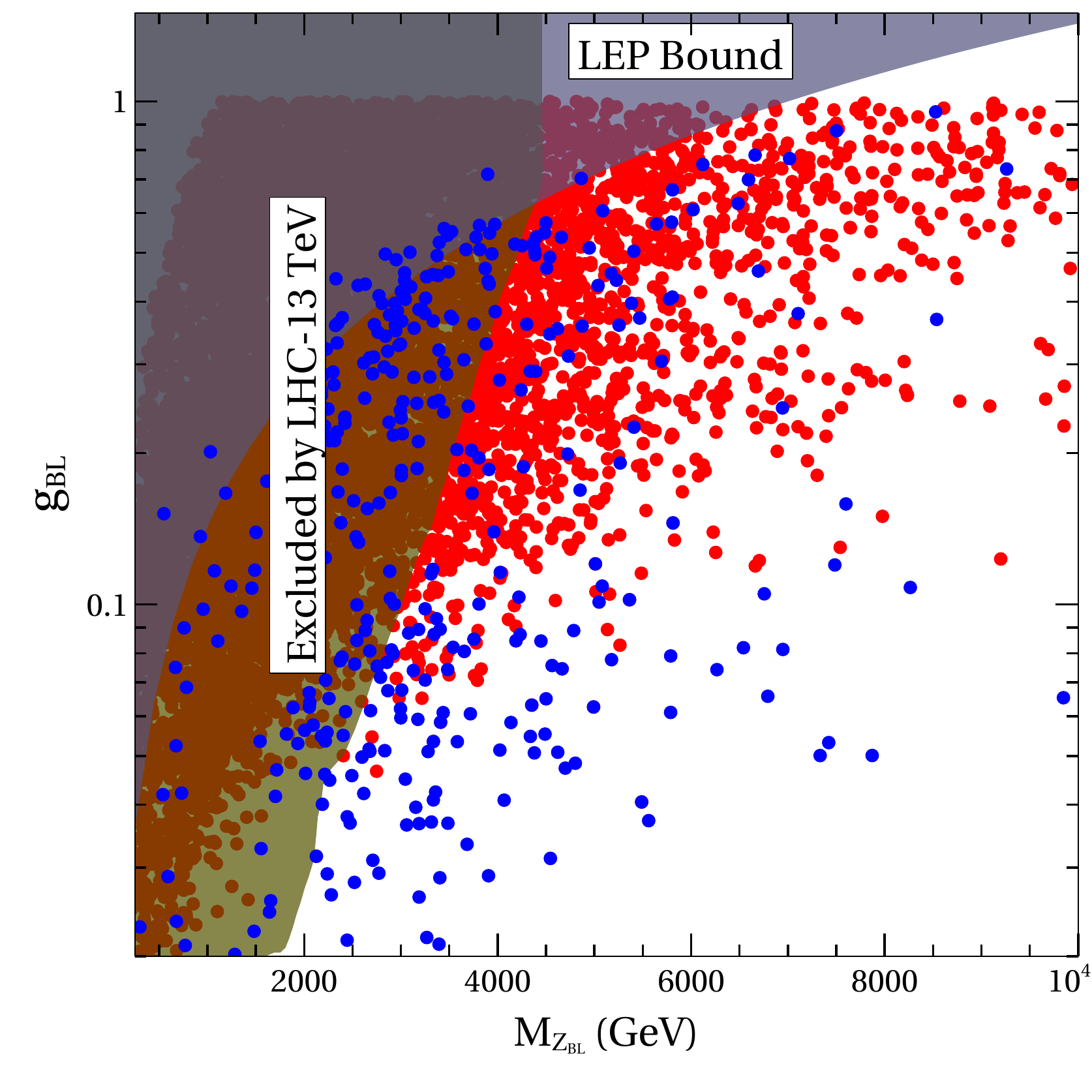}
\caption{Summary plot showing the allowed points with and without
applying the direct detection bounds from XENON1T experiment.}
\label{Fig:DD2}
\end{figure}

In figure \ref{Fig:DD}, we show the spin independent DM-nucleon scattering cross section for individual DM candidates as functions of their mass. All the points satisfy the total DM relic, perturbativity, bounded from below criteria of the scalar potential. {We have also shown the projected sensitivities of XENONnT (blue region) and DARWIN (grey region) experiments in both the plots which clearly show that these two future experiments can probe a large region of parameter space of the model.} As mentioned earlier, the actual scattering cross section is multiplied by individual relative number densities $n_{\xi_{1,2}}/(n_{\xi_1}+n_{\xi_2})$ in order to compare with the XENON1T bounds \cite{Aprile:2018dbl} derived for single DM component. This clearly shows that the model remains very much sensitive to the direct detection experiments with many parts of parameter space already being ruled out. We then superimpose the collider bounds also on these points and show the resulting parameter space in figure \ref{Fig:DD2}. The severe impact of XENON1T constraints is clearly visible here with a very few points allowed compared to the points allowed without applying the XENON1T bounds. Although the individual DM-nucleon scattering bound allowed several points in the parameter space to survive, as seen from figure \ref{Fig:DD}, when we impose the condition that both the DM candidates must satisfy the XENON1T bounds, it results in a much smaller allowed parameter space, seen in figure \ref{Fig:DD2}. 
{This may be due to the fact that
the elastic scatterings of both dark matter components ($\xi_1$ and $\xi_2$)
with the detector nuclei occur predominantly through $Z_{BL}$ exchange and hence, $\sigma^{\rm SI}_{\xi_i}$ ($i=1$, 2)
are extremely sensitive to both $g_{BL}$ and $M_{Z_{BL}}$. On the other hand,
$Z_{BL}$ as well as scalar bosons mediated annihilation processes have significant
impact in the relic densities of $\xi_1$ and $\xi_2$. Therefore,
in the latter case we have more parameters (Yukawa couplings, $g_{BL}$ etc.)
and consequently the large portion of $g_{BL}-M_{Z_{BL}}$ plane is allowed.}
The allowed points in the chosen range of random scan will face further constraints from future LHC runs as well as direct detection experiments like LZ~\cite{Akerib:2015cja}, XENONnT~\cite{Aprile:2015uzo}, DARWIN~\cite{Aalbers:2016jon} and PandaX-30T~\cite{Liu:2017drf}.

\section{LHC signatures of fermion triplets}
\label{sec:LHC}
Although typical LHC signatures of a $U(1)_{B-L}$ model is via search for dilepton resonance mediated by $Z_{BL}$ mentioned before, the present model can have additional prospects of being discovered at the LHC due to the presence of triplet fermions. While the neutral components of fermion triplets play the role of generating light neutrino masses through type III seesaw mechanism, the charged components can leave interesting signatures at colliders. {Although a detailed study of collider aspects of such fermion triplets is beyond the scope of this present work and can be found elsewhere, we briefly highlight the additional advantage of having such triplets in a gauged $B-L$ model. Detailed collider studies of charged components of fermion triplet can be found in the context of long lived wino in supersymmetric models; for example, see \cite{Rolbiecki:2015gsa} and references therein. In the context of non-supersymmetric models, collider studies of such long lived fermion triplets can be found in \cite{Kadota:2018lrt, Jana:2019tdm} and references therein.} We consider the third fermion triplet and its charged component as it can have enhanced production cross section at colliders due to larger $B-L$ charge. It should be noted that unlike in usual type III seesaw model, here the charged components of fermion triplets can be produced at the LHC via both $Z$ and $Z_{BL}$ gauge bosons, apart from usual photon mediation. The charged components can then decay into (i) charged lepton and Higgs boson or (ii) neutral fermion triplet plus on or off-shell $W$ boson depending upon the mass splitting $\Delta M \gtrsim 80~\rm GeV$ or $\Delta M \lesssim 80~\rm GeV$. Although the components of a particular fermion triplet have degenerate masses ($M_{k0}$) at tree level, one can make the
charged components $\psi^{\pm}_k$ heavier by considering one loop
electroweak radiative corrections \cite{Cirelli:2005uq, Ma:2008cu}, which result in
a mass splitting $\Delta M \sim 166$ MeV between $M_{\psi^{\pm}_k}$ and
$M_{\psi^0_k}$ for $M_{k0}\gtrsim1\,{\rm TeV}$. Thus, in the second possible decay channel mentioned above, only the off-shell $W$ boson is possible. The first decay mode is typically sub-dominant by the constraints from light neutrino masses, which require TeV scale fermion triplet Yukawa couplings with the SM leptons to be as small as around $10^{-6}-10^{-5}$. Therefore we consider the second decay mode only where the charged component of triplet decays into the neutral component and an off-shell $W$ boson. Due to the small mass splitting, the dominant decay mode has final states $\psi^0_k, \pi^{\pm}$. The decay width of $\psi_3^\pm$ to $\psi_3^0$ and $\pi^\pm$ can be written as 
\begin{eqnarray}
\Gamma_{\psi_3^\pm \rightarrow \psi_{3}^0 \pi^\pm}&=&\dfrac{g^4\,f^2_{\pi}\,\,V^2_{ud}}
{128\pi\,M^4_W\,M_{\psi^\pm_3}} \Delta{M^2}
\left((M_{\psi_{3}^0}+M_{\psi_3^\pm})^2
-M^2_{\pi}\right)\times \nonumber \\ &&
\sqrt{1-\dfrac{(M_{\psi_{3}^0}-M_{\pi})^2}{M^2_{\psi_3^\pm}}}
\sqrt{1-\dfrac{(M_{\psi_{3}^0}+M_{\pi})^2}{M^2_{\psi_3^\pm}}}\,\,,
\label{psidecay}
\end{eqnarray}
where $\Delta{M}=M_{\psi_3^\pm}-M_{\psi_{3}^0}\simeq$ 166 MeV,
$g$ is the SU(2)$_{\rm L}$ gauge coupling, $f_{\pi}=131$ MeV \cite{Cirelli:2005uq} is the pion decay constant, and the first diagonal element of the CKM matrix $V_{ud}\simeq0.974$ respectively. Such tiny decay width keeps the lifetime of $\psi^{\pm}_3$ considerably long enough so that it can reach the detector before decaying. In fact, the ATLAS experiment at the LHC has already searched for such long-lived charged particles with lifetime ranging from 10 ps to 10 ns, with maximum sensitivity around 1 ns \cite{Aaboud:2017mpt}. In the decay $\psi_3^{\pm}\rightarrow \psi^0_3\,\pi^\pm$, the final state pion typically has very low momentum and it is not reconstructed in the detector. On the other hand  $\psi^0_3$ is a long lived particle as it can decay to SM leptons through the small mixing with the other triplet fermions and leaves the detector without interacting. Therefore, it gives rise to a signature where a charged particle leaves a track in the inner parts of the detector and then disappears leaving no tracks in the portions of the detector at higher radii.

\begin{figure}[h!]
\centering
\includegraphics[scale=0.45]{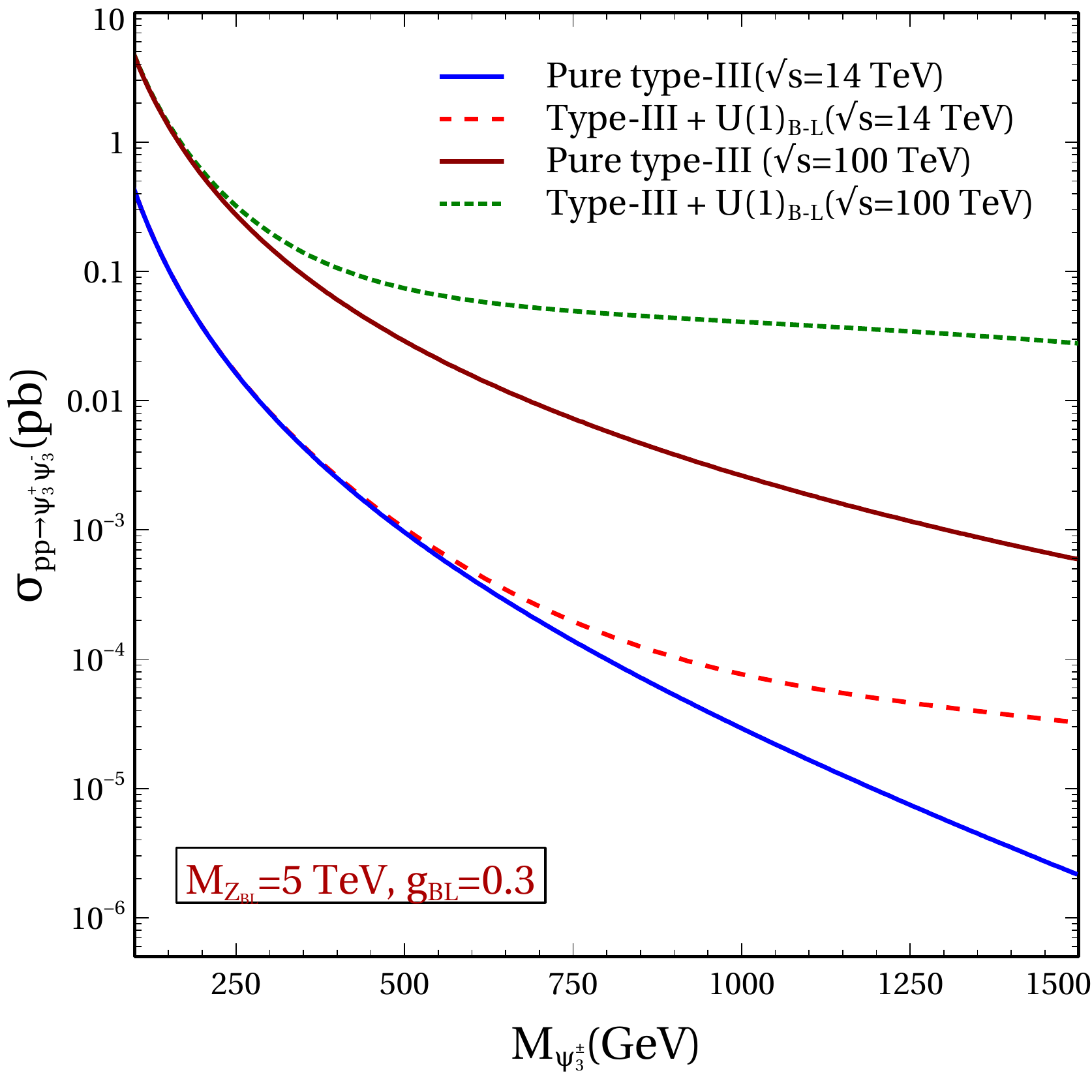}
\,\,
\includegraphics[scale=0.45]{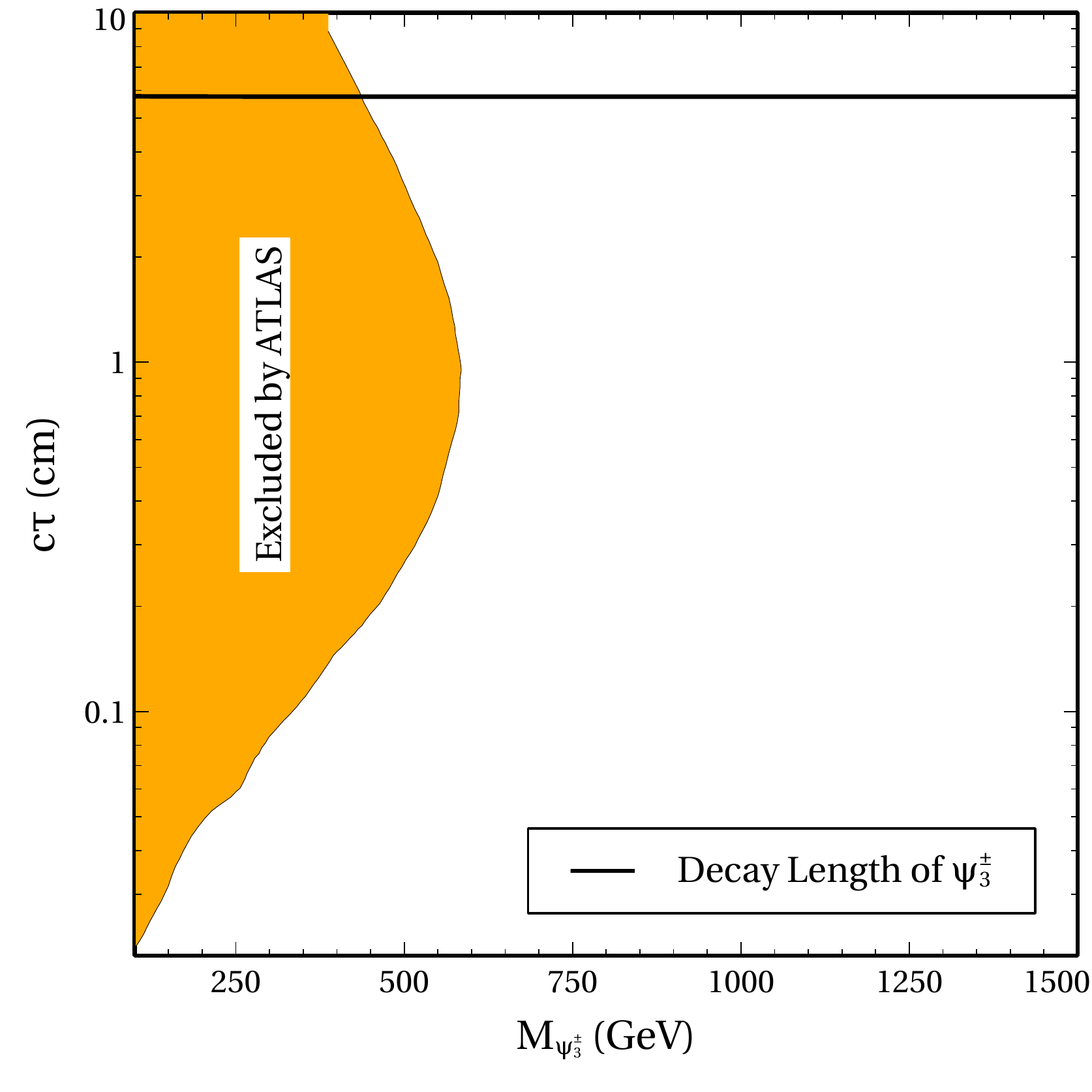}
\caption{Left panel: Plot showing improvement in production cross-section of the $\psi^{\pm}_3$ pairs due to $Z_{BL}$ mediation for two choices of centre of mass energies in proton proton collisions. Right panel: Decay length of $\psi_3^{\pm}$ versus its mass compared with the ATLAS bound on disappearing charge track searches at 13 TeV centre of mass energy.}
\label{Fig:LHC}
\end{figure}

Another crucial difference from usual type III seesaw extension of the SM is that here the production cross section of fermion triplets gets enhanced. The presence of $Z_{BL}$ significantly increases the production cross-section of the charged $\psi_3^{\pm}$ pairs at the collider. This is due to the fact that the decay of the BSM neutral gauge boson to the $\psi_3^{\pm}$ now happens on-shell in contrast to models where this decay takes place off-shell via SM $Z$ boson and photon. Also, as there is no negative interference between the $Z$ and $Z_{BL}$ mediated charged $\psi_3^{\pm}$ production channels, hence the addition of new channel always improves the production cross-section. In order to illustrate this improvement, we first show the variation of production cross-section $\sigma_{pp\to\psi^+_3 \psi^-_3}$ in the left panel of figure \ref{Fig:LHC} for two different choices of centre of mass energy. As can be seen, the improvement in production cross section is more significant in 100 TeV centre of mass energies of proton proton collisions. Once produced, $\psi_3^{\pm}$ can give rise to disappearing charged track signatures mentioned above. The ATLAS experiment at the LHC put constraints on such disappearing charged track signatures for a long lived chargino decaying into a pion and wino dark matter, which is shown as the solid black line in right panel of figure \ref{Fig:LHC}. Although $\psi^0_3$ is not the DM candidate in our model, it effectively behaves like one at the LHC due to its long life. It can be seen that the existing LHC constraint can already rule out $\psi^{\pm}_3$ masses below 500 GeV from its searches for disappearing charged tracks, keeping the parameter space considered in this study within near future sensitivity. Also, the bounds will get slightly tighter in our model due to enhanced production cross section. We leave a detailed study of such type III seesaw signatures at the LHC in the framework of a $B-L$ gauge model.

\begin{table}[h]
\centering

\begin{tabular}{|c||c|}
 \hline
 \multicolumn{2}{|c|}{{\bf ${\rm M_{Z_{BL}}=5 TeV\, , g_{BL}=0.3\, , M_{\psi_1}=2.5 TeV\,,M_{\psi_2}=2TeV\,, M_{s_i}=1 TeV\, , M_{A_{1}}=10 TeV}$}} \\
 \hline
Case - I & ${\rm M_{\psi_3}=M_{\xi_i}=(600 GeV - 2500 GeV)}$ \\
\hline
Case - II & ${\rm M_{\xi_i}=3 TeV}$,  ${\rm M_{\psi_3}=(600 GeV - 2500 GeV)}$ \\
\hline
Case - III & ${\rm M_{\psi_3}=2.6 TeV}$, ${\rm M_{\xi_i}=(600 GeV - 2500 GeV)}$ \\
\hline
\end{tabular}
\caption{Benchmark scenarios considered for calculating decay widths of $Z_{BL}$.}
\label{tab:2A}

\end{table}

\begin{table}[H]
\centering
\begin{tabular}{|l||l|l|l|}
 \hline
     \multirow{2}{*}{Mass (GeV)} &
 \multicolumn{3}{|c|}{Total Decay Width $ \Gamma_{Z_{BL}}$ (GeV)}  \\
 \cline{2-4}

 & {\bf Case-I} & {\bf Case -II} & {\bf Case -III}\\
\hline
600 & 249.2 & 181.2& 177.6\\ 
\hline
700 & 249.0 & 181.1& 177.6\\ 
\hline
800 & 248.8 & 181.0& 177.4\\ 
\hline
900 & 248.4 & 180.8& 177.3\\ 
\hline
1000 &  247.9 &180.6 & 177.0\\ 
\hline
1100 & 247.2 & 180.2& 176.7\\ 
\hline
1200 & 246.3 & 179.7 & 176.2\\ 
\hline
1300 & 245.1& 179.1 & 175.6\\ 
\hline
1400 & 243.5& 178.3 & 174.9\\ 
\hline
1500 & 241.5& 177.3 & 173.9\\ 
\hline
1600 & 239.0& 176.0& 172.7\\ 
\hline
1700 & 235.7& 174.3 & 171.1\\ 
\hline
1800 & 231.7 & 172.3 & 169.1\\ 
\hline
1900 & 226.7 & 169.7 & 166.7\\ 
\hline
2000 & 220.3 & 166.4 & 163.6\\ 
\hline
2100 & 212.2 & 162.2 & 159.6\\ 
\hline
2200 & 201.7 & 156.9 & 154.5\\ 
\hline
2300 & 187.6 & 149.6 & 147.6\\ 
\hline
2400 & 166.8 & 139.0 & 137.5\\ 
\hline
2500 & 109.7 & 109.7 & 109.7\\ 
\hline
\end{tabular}
\caption{Decay widths of $Z_{BL}$ for different benchmark
values of DM and fermion triplet masses.}
\label{tab:2B}
\end{table}

{It should be noted that here we have adopted the narrow width approximation for $Z_{BL}$ which implies $\Gamma_{Z_{BL}}/M_{Z_{BL}} \ll 1$. To justify this, we consider three benchmark scenarios namely, (i) $Z_{BL}$ can decay into both the DM candidates as well as the fermion triplet, (ii) $Z_{BL}$ can decay into the fermion triplet but not to the DM candidates, (iii) $Z_{BL}$ can decay into both the DM candidates but not to the fermion triplet. The corresponding widths, considering ${\rm M_{Z_{BL}}=5 TeV\, , g_{BL}=0.3}$ are shown below in Tables \ref{tab:2A} and \ref{tab:2B} respectively. As can be seen from these results, even for the maximum possible decay width of $Z_{BL}$ we can have $\Gamma_{Z_{BL}}/M_{Z_{BL}} < 5\%$. As discussed in several earlier works including \cite{Accomando:2013sfa, Accomando:2019ahs} and references therein, narrow width approximation remains valid for such small decay widths.}

\section{Conclusion}
\label{sec:conclude}
We have proposed a gauged $U(1)_{B-L}$ version of type III seesaw model which naturally predicts a two component Dirac fermion dark matter scenario due to the requirements of anomaly cancellation. Unlike type I seesaw scenario in $U(1)_{B-L}$ model where three right handed singlet neutrinos with $B-L$ charge $-1$ each leads to cancellation of all anomalies without any need of additional chiral fermions, type III seesaw implementation leads to additional anomalies due to the non-trivial $SU(2)_L$ structure of triplet fermions. We show that a gauged $U(1)_{B-L}$ model with three fermion triplets required for type III seesaw can be anomaly free due to the presence of two neutral Dirac fermions having fractional $B-L$ charges. Both of these fermions are naturally stable due to a remnant $\mathbb{Z}_2 \times \mathbb{Z}'_2$ symmetry. We study the DM phenomenology of the two component DM scenario in the model after incorporating all relevant theoretical and experimental bounds. Among the theoretical bounds, the bounded from below criteria of the scalar potential plays a crucial role restricting the parameter space. The relic abundance of the DM candidates are primarily dictated by their annihilations into SM particles mediated by $Z_{BL}$ as well as additional singlet scalars. We find the parameter space allowed from total relic abundance criteria and then apply the bounds from direct detection experiments on individual DM candidates. We find that all these constraints tightly constrain the parameter space which we scan through in our study, leaving a small region which can be probed at experiments operational at different frontiers. Finally we comment upon one interesting prospect of probing this model through production of charged components of fermion triplets at colliders. The presence of additional neutral gauge boson $Z_{BL}$ can significantly enhance the production cross section of charged triplets at LHC or 100 TeV future proton proton collider compared to usual type III seesaw extension of standard model. The charged components can then decay into the neutral ones and a pion with a long decay lifetime, leaving a disappearing charged track signature. Comparing with the ATLAS bounds on such signatures, we find that such triplets with masses upto 500 GeV or so can already be ruled out. While the charged components of fermion triplets can be probed via such disappearing charged track signatures, the neutral components, if long lived enough could be probed at proposed future experiments like MATHUSLA \cite{Curtin:2018mvb}. These triplets, however, do not affect the DM phenomenology much apart from opening another DM annihilation channel to triplet fermions whenever allowed kinematically.

\acknowledgements
DB acknowledges the support from IIT Guwahati start-up grant (reference number: xPHYSUGI-ITG01152xxDB001), Early Career Research Award from DST-SERB, Government of India (reference number: ECR/2017/001873) and Associateship Programme of Inter University Centre for Astronomy and Astrophysics (IUCAA), Pune. DN  would  like  to  thank Rashidul Islam and Claude Duhr for various discussions regarding the implementation of the model in FeynRules. He would also like to thank Lopamudra Mukherjee and Suruj Jyoti Das for for providing some computational resources.
\newpage
\appendix
\section{Diagonalisation of mass matrix of real scalars}
\label{rs_matrix_diagonalisation} 
In this section we have listed all the elements
of the diagonalising matrix $\mathcal{O}$ of $\mathcal{M}_{rs}$ in terms
of mixing angles under the assumption that all the mixing angles
among the singlet scalars are identical and equal to $\theta_{24}$:
\begin{eqnarray*}
\mathcal{O}_{11}&=& c_{12}\ c_{13}\ c_{14},\\
\mathcal{O}_{12}&=& c_{13}\ c_{14}\ s_{12}, \\
\mathcal{O}_{13}&=& c_{14}\ s_{13},\\
\mathcal{O}_{14}&=& s_{14},\\
\mathcal{O}_{21}&=& -c_{12}\ c_{13}\ s_{14}\ s_{24}-c_{12}\ c_{24}\ s_{13}\ s_{24}-c_{24}^2\ s_{12},\\
\mathcal{O}_{22}&=& c_{12}\ c_{24}^2-c_{13}\ s_{12}\ s_{14}\ s_{24}-c_{24}\ s_{12}\ s_{13}\ s_{24},\\
\mathcal{O}_{23}&=& c_{13} \ c_{24} \ s_{24}-\ s_{13} \ s_{14} \ s_{24}, \\
\mathcal{O}_{24}&=& c_{14} \ s_{24},\\ 
\mathcal{O}_{31}&=& -c_{12} \ c_{13}\ c_{24}\ s_{14}\ s_{24}-c_{12}\ c_{24}^2\ s_{13}
+c_{12}\ s_{13}\ s_{24}^3+c_{24}\ s_{12}\ s_{24}^2+c_{24}\ s_{12}\ s_{24},\\ 
\mathcal{O}_{32}&=& -c_{24}\ \left(c_{12} \ s_{24}^2+c_{24}\ s_{12}\ s_{13}\right)-c_{12}\ c_{24}\ s_{24}
+s_{12}\ s_{24} \left(s_{13}\ s_{24}^2- c_{13}\ c_{24}\ s_{14}\right),\\
\mathcal{O}_{33}&=& c_{13}\ c_{24}^2-c_{13}\ s_{24}^3-c_{24}\ s_{13}\ s_{14}\ s_{24},\\
\mathcal{O}_{34}&=& c_{14}\ c_{24}\ s_{24},\\
\mathcal{O}_{41}&=&s_{12}\ (c_{24}^2\ s_{24}-s_{24}^2)+c_{12}\ (-c_{13}\ c_{24}^2\ s_{14}+s_{13}\ (c_{24}\ s_{24}+c_{24}\ s_{24}^2)),\\
\mathcal{O}_{42}&=&-c_{13}\ c_{24}^2\ s_{12}\ s_{14}+c_{12}\ (-c_{24}^2\ s_{24}+s_{24}^2)+s_{12}\ s_{13} (c_{24}\ s_{24}+c_{24}\ s_{24}^2),\\
\mathcal{O}_{43}&=&-c_{24}^2\ s_{13}\ s_{14}-c_{13} (c_{24}\ s_{24}+c_{24}\ s_{24}^2),\\
\mathcal{O}_{44}&=&c_{14}\  c_{24}^2\,.
\end{eqnarray*}
Here, we have denoted $c_{ij}\equiv \cos \theta_{ij}$ and
$s_{ij}\equiv \sin \theta_{ij}$ respectively.
Now, using the elements of the diagonalising matrix, the physical scalar
fields can be expressed as a linear combinations of unphysical
states as 
\begin{eqnarray}
h=\mathcal{O}_{11} h^{\prime} + \mathcal{O}_{21} s_{1}^\prime + \mathcal{O}_{31} s_{2}^\prime
+ \mathcal{O}_{41} s_{3}^\prime\,,\\
s_{1}=\mathcal{O}_{12} h^{\prime} + \mathcal{O}_{22} s_{1}^\prime + \mathcal{O}_{32} s_{2}^\prime
 + \mathcal{O}_{42} s_{3}^\prime\,,\\
s_2=\mathcal{O}_{13} h^{\prime} + \mathcal{O}_{23} s_{1}^\prime + \mathcal{O}_{33} s_{2}^\prime 
+ \mathcal{O}_{43} s_{3}^\prime\,,\\
s_3=\mathcal{O}_{14} h^{\prime} + \mathcal{O}_{24} s_{1}^\prime + \mathcal{O}_{34} s_{2}^\prime 
+ \mathcal{O}_{44} s_{3}^\prime\,.
\end{eqnarray}
\section{Diagonalisation of mass matrix of pseudo scalars}
\label{ps_matrix_diagonalisation}
The pseudo scalar mass matrix with respect to
the basis state $\frac{1}{\sqrt{2}}
\left(A^\prime_1 \,\,A^\prime_2\,\,A^\prime_3\right)^{T}$
is given by
{\footnotesize{
\begin{eqnarray*}
\mathcal{M}_{ps} =
\left(
\begin{array}{ccc} 
-2u\,\left(u\,\beta + \sqrt{2}\,\delta\right) &  u^2\,\beta &
u\left(-u\,\beta +\sqrt{2}\,\delta\right)\\
 u^2\,\beta & -\dfrac{u}{2}\left(u\,\beta + \sqrt{2}\,\zeta \right) &
 \dfrac{u}{2}\left(u\,\beta + 2\sqrt{2}\,\zeta \right)\\
u\left(-u\,\beta +\sqrt{2}\,\delta\right) & \dfrac{u}{2}\left(u\,\beta
+ 2\sqrt{2}\,\zeta \right) & -\dfrac{u}{2}\left(u\,\beta + \sqrt{2}(\delta + 4\,\zeta)
\right)\\
\end{array}
\right)\,.
\end{eqnarray*}}}
As expected, the pseudo scalar mass matrix is also real symmetric.
Therefore, the matrix $\mathcal{M}_{ps}$ can be diagonalised by
a $3\times 3$ orthogonal matrix and the corresponding eigenvalues
are given below,
\begin{eqnarray}
m^2_{A_1} &=& 0\,, \nonumber \\
m^2_{A_2} &=&   \frac{-6 \beta  u^4-5 \sqrt{2} u^3 (\delta +\zeta
   )+\sqrt{2} \sqrt{u^6 \left(25 \delta ^2-34\,\delta \zeta
   +25 \zeta ^2+18 \beta ^2 u^2-12 \sqrt{2} \beta  u
   (\delta +\zeta )\right)}}{4 u^2}\,,\nonumber \\
m^2_{A_3} &=& -\frac{6 \beta  u^4+5 \sqrt{2} u^3 (\delta +\zeta
   )+\sqrt{2} \sqrt{u^6 \left(25 \delta ^2-34\,\delta\zeta
   +25 \zeta ^2+18 \beta ^2 u^2-12 \sqrt{2} \beta u
   (\delta +\zeta )\right)}}{4 u^2}\,. \nonumber \\ 
\end{eqnarray}  
\section{Collision term of the Boltzmann equation for dark matter conversion
process ${\xi}_2 \bar{\xi}_2 \rightarrow {\xi}_1 \bar{\xi}_1$}
\label{coupled-term-BE}
In this section, we have derived the Boltzmann equation for an interaction
which converts one type of dark matter into another i.e.
${\xi}_2 \bar{\xi}_2 \rightarrow {\xi}_1 \bar{\xi}_1$. In this
process, both initial and final states particles are dark matter candidates
which are not in thermal equilibrium during and after their freeze-out. The
situation is completely different when each type of dark matter annihilates
into the SM particles in thermal equilibrium. The collision term of
the Boltzmann equations depends strongly on whether the final state particles
in a scattering process are in thermal contact with thermal bath of the
Universe or not. This actually modifies the form of the collision term
which reduces to the known form when outgoing particles are in thermal
equilibrium. The contribution of the process
${\xi}_2 \bar{\xi}_2 \rightarrow {\xi}_1 \bar{\xi}_1$
to the collision term of $\xi_2$ is given by
\begin{eqnarray}
&&- \dfrac{g_{\xi_2}}{(2\pi)^3}\,\int \dfrac{\mathcal{C}[f_{\xi_2}]}{E_{\xi_2}}
\,d^3\vec{p}_{\xi_2}\,, \\
&=& -\int d \Pi_{\xi_2} d \Pi_{\bar{\xi}_2}
d \Pi_{\xi_1} d \Pi_{\bar{\xi}_1} (2\pi)^4\,
\delta^4(p_{\xi_2}+p_{\bar{\xi}_2}-p_{\xi_1}-p_{\bar{\xi}_1})
\left|\mathcal{M}_{{\xi}_2 \bar{\xi}_2 \rightarrow {\xi}_1 \bar{\xi}_1}
\right|^2 \times \nonumber \\
&& \Bigg(f_{\xi_2}(\left|\vec{p}_{\xi_2}\right|, T)\,
f_{\bar{\xi}_2}(\left|\vec{p}_{\bar{\xi}_2}\right|, T)
- f_{\xi_1}(\left|\vec{p}_{\xi_1}\right|, T)\,
f_{\bar{\xi}_1}(\left|\vec{p}_{\bar{\xi}_1}\right|, T)
\Bigg) \,,
\end{eqnarray}
where, $p_i$ is the four momentum of species $i$ while the
corresponding three momentum and energy are denoted by
$\vec{p}_i$ and $E_i$ respectively. The
phase space factor $d\Pi_i \equiv
\dfrac{g_i\,d^3\vec{p}_{i}}{(2\,\pi)^3E_i}$ with $g_i$ is the
internal degrees of freedom of species $i$ having distribution
function $f_i(\left|\vec{p}_{i}\right|, T)$. Here we have neglected
the factors related to Pauli blocking for fermions (dark matter candidates)
as those are not significant enough for the present context. Furthermore,
$\left|\mathcal{M}_{{\xi}_2 \bar{\xi}_2 \rightarrow {\xi}_1 \bar{\xi}_1}
\right|^2 $ is the Feynman amplitude square of the concerned process
${\xi}_2 \bar{\xi}_2 \rightarrow {\xi}_1 \bar{\xi}_1$ and
it is spin-averaged over both initial and final states particles.
The above collision term can also be as
\begin{eqnarray}
&& -\int d \Pi_{\xi_2} d \Pi_{\bar{\xi}_2}
\int d \Pi_{\xi_1} d \Pi_{\bar{\xi}_1} (2\pi)^4\,
\delta^4(p_{\xi_2}+p_{\bar{\xi}_2}-p_{\xi_1}-p_{\bar{\xi}_1})
\left|\mathcal{M}_{{\xi}_2 \bar{\xi}_2 \rightarrow {\xi}_1 \bar{\xi}_1}
\right|^2  \exp\left(-\frac{E_{\xi_2} + E_{\bar{\xi}_2}}{T}\right) \times \nonumber \\
&& \Bigg(\dfrac{f_{\xi_2}(\left|\vec{p}_{\xi_2}\right|, T)}{ \exp\left(-\frac{E_{\xi_2}}{T}\right)}\,
\dfrac{f_{\bar{\xi}_2}(\left|\vec{p}_{\bar{\xi}_2}\right|, T)}
{\exp\left(-\frac{E_{\bar{\xi}_2}}{T}\right)}
- \dfrac{f_{\xi_1}(\left|\vec{p}_{\xi_1}\right|, T)}
{\exp\left(-\frac{E_{\xi_1}}{T}\right)}\,
\dfrac{f_{\bar{\xi}_1}(\left|\vec{p}_{\bar{\xi}_1}\right|, T)}
{\exp\left(-\frac{E_{\bar{\xi}_1}}{T}\right)}
\Bigg) \,.
\label{intrmid1}
\end{eqnarray}
In the denominator of the last term, we have used the energy conservation
condition i.e. $E_{\xi_2} + E_{\bar{\xi}_2} = E_{\xi_1} + E_{\bar{\xi}_1}$. 
The distribution function of a dark matter species $\xi_i$ which
not in equilibrium (both thermally as well as chemically) can be
written as $f_{\xi_i} = \exp\left(\dfrac{\mu_{\xi_i}-E_{\xi_i}}{T}\right)$
with chemical potential $\mu_{\xi_i} \neq 0$ and $\mu_{\xi_i} =0$ represents
the equilibrium distribution function of $\xi_i$ which is just
$f^{\rm eq}_{\xi_i} =\exp\left(-\dfrac{E_{\xi_i}}{T}\right)$.
Therefore, one can easily show that $\dfrac{f_{\xi_i}}{f^{\rm eq}_{\xi_i}}=
\dfrac{n_{\xi_i}}{n^{\rm eq}_{\xi_i}}$, where the number density $n^{({\rm eq})}_{\xi_i}$
is defined as
\begin{eqnarray} 
n^{({\rm eq})}_{\xi_i} =
\dfrac{g_{\xi_i}}{(2\pi)^3} \int f^{({\rm eq})}_{\xi_i}\,d^3\vec{p}_{\xi_i}.
\end{eqnarray}
Using this, the terms within the brackets in Eq.\,\,(\ref{intrmid1})
can be replaced by number densities of $\xi_1$ and $\xi_2$ as  
\begin{eqnarray}
&& -\int d \Pi_{\xi_2} d \Pi_{\bar{\xi}_2}
\left\{
\int d \Pi_{\xi_1} d \Pi_{\bar{\xi}_1} (2\pi)^4\,
\delta^4(p_{\xi_2}+p_{\bar{\xi}_2}-p_{\xi_1}-p_{\bar{\xi}_1})
\left|\mathcal{M}_{{\xi}_2 \bar{\xi}_2 \rightarrow {\xi}_1 \bar{\xi}_1}
\right|^2 \right \} \times \nonumber \\ && 
\exp\left(-\frac{E_{\xi_2} + E_{\bar{\xi}_2}}{T}\right)
\Bigg(\dfrac{n_{\xi_2}}{n^{\rm eq}_{\xi_2}}\,
\dfrac{n_{\bar{\xi}_2}}{n^{\rm eq}_{\bar{\xi}_2}}
- \dfrac{n_{\xi_1}}{n^{\rm eq}_{{\xi}_1}}\,
\dfrac{n_{\bar{\xi}_1}}{n^{\rm eq}_{\bar{\xi}_1}}
\Bigg) \,.
\label{intrmid2}
\end{eqnarray}
Now, the quantity within the curly brackets is equal to $4\,\sigma {\rm v}\,
E_{\xi_2}E_{\bar{\xi}_2}$ with $\sigma$ being the annihilation cross section
of ${\xi}_2 \bar{\xi}_2 \rightarrow {\xi}_1 \bar{\xi}_1$. Substituting this in
the above we get
\begin{eqnarray}
&& -\left\{\dfrac{1}{{n^{\rm eq}_{\xi_2}\,n^{\rm eq}_{\bar{\xi}_2}}}
\int \dfrac{d^3\vec{p}_{\xi_2}}{(2\pi)^3}
\dfrac{d^3\vec{p}_{\bar{\xi}_2}}{(2\pi)^3} \left(\sigma {\rm v}
\right)_{{\xi}_2 \bar{\xi}_2 \rightarrow {\xi}_1 \bar{\xi}_1}
\exp\left(-\frac{E_{\xi_2} + E_{\bar{\xi}_2}}{T}\right)\right\} \times \nonumber \\
&& \,\,\,\,\,\,
\Bigg(n_{\xi_2}\,n_{\bar{\xi}_2}
- \dfrac{n^{\rm eq}_{\xi_2}}{n^{\rm eq}_{{\xi}_1}}\,
\dfrac{n^{\rm eq}_{\bar{\xi}_2}}{n^{\rm eq}_{\bar{\xi}_1}} n_{\xi_1}\,n_{\bar{\xi}_1}
\Bigg) \,,
\label{intrmid3}
\end{eqnarray}
where, the quantity within curly brackets is the thermal averaged cross section
$\langle {\sigma {\rm v}}_{{\xi}_2 \bar{\xi}_2 \rightarrow {\xi}_1 \bar{\xi}_1} \rangle$.
Therefore, in terms of the thermal averaged cross section, the collision term
of $\xi_2$ due to dark matter conversion process is given by
\begin{eqnarray}
 - \langle {\sigma {\rm v}}_{{\xi}_2 \bar{\xi}_2
\rightarrow {\xi}_1 \bar{\xi}_1} \rangle \Bigg(n_{\xi_2}\,n_{\bar{\xi}_2}
- \dfrac{n^{\rm eq}_{\xi_2}}{n^{\rm eq}_{{\xi}_1}}\,
\dfrac{n^{\rm eq}_{\bar{\xi}_2}}
{n^{\rm eq}_{\bar{\xi}_1}} n_{\xi_1}\,n_{\bar{\xi}_1} \Bigg) \,.
\end{eqnarray}
In this work, we have assumed that there is no asymmetry between
the number densities of dark matter candidate $\xi_i$ and its
antiparticle $\bar{\xi}_i$ and therefore, $n^{({\rm eq})}_{\xi_i}
= n^{({\rm eq})}_{\bar{\xi}_i}$. Hence, the contribution of
${\xi}_2 \bar{\xi}_2 \rightarrow {\xi}_1 \bar{\xi}_1$ to the
collision term of $\xi_2$ is given by
\begin{eqnarray}
 - \langle {\sigma {\rm v}}_{{\xi}_2 \bar{\xi}_2
\rightarrow {\xi}_1 \bar{\xi}_1} \rangle \Bigg\{n_{\xi_2}^2
- \left(\dfrac{n^{\rm eq}_{\xi_2}}{n^{\rm eq}_{{\xi}_1}}\right)^2
n_{\xi_1}^2 \Bigg\} \,,
\end{eqnarray}
while the collision term of $\xi_1$ also has an additional contribution
which is exactly identical to the above except the overall $-$ve sign will be
replaced by a $+$ve sign as the dark matter conversion process
${\xi}_2 \bar{\xi}_2 \rightarrow {\xi}_1 \bar{\xi}_1$ increases
the number densities of both $\xi_1$ and $\bar{\xi}_1$.


\providecommand{\href}[2]{#2}\begingroup\raggedright\endgroup

\end{document}